\documentclass[]{aa}
\usepackage{graphicx}
\usepackage{txfonts}
\usepackage[figuresright]{rotating}

\usepackage{natbib}
\bibpunct{(}{)}{;}{a}{}{,}
\begin{document}
   \title{The UV spectrum of HS~1700+6416}

   \subtitle{I. Predicting the metal line content of the far UV spectrum}

   \author{C. Fechner
          \inst{1}
          \and
          D. Reimers\inst{1}
	  \and
	  A. Songaila\inst{2}
	  \and
	  R.~A. Simcoe\inst{3}
	  \and
	  M. Rauch\inst{4}
	  \and
	  W.~L.~W. Sargent\inst{5}
          }

   \offprints{C. Fechner}

   \institute{Hamburger Sternwarte, Universit\"at Hamburg,
              Gojenbergsweg 112, 21029 Hamburg, Germany,\\
              \email{[cfechner,dreimers]@hs.uni-hamburg.de}
\and
	      Institute for Astronomy, University of Hawaii,
	      2680 Woodlawn Drive, Honolulu, HI 96822, USA
\and
	      MIT Center for Space Research, 
	      77 Massachusetts Ave. 37-664B, Cambridge, MA 02139, USA
\and
	      Carnegie Observatories,
	      813 Santa Barbara Street, Pasadena, CA 91101, USA
\and
	      Palomar Observatory, California Institute of Technology,
	      Pasadena, CA 91125, USA
             }

   \date{Received November 2, 2005; accepted May 9, 2006}

   \abstract{
In preparation of a study of the \ion{He}{ii}/\ion{H}{i} ratio towards the bright QSO \object{HS~1700+6416}, we predict the metal line content of the far-UV spectral range by modelling 18 metal absorption line systems with redshifts $0.2 < z < 2.6$ identified in the spectrum of this quasar.
For that purpose, we investigate the spectral energy distribution of the metagalactic ionizing radiation field.
Simple photoionization models based on 8 different shapes of the ionizing background are tested for each system.
The adopted energy distributions comprise the Haardt \& Madau (2001, HM) model of  metagalactic UV background as well as typical spectra of AGN and starburst galaxies.
The models are evaluated and the favoured one is estimated.
We find that the majority of the considered systems is best reproduced with a HM-like ionizing radiation, where the \ion{He}{ii} break, formally located at 4\,Ryd, is shifted to lower energies ($\sim$ 3\,Ryd), probably due to the opacity of the higher \ion{He}{ii} Lyman series lines.
The remaining systems can be reasonably described with models based on the unmodified HM background or the spectra of AGN or starburst galaxies.
This finding supports the idea that the UV background is spatially variable due to both IGM opacity variations and to local sources.
In comparison to an unmodified HM background, the resulting ionizing spectrum leads to carbon abundances lower by $\sim$ 0.5\,dex.
Furthermore, if the ionizing radiation field as determined from metal line systems was typical for the IGM, the expected \ion{He}{ii}/\ion{H}{i} ratio would be $150$ to $190$.

}
   \maketitle
%

\section{Introduction}

Observations with the Far UV Spectroscopic Explorer (FUSE) for the first time resolved the \ion{He}{ii} Ly$\alpha$ forest ($\lambda_0 = 303.7822\,\mathrm{\AA}$) towards two lines of sight, HE~2347-4342 \citep{krissetal2001, shulletal2004, zhengetal2004} and HS~1700+6416 \citep{reimersetal_fuse}.
Comparison to the \ion{H}{i} Ly$\alpha$ forest revealed strong variations of the \ion{He}{ii}/\ion{H}{i} ratio.
Since \ion{He}{ii}/\ion{H}{i} depends on the shape of the ionizing radiation, a strongly fluctuating metagalactic UV background is implied.
The radiation background is believed to be generated by the light of quasars and stellar sources such as star-forming galaxies reprocessed in the clumpy matter of the intergalactic medium \citep[IGM; e.g.][]{haardtmadau1996, fardaletal1998, haardtmadau2001}.
Due to the evolution of the sources, cosmic expansion, and the process of structure formation, even the UV background evolves with redshift.
Deviations from the mean metagalactic background may also be due to the spatial vicinity of local sources.
Observational hints for such effects are found by \citet{boksenbergetal2003} and \citet{levshakovetal2003a}, supported by recent theoretical work of \citet{schaye2004} and \citet{miraldaescude2005}.

Possible sources affecting the ionization conditions of intergalactic absorbers are quasars along the line of sight of the background QSOs as well as star-forming galaxies.
Searches for galaxies close to the sight lines of background quasars find a correlation between the Ly$\alpha$ forest and the galaxy density in the low redshift universe \citep[e.g.][]{bowenetal2002} as well as at high redshifts \citep{adelbergeretal2003}.
The authors conclude that both classes of objects trace the same large scale structure.
The vicinity of an absorber to a galaxy is supposed to make the galaxy's radiation dominate over the general UV background, at least if there are enough photons escaping from the galaxy \citep{bianchietal2001, steideletal2001}.

QSOs close to the line of sight are expected to produce a transverse proximity effect resulting in a reduction in the strength of the Ly$\alpha$ forest absorption due to the hard radiation of the QSO \citep{bajtliketal1988}.
This effect has been measured in the spectra of QSOs close to the emission redshift \citep[e.g.][]{scottetal2000}, but except in the quasar Q0302-003 \citep{dobrzyckibechtold1991, jakobsenetal2003} no transverse proximity effect could be detected so far \citep{schirberetal2004, croft2004}. 
The authors argue about an increase of the gas density in the vicinity of QSOs , anisotropy of the QSO emission, and time variability in order to explain the absence of the proximity effect.

In this paper, we present a study of the rich metal line spectrum of the QSO HS~1700+6416 ($z = 2.73$, $\alpha (2000.0) = 17^\mathrm{h}01^\mathrm{m}00\,\fs 6$, $\delta (2000.0) = +61\degr 12\arcmin 09\arcsec$).
This object is exceptionally bright in the optical ($V=16.1$).
Although it provides 7 optical thin Lyman limit systems (LLS) along the line of sight and is variable in the UV \citep{reimersetal2005b}, HS~1700+6416 is one of the few quasars, where \ion{He}{ii} is observable \citep{davidsenetal1996, reimersetal_fuse}.
Due to its brightness, HS~1700+6416 has been addressed in several analyses dealing with metal line systems \citep[e.g.][]{reimersetal1992, vogelreimers1993, vogelreimers1995, koehleretal1996, petitjeanetal1996, trippetal1997, simcoeetal2002, simcoeetal2005}, and due to its rich metal line spectrum, it has been target of deep direct observations in several spectral ranges aiming to identify objects which give rise to the Lyman limit absorption \citep{reimersetal1995, reimersetal1997b, teplitzetal1998, erbetal2003, shapleyetal2005}.
Our main objective is to predict the metal line content of the FUSE spectral range in the spectrum of this quasar, which will be considered in the analysis of the \ion{He}{ii} Ly$\alpha$ forest towards this quasar.
Since the available \ion{He}{ii} data is rather noisy ($S/N \approx 5 - 7$), the identification of narrow absorption features due to transitions of metal ions is nearly impossible.
Nevertheless, simple tests with simulated data indicate that the presence of unrecognized metal line absorption may bias the derived \ion{He}{ii}/\ion{H}{i} ratio \citep{fechnerreimers_fuse}.
Thus, an investigation of the metal line content in the corresponding spectral region ($1000-1180\,\mathrm{\AA}$) will improve the analysis of the \ion{He}{ii} forest. 

In order to model metal absorption line systems, photoionization calculations are performed.
One basic assumption concerns the shape of the ionizing energy distribution.
We use the modelling procedure to study the potential and restrictions of observed metal absorption systems with the aim to constrain the shape of the UV ionizing background.
Because of the numerous metal line systems in the spectrum and the many observations already made, HS~1700+6416 is highly eligible for this task.
As a first step, we compute simple photoionization models based on different ionizing radiation backgrounds for each system.
The resulting models are compared to the observed features and checked for the plausibility of the model parameters, deciding which one of the presumed energy distributions leads to the best description of the system.

Having presented the observations in Sect. \ref{observations} we describe the modelling procedure and its limitations in Sect. \ref{procedure} and \ref{limitations}.
The observed systems are outlined individually in Sect. \ref{systems} also presenting the models.
The results from the analysis of the whole sample and their implications are discussed in Sect. \ref{discussion}. 
Finally, the prediction for the metal lines in the FUSE spectral range is presented.

\section{Observations}\label{observations}

The optical data have been taken with the HIRES spectrograph at the Keck telescope.
We have got two datasets already published by \citet{songaila1998} and \citet{simcoeetal2002}.
These two datasets are co-added resulting in a spectrum with a total exposure time of 84\,200 s and a signal-to-noise of $S/N \sim 100$ at 4000\,\AA.
The co-added spectrum covers the wavelength range $3680 - 5880$\AA\, with a resolution of $R \sim 38\,000$.
The wavelength coverage of the \citet{simcoeetal2002} data goes down to $\sim 3220\,\mathrm{\AA}$ with decreasing $S/N$ and reaches upto $\sim 6140\,\mathrm{\AA}$.
To identify and model the metal line systems, we use the whole wavelength range.

The UV data are taken from the Hubble archive.
HS~1700+6416 was observed with STIS using the Echelle E140M grating ($R \sim 45\,800$) in 1998 July 23 -- 25.
The data cover the wavelength range $1142 - 1710\,\mathrm{\AA}$.
The exposure time of $\sim$ 76\,500\,s leads to a poor signal-to-noise of $\sim 3$.
Therefore, we only use the best portion of the spectrum ($1230 - 1550\,\mathrm{\AA}$).
Due to its rather limited quality, the STIS spectrum is used predominately for consistency checks.
Nevertheless, in case of the low redshift systems ($z \lesssim 1$), the vast majority of the absorption lines is located in the UV, and a more quantitative way to consider the corresponding transitions is necessary.
Therefore, the data are smoothed applying a Savitzky-Golay filter \citep[see e.g.][p. 650]{numericalrecipes} and fitted as described below.
However, we avoid to use the derived numbers, if possible.

The wavelength scales of the optical and the UV data have been checked to be well aligned.
Interstellar absorption of \ion{Na}{i} is detected in the optical at $v = (-35.0 \pm 0.1)\,\mathrm{km\,s}^{-1}$, in good agreement with interstellar absorption of \ion{S}{ii}, \ion{O}{i}, \ion{Si}{ii}, and \ion{Fe}{ii} measured in the UV. 

The HST archive contains further UV data taken with the instruments FOS and GHRS, respectively, presented by \citet{vogelreimers1995} and \citet{koehleretal1996}.
These datasets are in general less noisy ($S/N \gtrsim 10$) but since they have only medium resolution ($R \approx 1300$ and slightly higher), we reject them for this analysis, except for using the optical depths of the LLS.

All metal line systems have been identified by looking for features like the doublets of \ion{C}{iv} and/or \ion{Mg}{ii} in the optical and then searching for further transitions expected to arise in the optical and UV from the list of \citet{verneretal1994}.
The results have been compared with the identifications made in former work \citep{vogelreimers1995, koehleretal1996, petitjeanetal1996, trippetal1997, simcoeetal2002}.
The line parameters of the absorption features are estimated using the line fitting program CANDALF developed by R. Baade, which performs simultaneously a Doppler profile line fit and the continuum normalization.
The derived column densities are given in Tables \ref{coldens}, \ref{coldens2}, \ref{coldens3}, and  \ref{coldens4}.

\section{Modelling procedure}\label{procedure}

\begin{figure}
  \centering
  \resizebox{\hsize}{!}{\includegraphics[bb=35 220 485 775,clip=]{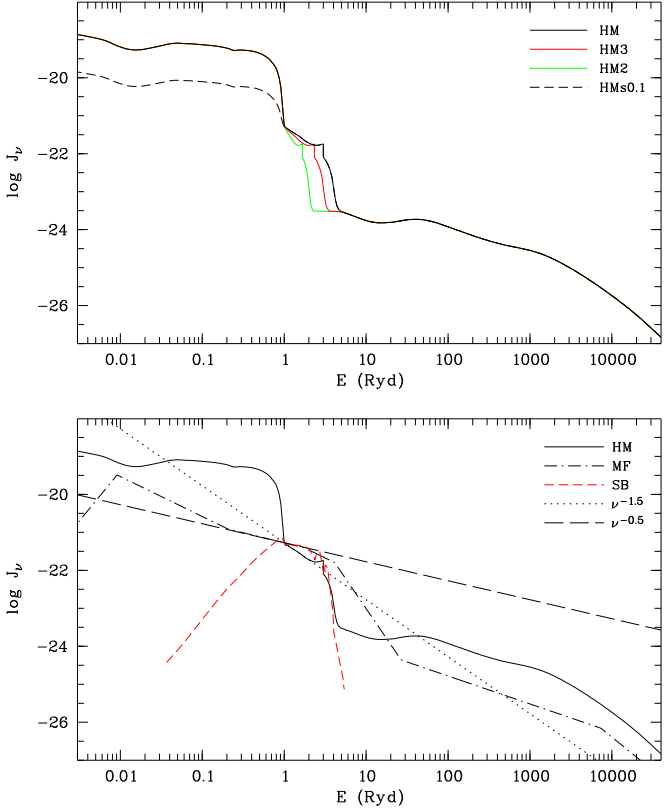}}
  \caption{The shapes of the ionizing radiation used for the presented investigation. The upper panel shows the UV background of \citet{haardtmadau2001} at $z \sim 2$ and the considered modifications with the 4\,Ryd break shifted to 3.0 and 2.0\,Ryd, respectively (solid lines). The dashed line represents the modification, where the low energy part ($<1$\,Ryd) is scaled by a factor 0.1.
The lower panel presents the energy distributions for a starburst galaxy \citep[][short dashed line]{bruzualcharlot1993}, an AGN \citep[][dot-dashed]{mathewsferland1987}, and power laws with $\alpha = -1.5$ (dotted) and $\alpha = -0.5$ (long dashed), respectively. The \citet{haardtmadau2001} background at $z \sim 2$ (solid line) is also given for a better orientation.
  }
  \label{spectra}
\end{figure}

In analyzing the observed systems we use the photoionization code CLOUDY \citep{ferland1997}. 
In order to investigate the shape of the ionizing radiation, eight different energy distributions are considered, among them the \citet{haardtmadau2001} background (HM) at the appropriate redshift.
Compared to the classical \citet{haardtmadau1996} radiation field the new version includes contributions of galaxies.
We use a model where the escape fraction of Lyman limit photons from a galaxy is $f_{\mathrm{esc}} = 0.1$.
In addition, we use 3 different types of modified HM continua.
One possible modification is a shift of the break at 4\,Ryd to lower energies. 
The filtered radiation might change this way, if absorption by the \ion{He}{ii} Lyman series becomes important in addition to the \ion{He}{ii} continuum at redshifts below 3.
We create modified Haardt-Madau energy distributions, shifting the 4\,Ryd break to 3.0 (HM3) and 2.0\,Ryd (HM2), respectively.
In a third modification, the plateau at energies $<1$\,Ryd is scaled with the factor 0.1 (HMs0.1).
Since at these low energies the background is dominated by the radiation of galaxies, a lower flux level mimics roughly a reduction of the fraction of the galaxy radiation.
All Haardt-Madau type spectra are shown in the upper panel of Fig. \ref{spectra} at a redshift $z \sim 2$.

Furthermore, typical spectral energy distributions of galaxies and quasars are used in order to examine the possible presence of local sources dominating the ionizing radiation.
Actually, the energy distribution of a local source and the general background should be merged to present a more realistic model \citep{boksenbergetal2003}.
Here, we start with the extreme assumption that the absorber is solely illuminated by the local source.
A model starburst galaxy with a constant starformation rate after 0.001\,Gyr given by \citet[][SB]{bruzualcharlot1993} is adopted as a pure galaxy spectrum.
The energy distribution of an AGN is taken from \citet[][MF]{mathewsferland1987} as it is implemented in CLOUDY.
For comparison pure power law models $f_{\nu} \propto \nu^{\,\alpha}$ with $\alpha = -1.5$ (PL15) and $-0.5$ (PL05), respectively, are adopted as well.

For each of the considered ionizing energy distributions and each system a grid of models is computed.
The model parameters are the observed \ion{H}{i} column density, the ionization parameter, which is chosen as $-4.0 \le \log U \le 0.5$, the metallicity in units of the solar value ($-3.0 \le \log (Z/Z_{\sun}) \le 0.0$), and the hydrogen density ($-5.0 \le \log n_{\mathrm{H}} \le 3.0$).
The models depend only weakly on the hydrogen density but its importance increases for high ionization parameters.
Solar metallicities are taken from \citet{grevessesauval1998} with the updates from \citet{holweger2001}.
The helium abundance is assumed to be primordial.

On the computed grids, the best fitting models are chosen by considering the most reliable observed column densities.
If possible we use the column density ratios of two ionizations stages of the same element, e.g. $N(\mathrm{\ion{C}{iv}})/\,N(\mathrm{\ion{C}{iii}})$.
The estimated grid parameters are the starting values for a detailed calculation, where the ionization parameter, the hydrogen density, and the metallicity are optimized, using the related CLOUDY command.
Deviations from the solar abundance pattern are also considered if suggested by the data with the purpose to recover the abundance pattern usually found for low metallicity gas.
Throughout this paper abundances are given in the common notation $\mathrm{[M/H]} = \log(\mathrm{M/H})_{\mathrm{obs}} - \log(\mathrm{M/H})_{\sun}$.

\begin{figure}
  \centering
  \resizebox{\hsize}{!}{\includegraphics[bb=28 105 550 770,clip=]{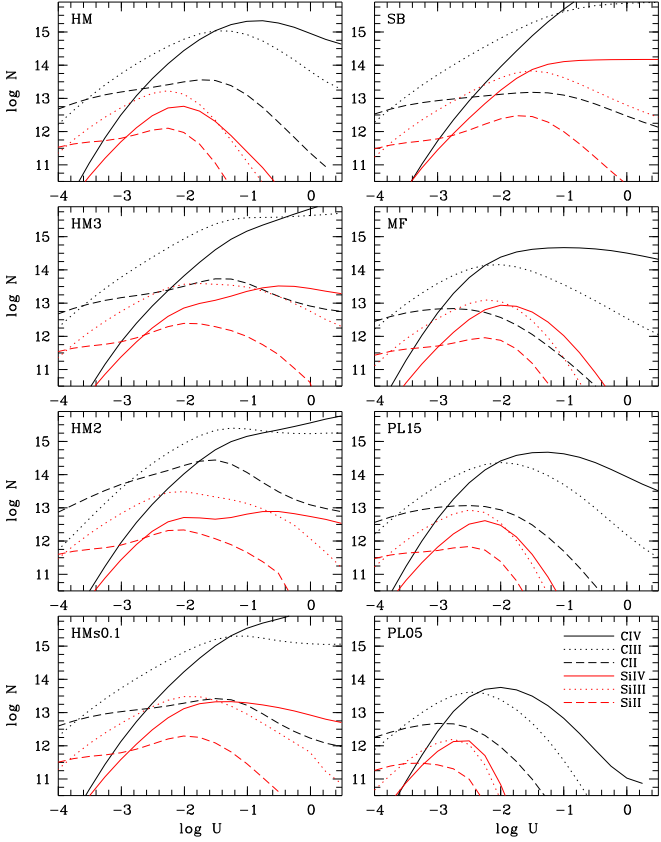}}
  \caption{Dependence of the column density on the ionization parameter $\log U$ of ions of carbon (\ion{C}{ii}, \ion{C}{iii}, \ion{C}{iv}) and silicon (\ion{Si}{ii}, \ion{Si}{iii}, \ion{Si}{iv}) for the ionizing spectra considered in the modelling procedure (see text). The legend is given in the lower right panel. The column density of neutral hydrogen in $\log N_{\ion{H}{i}} = 16.83$, the metallicity is $\mathrm{[M/H]} = -2.0$ and density $\log n_{\mathrm{H}} = -2.0$ for all models.
  }
  \label{colden_models}
\end{figure}
For illustrating the importance of the UV background, Fig. \ref{colden_models} presents how the column density of ions of carbon (\ion{C}{ii}, \ion{C}{iii}, \ion{C}{iv}) and silicon (\ion{Si}{ii}, \ion{Si}{iii}, \ion{Si}{iv}) changes with the ionization parameter for the considered ionizing energy distributions.
Particularly for $\log U \gtrsim -2.5$ significant differences are present, even if the UV background is only moderately changing as given in case of the HM-like spectra (left panels of Fig. \ref{colden_models}).  

As noted above, the spectra of the local sources realistically should be merged with a general background.
In order to estimate the uncertainties which are introduced by our extreme assumption of a pure SB energy distribution, the distribution of column densities for a pure SB model, as shown in the upper right panel of Fig.\ \ref{colden_models}, are compared to a \citet{haardtmadau2001} background merged with the energy distribution of a starburst galaxy (HM+SB).
We find that the column density distribution for the HM+SB energy distribution clearly resembles that of a pure SB continuum.
This is true especially for ions with ionization potentials $\lesssim$ 4\,Ryd at ionization parameters $\log U \lesssim -1.0$ which are realistically expected for intervening absorbers.
However, above 4\,Ryd the combined HM+SB energy distribution equals the pure HM background. 
Consequently, ions with higher ionization potentials, e.g. \ion{N}{v} and \ion{O}{vi} will follow a column density distribution very similar to the pure HM model if the absorber is exposed to a HM+SB radiation field. 
As a consequence our pure SB models will fail if highly ionized gas is involved.
The detailed discussion of the individual systems and models in Section \ref{systems} will show that there are only a few systems which favour the SB energy distribution as ionizing radiation.
We will discuss the reasonability of those models for the corresponding systems.

A synthetic spectrum based on the resulting parameters is computed for comparing the model with the data and evaluate the quality of the fit. 
For each transition, Doppler profiles are calculated, adopting the column density directly from the model result.
The Doppler parameter $b$ is derived from the model temperature as follows:
Using the Doppler parameter of one unblended well-measured line, the turbulent part of the $b$-parameter is computed using $b^2 = (2kT/m) + b_{\mathrm{turb}}^2$, where $T$ is the temperature obtained for the model, $m$ the mass of the considered element and $k$ Boltzmann's constant.
The estimated turbulent part $b_{\mathrm{turb}}$ is fixed for all lines, while the thermal contribution is re-computed for each element. 
The exact redshift of the lines is also adopted from one unblended, well-measured transition.
The resulting spectrum is then broadened to the resolution of the observed data.
Thus, a quantitative comparison with the Keck data can be performed.
Furthermore, the models are checked for consistency with the STIS data.

The evaluation of the goodness of fit is performed by estimating a $\chi^2$ as a quantitative indicator.
We use
\begin{equation}
  \chi^2 = \frac{1}{n}\sum_{i=1}^{n}\left(\frac{f_{\mathrm{obs},\,i} - f_{\mathrm{mod},\,i}}{\sigma_{i}}\right)^{2}\,\mbox{,}
\end{equation} 
where $f_{\mathrm{obs},\,i}$ and $f_{\mathrm{mod},\,i}$ is the observed and modelled flux at pixel $i$, respectively, $\sigma_{i}$ is the error of the observed flux, and $n$ is the number of pixels considered for the $\chi^2$-estimation.
Since lines arising in the STIS part of the spectrum would dominate the $\chi^2$, only features observed in the Keck spectrum are regarded. 
However, the actual $\chi^2$-value is often spoiled by a few pixels that are contaminated by problems with the noise array or simply line blends \citep[formerly noticed also by e.g.][]{charltonetal2003, dingetal2003, dingetal2003b, zonaketal2004}.
Even though we try to minimize the influence of bad pixel on the $\chi^2$ by selecting the lines and intervals appropriate to be used for each system individually, a visual inspection is needed.
Therefore, we will discuss the contribution of the individual profiles and the decide for a preferred model taking into account the $\chi^2$-values and the visual comparison of the models.

\section{Limitations}\label{limitations}

Many metal absorption line systems are multi-phase absorbers, which means low and high ionized absorption features are not formed in the same volume \citep[examples of well studied low redshift absorption systems can be found e.g. in][]{charltonetal2003, dingetal2003, dingetal2003b, dingetal2005, zonaketal2004, masieroetal2005}.
Especially \ion{O}{vi} is believed to arise in a spatially more extended, low density gas \citep[e.g.][]{lopezetal1999, simcoeetal2002}. 
Thus, it is clear that our models are only a first step.
More realistical descriptions should take into account the probably multi-phase nature of the absorbers.
In the simple models presented here, the modelled ionization levels partly depend on the strategy of the analysis.
Depending on the available observed line ratios either the low or the high ionization phase is reproduced.
The choice of the constraining ions depends on the observed spectrum.
In most of the cases \ion{C}{iii}/\ion{C}{iv} or \ion{Si}{iii}/\ion{Si}{iv} can be used since mostly \ion{C}{iv} and \ion{Si}{iv} are unambiguously detected and the column density can be estimated reliably even if the features are rather weak.
Contrarily, the features of the low ionization phase are more often located in the Ly$\alpha$ forest, where blending limits the measurability of column densities.
Nevertheless, in a few cases ratios like \ion{C}{ii}/\ion{C}{iii} or \ion{Si}{ii}/\ion{Si}{iii} are more prominent than the highly ionized species and therefore the low ionization component is modelled.
There are also systems, where \ion{C}{iii} and/or \ion{Si}{iii} cannot be detected because they are blended with strong or saturated Ly$\alpha$ forest lines.
In this case ratios like \ion{C}{ii}/\ion{C}{iv} or \ion{Si}{ii}/\ion{Si}{iv} have to constrain the model, which is extremely problematic since the ions may originate in distinct volumes.
We will discuss the implications of approaches like this in more detail when the respective systems are presented.
Furthermore, we will specify the validity of the models with regard to the ionization phase.

Another problem is the considered amount of neutral hydrogen.
Usually, the Ly$\alpha$ line is saturated and only in a few cases substructures in the \ion{H}{i} features become visible in higher order Lyman series.
For systems at $z < 1.5$, Ly$\alpha$ is located beyond the observed spectral range (except the $z=0.2140$ system), thus we adopt the column densities measured by \citet{vogelreimers1995} using the Lyman limit break.
Anyhow, following the described modelling procedure it is necessary to distribute the total \ion{H}{i} column density over the subcomponents observed in the metal lines.
This is done by following the distribution of appropriate metals.
If possible all ionization stage are considered, e.g. $N(\mathrm{C}_{\mathrm{tot}}) = N(\ion{C}{ii})+N(\ion{C}{iii})+N(\ion{C}{iv})$ or $N(\mathrm{Mg}_{\mathrm{tot}}) = N(\ion{Mg}{i})+N(\ion{Mg}{ii})$.
Due to incomplete coverage of the ionization stages, the distribution of neutral hydrogen according to this method is a considerable source of systematic error.
If only part of the measured \ion{H}{i} absorption is associated to the metal absorption, the absolute elemental abundances are underestimated.
However, the relative abundances are probably not affected.

But also handling relative abundances may cause problems since they depend on the shape of the ionizing background. 
The value of [Si/C] derived from observations is typically $\mathrm{[Si/C]} \lesssim 0.5$, consistent with the predictions in theoretical supernova yields from various progenitor properties \citep{woosleyweaver1995, hegerwoosley2002, chieffilimongi2004, umedanomoto2002, umedanomoto2005}.
However, since ionization corrections are needed, the $\mathrm{[Si/C]}$ derived from observations depends on the adopted UV background. 
\citet{aguirreetal2004} pointed out that $\mathrm{[Si/C]}$ gets higher for harder ionizing radiation.
They found $\mathrm{[Si/C]} \sim 1.5$ for an unrealistic hard UV background.
However, \citet{qianwasserburg2005} predict that $\mathrm{[Si/C]}$ can be as high as $\sim 1.3$, if the metals are produced by very massive stars ($>100\,M_{\sun}$) adopting a model of \citet{umedanomoto2002}.

Furthermore, only 8 discrete shapes of the UV background are considered.
This means, we follow a very simple approach to constrain the shape of the ionizing background from observed data.
A real fit would require some kind of algorithm \citep[a possible method has recently been introduced by][]{agafonovaetal2005}. 
Thus, our results do not represent the ionizing radiation most probably illuminating the absorbing gas.
But they provide models, which fit the observed data more accurately and reasonably than the models based on the other energy distributions considered here.

\section{Observed systems}\label{systems}

The spectrum of HS~1700+6416 is characterized by a huge amount of metal line absorption features.
This rich metal line spectrum has been studied in detail with medium resolution in the UV \citep{reimersetal1992, vogelreimers1995, koehleretal1996} and with high resolution in the optical \citep{petitjeanetal1996, trippetal1997, simcoeetal2002}.

Including the 7 LLS, we identify 25 metal line systems in the spectrum of HS~1700+6416. 
Three metal line systems located at $z=2.7124, 2.7164$, and $2.7443$ are apparently associated with the QSO. 
Each of them show a multicomponent velocity structure.
In 2 of them, \ion{N}{v}, a typical transition observed in associated systems, is detected and signs of partial coverage are seen.
The intergalactic absorption systems cover a redshift range $0.2 \lesssim z \lesssim 2.6$.
The 7 Lyman limits systems are located at $z=0.8643, 1.1573, 1.7241, 1.8450, 2.1680, 2.3155$, and $2.4331$.
Four of the intervening systems (at $z=2.0211,\, 2.1278,\, 2.1989,$ and $2.3079$) show only metal absorption features of \ion{C}{iv}.
Thus, it is impossible to constrain models.
However, the expected \ion{C}{iv} $\lambda\lambda 312.5,312.4$ lines can be computed directly from the estimated line parameters, and will be considered in the predicted metal line spectrum. 

In the following, we present the observations of the identified metal line systems.
The models are discussed and the most appropriate one is derived.
Figures of the data and the models can be found in Figs.\ \ref{z0.2140fig} to \ref{z2.5786fig}. 

\subsection{System $z = 0.2140$}

The lowest redshifted metal line absorber is located at $z = 0.2140$.
According to \citet{reimersetal1989, reimersetal1997b}, the X-ray cluster Abell2246 ($z =0.25$) gives rise to this system.
Because of the low redshift, \ion{Mg}{ii} is visible in the optical.
\ion{Mg}{i} may be present as well, but its identification is questionable.
The magnesium features reveal no velocity substructure.
In the UV besides \ion{H}{i} Ly$\alpha$, features of \ion{C}{ii}, \ion{Si}{ii}, and \ion{Si}{iii} are clearly detected.
Probably \ion{N}{ii} is also present.

Since the models are based on the noisy STIS observations, a quantitative statement is difficult. 
\ion{Mg}{ii} is used to fix the $b$-parameter, therefore it is reproduced well with all models, even though super-solar abundances are needed in any case.
But the absorption feature claimed for \ion{Mg}{i} is underpredicted.
Notable but still insignificant features are produced by the MF, SB, HMs0.1, and PL05 model.
In the UV, the \ion{C}{ii} and silicon features are well reproduced using HM, HM2, SB, or PL05.
All other models underestimate the strength of at least one ion.
The strongest \ion{N}{ii} feature is predicted by the HM2 model.

Since only magnesium is observed in the optical part of the spectrum, it is impossible to quantify the goodness of the models by the $\chi^2$-estimate introduced in Sect. \ref{procedure}.
Therefore, the evaluation of the results is only based on the discussion above.
The favoured models are then HM, HM2, and PL05.
Considering the model parameters, HM and PL05 require unphysically high or low densities ($\log n_{\mathrm{H}} = 2.5$ or $-6.8$), respectively, and very high abundances (the metallicity derived for the PL05 model exceeds the solar values by more than 1\,dex).
Whereas, using the HM2 background leads to a cold ($T = 10^{3.63}\,\mathrm{K}$) absorber with somewhat super-solar abundances ($\mathrm{[M/H]} = 0.19$) and [Si/C] = 0.17.
The ionization parameter is $\log U = -2.97$ consistent with the detection of low ionization stages only.

\subsection{System $z = 0.7222$}

The absorber at $z = 0.7222$ might be an additional Lyman limit system. 
Even though the FOS data analyzed by \citet{vogelreimers1995} reveals no Lyman limit, the STIS data indicate a break at the corresponding position.
Furthermore, \citet{vogelreimers1995} measure a \ion{H}{i} column density of $\log N = 16.20$, which may give rise to an optically thin LLS.
In the Keck data, we detect \ion{Mg}{ii} in 3, probably 4 subcomponents spread over a velocity range of $\sim 50\,\mathrm{km\,s}^{-1}$.
The $\lambda 2796.4$ doublet component is blended with \ion{Si}{iv} of the $z = 2.4331$ system.
At the velocity of the strongest \ion{Mg}{ii} component, \ion{Mg}{i} and \ion{Fe}{ii} are present as well.
The STIS data indicates the presence of several additional ions of nitrogen, oxygen, and sulphur.
The column densities are estimated assuming the \ion{Mg}{ii} velocity structure and fitting line profiles of the same ion simultaneously.

The models are based on the ions observed in the optical, i.e. on magnesium and iron.
The ionization parameter and density as well as the metallicity are fixed considering \ion{Mg}{ii} and \ion{Fe}{ii}. 
Thus, all other lines are predicted by the models.
For all ions of a single element, an offset in line strength may be given, if deviations from the solar abundance pattern are present.

Using \ion{Mg}{ii} and \ion{Fe}{ii} to constrain the model parameters means that we model the low ionization phase.
An additional high ionization phase is neglected, even though there are indications for its presence (possible detection of \ion{S}{v}).

Discussing the models, we concentrate on the strongest component.
The ratio \ion{Mg}{i}/\ion{Mg}{ii} is modelled well with HM, HMs0.1, MF, PL15, and PL05. 
\ion{Mg}{i} is overproduced by SB, while it is underestimated by HM3 and HM2.
A significant amount of nitrogen (\ion{N}{iii} as well as \ion{N}{iv}) is produced by the hard spectra of MF and both the power laws, and all models predict \ion{N}{ii} absorption consistent with the FUSE data.
\ion{O}{ii} is overestimated for the PL05 model.
Very little \ion{O}{iii} and no detectable \ion{O}{iv} at all is predicted by the softer ionizing backgrounds (all the HM-like radiation fields and SB), while the models with harder spectra reveals \ion{O}{iii} absorption consistent with the STIS data (MF) or in case of the power laws even more.
Little \ion{O}{iv} absorption is predicted by the MF and PL15 models.
The PL05 model leads to prominent \ion{O}{iv} features absolutely consistent with the data.
The higher ionization stages of sulphur (\ion{S}{iv} and \ion{S}{v}) are only present for the power law models. 
The prediction by PL05 fits the data well, but the lower ionization stages \ion{S}{ii} and \ion{S}{iii} are overestimated.
The hard models tend to overpredict \ion{S}{iii}, while it is consistent with the data for all other models.

Since we observe 3 transitions in the optical, $\chi^2$-values can be derived.
We consider \ion{Mg}{i}, \ion{Mg}{ii} $\lambda 2084$, and \ion{Fe}{ii} $\lambda 2344$ and find HM to lead to the best model ($\chi^2 = 10.7$).
The models based on the hard ionizing spectra (MF, PL15, and PL05) fit slightly worse.
Since the power law models overestimate some oxygen and sulphur ions, we look for the best model among HM and MF.
From the appearance of the modelled features, under the assumption that as much observed absorption as possible originates from the system, HM is preferred, which is also consistent with the estimated $\chi^2$.
Adopting the HM model, the ionization parameter for each of the 3 components is $\log U \sim -3.7$, i.e.\ the low ionization phase is modelled.

\subsection{System $z = 0.8643$}

The absorber at $z = 0.8643$ gives rise to an optically thin LLS.
\ion{Mg}{ii}, located in the optical, shows a complex velocity structure with 7 subcomponents spread over $\sim 150\,\mathrm{km\,s}^{-1}$.
We use the \ion{Mg}{ii} velocity splitting to fit the ions arising in the STIS portion of the spectrum (\ion{N}{iii}, \ion{N}{iv}, \ion{O}{iii}, \ion{O}{iv}, \ion{S}{iii}, \ion{S}{iv}, and \ion{S}{v}), where the weakest component is neglected.
Thus, we concentrate on the 6 main features. 
Continuum windows are present for \ion{Mg}{i}, \ion{Al}{iii} and \ion{Fe}{ii}.

Because of only one ion observed in the optical, the models have to be constrained on the basis of the less confident UV lines.
We adopt \ion{S}{iii} and \ion{S}{iv} as suitable lines, which therefore are fitted well by all models.
Thus, \ion{S}{ii} and \ion{S}{v} can be used to constrain the model.
While \ion{S}{ii} is predicted consistent with the data by all models, \ion{S}{v} appears to be overpredicted in any case except the SB model.
The nitrogen abundance is scaled to match the observed feature of \ion{N}{iv}. 
The models HM3, HMs0.1, and SB predict slightly too much \ion{N}{iii} absorption, while \ion{N}{ii} is modelled consistently for all models.
We left the carbon abundance unscaled obtaining a prediction of \ion{C}{ii}.
Too much \ion{C}{ii} absorption is produced by the models HM3, HM2, and both the power laws.

Following this discussion, the best models are HM, HMs0.1, and SB.
Evaluation of the derived model parameters reveals that they are spread over a wide range since six subcomponents are concerned.
Because of the similarity of the ionizing spectra, the distribution of the elemental abundances over the subcomponents obtained is very similar regarding the HM and HMs0.1 model.
[S/H] is roughly constant for all components with $\sim -0.35$ for HM and $\sim -0.68$ for HMs0.1.
Unlike this, the sulphur abundances are spread over a range of roughly 0.34\,dex for the SB model, while here $\mathrm{[Mg/H]}\sim -1.03$ is nearly constant for all subcomponents.
Since the assumption of a pure SB energy distribution as ionizing background is an oversimplification, the results for ions with ionization potentials above 4\,Ryd (\ion{N}{iv}, \ion{O}{iv}, \ion{S}{v}) may be biased.
A combination of the HM and SB radiation could possibly improve the model.
However, since our main purpose is to derive a prediction of the metal line spectrum in the FUSE spectral range, we stop the discussion here and keep HM, HMs0.1, and SB as equivalent models for the $z=0.8643$ absorber.

A further uncertainty for this systems is introduced by the modelled ionization phase.
Since the models are constrained by \ion{S}{iii} and \ion{S}{iv}, which have ionization potentials of 2.6\,Ryd and 3.5\,Ryd, respectively, an intermediate state is represented.
Consequently, the derived metallicities are probably biased.
However, as indicated by only weak features of e.g. \ion{Mg}{ii} and \ion{S}{ii} as well as the absence of \ion{Al}{iii} and \ion{Fe}{ii} absorption, the absorbing medium does not appear to be dominated by a low ionization phase.
Therefore, our results will probably reflect the correct tendencies.

\subsection{System $z = 1.1573$}

The Lyman limit system at $z = 1.1573$ shows \ion{Mg}{ii} absorption in the optical.
We assume only one component even though asymmetries in the \ion{Mg}{ii} profiles indicate the presence of at least two unresolved components.
Furthermore, \ion{Fe}{ii} and \ion{Al}{ii} can be detected in the Keck part of the spectrum. 
\ion{Al}{iii} might be present, but cannot be identified because of blending with Ly$\alpha$ forest lines.
\ion{Si}{ii} $\lambda 1526.7$ may be identified as well, but a reasonable profile fit was impossible.
\ion{C}{iv} could in principle be observed in the optical part of the spectrum at this redshift, but the expected features would arise in the Ly$\gamma$ absorption troughs of the $z \approx 2.433$ absorption complex.
Thus, \ion{C}{iv} might be present but its column density cannot be measured.
The red wing of the absorption troughs at the expected position of \ion{C}{iv} might suggest the presence of an high ionization phase shifted by $\sim 40\,\mathrm{km\,s}^{-1}$.
This component may be also present in oxygen (\ion{O}{iii}, \ion{O}{iv}, \ion{O}{v}), even though due to saturation the  identification is questionable.
In the UV, we measure column densities of \ion{N}{ii}, \ion{N}{iii}, \ion{O}{iii}, \ion{O}{v}, \ion{S}{iii}, and \ion{S}{iv}. 
Additionally, \ion{He}{i} is detected.

For constraining the model parameters, we use \ion{Mg}{ii} and \ion{Fe}{ii}. 
Furthermore, a solar abundance pattern is assumed.
In this case, the low ionization stages like \ion{C}{ii}, \ion{Si}{ii}, \ion{N}{ii}, \ion{O}{ii}, and \ion{Al}{ii} are fitted well with the models based on soft ionizing radiation.
The harder UV backgrounds (MF, PL15, and PL05) overestimate the low ionization stages but produce significant features of transitions from higher ionized elements, which appear only sparsely using the softer spectra.
This is expected since with \ion{Mg}{ii} and \ion{Fe}{ii} the ionization parameter of the low ionization phase is constrained.
Of course, each model could be further optimized by scaling the elemental abundances.
However, the best fitting model found this way is HMs0.1 leading to the best description of the low ionization stages 
with $\log U = -3.55$.

Lines of high ionization stages of neon are expected in the FUSE spectral range.
However, if present, the \ion{Ne}{iv} $\lambda\lambda\lambda 541, 542, 543$ triplet would by located above the \ion{He}{ii} emission redshift, and \ion{Ne}{v} $\lambda 480$ would be blended with complex \ion{C}{iv} absorption of the $z=2.3155$ LLS.
The only feature that might be underestimated, is \ion{Ne}{vii} $\lambda 465$, expected to arise at $1003.6\,\mathrm{\AA}$.
If the analysis of the \ion{He}{ii} Ly$\alpha$ forest revealed a high value of $\eta$ at this redshift, it might originate from unidentified additional absorption of \ion{Ne}{vii}.

\subsection{System $z = 1.4941$}

The $z = 1.4941$ system shows two strong and one weak subcomponent of the \ion{C}{iv} doublet in the optical.
Weak \ion{Si}{iv} absorption may also be present, but only the $\lambda 1393.8$ component is unblended. 
If this feature is misidentified, the derived column density serves as an upper limit.
Additional upper limits can be derived for \ion{C}{ii}, \ion{Al}{ii}, \ion{Al}{iii}, and \ion{Fe}{ii}.
The absence of lines arising from low ionization stages indicates a high or intermediate ionization phase.
In the UV, we clearly detect \ion{O}{iv}, which is fitted using the velocity spread observed in \ion{C}{iv}.
Furthermore, this system may show \ion{He}{i} absorption.

The ionization parameter for each model is fixed using \ion{C}{iv} and \ion{Si}{iv} and assuming a solar abundance pattern. 
Since the column density of \ion{Si}{iv} suffers from significant uncertainties due to the weakness of the absorption features, the derived models are very preliminary.
However, the model HM2 heavily overproduces \ion{He}{i}, \ion{C}{ii}, and \ion{O}{iii}.
Similar problems arise in case of HM3, HMs0.1, and SB, while the prediction of \ion{He}{i} and \ion{O}{iii} absorption made with the hard radiation fields (MF, PL15, and PL05) is consistent with the data.

Even though only a few lines are observed in the optical, it is possible to derive a $\chi^2$ considering \ion{C}{ii} $\lambda 1335$, both components of \ion{C}{iv}, \ion{Al}{ii} $\lambda 1671$ and \ion{Si}{iv} $\lambda 1394$.
The best value reveals the model PL15 (0.99), which is consistent with the discussion above as well.
In this case, the derived parameters are $\log U \sim -2.3$, $-2.1$, $-2.5$ for the 3 components.

Due to the few lines observed for this system, one should keep in mind that the model is rather ambiguous.
Considering deviations from the solar abundance pattern, e.g. a modification of the oxygen abundance, may lead to different results and  possibly to prefer another model background.

\subsection{System $z = 1.7241$}

Another Lyman limit system is located at $z = 1.7241$.
The \ion{H}{i} Ly$\alpha$ transition can be observed in the Keck spectrum.
Features of \ion{C}{iv} and \ion{Si}{iv} are clearly detected, but since they are located in the Ly$\alpha$ forest, it is difficult to determine a velocity structure.
Possibly, there is a second component of \ion{C}{iv} at $\sim 40\,\mathrm{km\,s}^{-1}$ visible also in \ion{C}{ii}.
However, we assume a single component, which is visible in \ion{C}{iv} as well as \ion{Si}{iv}.
Therefore the metallicity of the absorber might be underestimated. 
The $\lambda 1393.8$ component of \ion{Si}{iv} is blended with the \ion{C}{ii} feature of the $z = 1.8450$ system. 
Only the $\lambda 1334.5$ component of \ion{C}{ii} is located in the optical.
Since the feature is weak, the derived column density is quite uncertain and may serve only as an upper limit.
If \ion{Si}{iii} is present, it is heavily blended and an estimate of its column density is impossible.
Upper limits can be derived for \ion{Si}{ii} and \ion{Al}{iii}.
In the STIS portion of the spectrum, we identify ions of neon and oxygen as well as \ion{He}{i} absorption.

The ionization parameter of each model is constrained using the ratio \ion{O}{iii}/\ion{O}{iv},
i.e. we model a high ionization phase which seems to be a reasonable strategy due to the absence of many ions in low ionization stages.
Then, the abundances of carbon and silicon have been scaled to match the observed column densities of \ion{C}{iv} and \ion{Si}{iv}, respectively.
Additionally, the marginally detected \ion{C}{ii} feature and the non-detection of \ion{Si}{ii} give requirements for the elemental abundances.
To evaluate the goodness of the models, the main characteristics are the strength of \ion{Si}{iv}, the width of the \ion{H}{i} Ly$\alpha$ feature, i.e. the temperature of the absorber, the amount of \ion{He}{i} absorption, and the strength of the \ion{C}{ii} feature.
\ion{C}{ii} is overestimated by HM2 and both the power law models.
The hard backgrounds (MF, PL15, and PL05) fit best the \ion{He}{i} absorption, while the soft spectra tend to overproduce it.
Concerning the width of the \ion{H}{i} absorption, the modified HM backgrounds (HM3, HMs0.1, and HM2) lead to the highest temperatures ($\sim 10^{4.6}\,\mathrm{K}$), and thus give the best fits. 
All models except SB underestimate \ion{Si}{iv}.
Since both components are at least partly blended, this might suggest that the detected features do not originate mainly from silicon. 

Enough transitions are detected in the optical spectral range to derive $\chi^2$-values.
Reasonable velocity intervals are chosen covering the features of \ion{H}{i} Ly$\alpha$, \ion{C}{ii}, \ion{C}{iv}, \ion{Al}{iii}, \ion{Si}{ii}, and \ion{Si}{iv}.
As expected, the contributions of \ion{H}{i}, \ion{C}{ii}, and \ion{Si}{iv} dominate $\chi^2$.
If \ion{Si}{iv} is considered in the estimate of $\chi^2$, the best fitting model is SB ($\chi^2 = 3.00$). 
If it is neglected, HM3 fits the system best ($\chi^2 = 1.60$).
Being aware of the uncertainty of the presence of \ion{Si}{iv}, we prefer model HM3,
which leads to an ionization parameter of $\log U = -0.52$.

\subsection{System $z = 1.8450$}

The Lyman limit system at $z = 1.8450$ shows complex absorption with at least 3 subcomponents spread over $\sim 65\,\mathrm{km\,s}^{-1}$.
The features are clearly detected in \ion{C}{iv}, where the strongest component is saturated in both the doublet components, and \ion{C}{ii}, which is blended with the \ion{Si}{iv} $\lambda 1394$ component of the $z = 1.7241$ system, as well as in \ion{Si}{iv}, \ion{Si}{iii}, and \ion{Si}{ii}. 
The red component of \ion{Si}{iii} appears to be blended with Ly$\alpha$ forest lines.
Furthermore, we detect \ion{Al}{ii} and \ion{Al}{iii}.
Even \ion{N}{v} is present.
The \ion{Mg}{ii} doublet was observed by \citet{trippetal1997}. 
Since it is located outside the coverage of the Keck data, we adopt their colum density value.
In the UV, we identify \ion{O}{iii}, \ion{Ar}{iv} and several ions of neon.
In this system \ion{He}{i} appears also to be present.

The variety of transitions observed in the optical should give good constraints of the models.
The ratios of the ions of carbon, silicon, and aluminium are supposed to restrict well-defined models.
Nevertheless, none of the backgrounds produce a model fitting the low ionized lines and the lines of highly ionized elements simultaneously.
For example, HM3 and HMs0.1 produce only very little \ion{C}{ii} absorption, but match well the observed \ion{C}{iv} column density.
Model PL15 and MF have problems to not overproduce \ion{Si}{ii} without underestimating \ion{Si}{iv},
which indicates a multi-phase absorber.

We compute $\chi^2$ considering appropriate velocity intervals of \ion{H}{i} Ly$\alpha$, \ion{C}{ii} $\lambda 1335$, \ion{C}{iv} (both doublet components), \ion{Al}{ii}, \ion{Al}{iii} (both components), \ion{Si}{ii} $\lambda 1527$, and \ion{Si}{iv} (both components).
The $\chi^2$-value is then strongly dominated by the carbon lines, especially by the \ion{C}{iv} doublet.
The best fitting model is that one based on the MF energy distribution.
Despite carbon, the next dominating lines for the $\chi^2$ are \ion{H}{i} Ly$\alpha$, where we concentrate on the blue wing of the observed feature, and silicon (especially \ion{Si}{iii} and \ion{Si}{iv}).
Considering silicon, the best fits are produce by the models SB and HM2.
While HM2 matches the blue wing of the \ion{H}{i} Ly$\alpha$ feature, SB reveals a lower absorber temperature and thus a narrower line.
 A further indicator for the goodness of the fit may be the amount of \ion{He}{i} absorption.
HM2 severely overproduces \ion{He}{i}.
Also the models HM and SB predict stronger \ion{He}{i} than consistent with the data.

Considering all arguments above, we come to the conclusion that the system at $z = 1.8450$ is a multi-phase absorber and should be analyzed applying a more sophisticated model.
Thus, the results from our crude modelling should be considered very preliminary.
However, our simple approach indicates ionization by a hard UV background.
This is in line with the detection of \ion{N}{v}, which is never seen in the `normal` IGM, but always requires a nearby AGN.
We derive consistent models adopting the MF energy distribution or a power law (PL15 and PL05).
The lowest $\chi^2$ is produced for MF ($155.6$).

\subsection{System $z = 2.1680$}

At $z = 2.1680$ another Lyman limit system arises. 
Ly$\alpha$ and Ly$\beta$ are detected in the Keck spectrum.
We clearly identify \ion{C}{iv}, \ion{Si}{iii}, and \ion{Si}{iv} in two subcomponents, where the stronger one appears to be an unresolved blend of at least two components.
\ion{O}{vi} absorption might be present but is blended.
Furthermore, the measured \ion{C}{ii} column density might be biased by unresolved blending with Ly$\alpha$ forest lines.
In this case, the derived values serve as upper limits.
The detection of \ion{Si}{ii} is very uncertain.
Thus, the given value may represent an upper limit.
Neon is identified in the UV, argon might be present as well.

All models lead to strong hydrogen lines, broader than the observed features, indicating a problem with the \ion{H}{i} distribution.
The best fitting models are then MF, PL05, PL15, and SB.
Another crucial transition is \ion{Si}{ii}.
Even if it is not present, there are upper limits that have to be considered.
The models based on hard ionizing spectra (both the power laws, MF, but also HMs0.1), overproduce \ion{Si}{ii}.
The feature claimed for as \ion{C}{ii} is reproduced only by PL05, the other models reveal nearly no \ion{C}{ii} absorption at all.

Thus, we compute $\chi^2$ on the basis of \ion{C}{iv} (both components), \ion{N}{v} $\lambda 1239$, \ion{Al}{ii}, \ion{Si}{ii} $\lambda 1527$ and $\lambda 1260$, and \ion{Si}{iv} $\lambda 1403$.
The dominating transition is then \ion{Si}{iv}, which is fitted best with PL05 and SB.
According to the overall $\chi^2$-value, the best fitting models are SB, HM3, and HM (3.67, 4.10, 4.38).
Considering the width of the \ion{H}{i} features, the SB model is preferred, but assuming a different amount of neutral hydrogen associated to the metal absorption, HM3 and HM provide consistent models as well.
Comparing the HM3 and HM model, HM leads to very high [Si/C] abundances ($\sim 2.0$), while HM3 reveals the more realistic value $\mathrm{[Si/C]} \sim 0.9$.
The SB model leads to $\mathrm{[Si/C]} \sim -0.4$ for the main component.

In order to constrain the models the ratio \ion{Si}{iii}/\ion{Si}{iv} has been adopted.
We find the ionization parameters $\log U = -0.98$ and $-0.70$ in case of the HM3 model and $\log U = -1.22$ and $-1.25$ for the SB model.
Since these lines may arise partly from a low and a high ionization phase, our results may be biased and the reasonability of a pure SB model might be questionable.
However, the presence of an unsaturated \ion{C}{iv} doublet as well as the detection of very weak ions from low ionization stages indicates that neither a low nor a high ionization phase is dominating.

\subsection{System $z = 2.2896$}

For the system at $z = 2.2896$, we detect the Lyman series down to Ly$\delta$, although Ly$\gamma$ and $\delta$ are very noisy. 
Weak \ion{C}{iv} is visible with 3 components spread over $\sim 30\,\mathrm{km\,s}^{-1}$ and weak \ion{Si}{iv} features are detected in the $\lambda 1393.8$ doublet component.
The corresponding \ion{C}{iii} absorption is identified in the optical, but the velocity structure cannot be resolved due to noise. 
Therefore, it is impossible to distribute the total column density of neutral hydrogen over the subcomponents reasonably.
We therefore concentrate on the main component assuming that it gives rise to the bulk of the \ion{H}{i} and \ion{C}{iii} absorption.

Since most of the observed features are weak, a conclusion about the dominated ionization phase is difficult.
Neither lines from low nor from highly ionized species are particularly strong.
Thus, it appears to be reasonable to use \ion{C}{iii} and \ion{C}{iv} to constrain the ionization parameter. 

All derived models are very similar to each other.
We specify a $\chi^2$ considering appropriate velocity intervals of \ion{C}{iii} $\lambda 977$, \ion{C}{iv} $\lambda 1551$, \ion{N}{ii} $\lambda 1084$, \ion{Al}{ii}, \ion{Si}{ii} $\lambda 1260$, and \ion{Si}{iv} $\lambda 1394$.
The features of \ion{H}{i} as well as \ion{Si}{iii} are neglected in lack of a reasonable choice of an velocity interval, although they are important indicators to estimate the best fitting model.
Considering \ion{Si}{iii}, absorption consistent with the observation is predicted by the models SB and PL15, while it is overestimated severely by the models HM2, MF, HM, and PL05.
Regarding the actual numbers of $\chi^2$, all models are of comparable quality. The smallest $\chi^2$-values are provided by HM3, HMs0.1, SB, and PL05 (0.86 for all of them).
PL05 is rejected since it overproduces \ion{Si}{iii}.
Comparing HM3, HMs0.1, and SB, the HM3 model is preferred since the derived [Si/C] abundance is more realistic ($0.89$ in comparison to $-0,17$ and $-0.44$) at a metallicity of $\mathrm{[M/H]} = -3.31$.
The ionization parameter is $\log U = -1.46$.

\subsection{System $z = 2.3155$}\label{section_z=2.3155}

\begin{figure*}
  \centering
  \resizebox{\hsize}{!}{\includegraphics[bb=35 165 545 780,clip=]{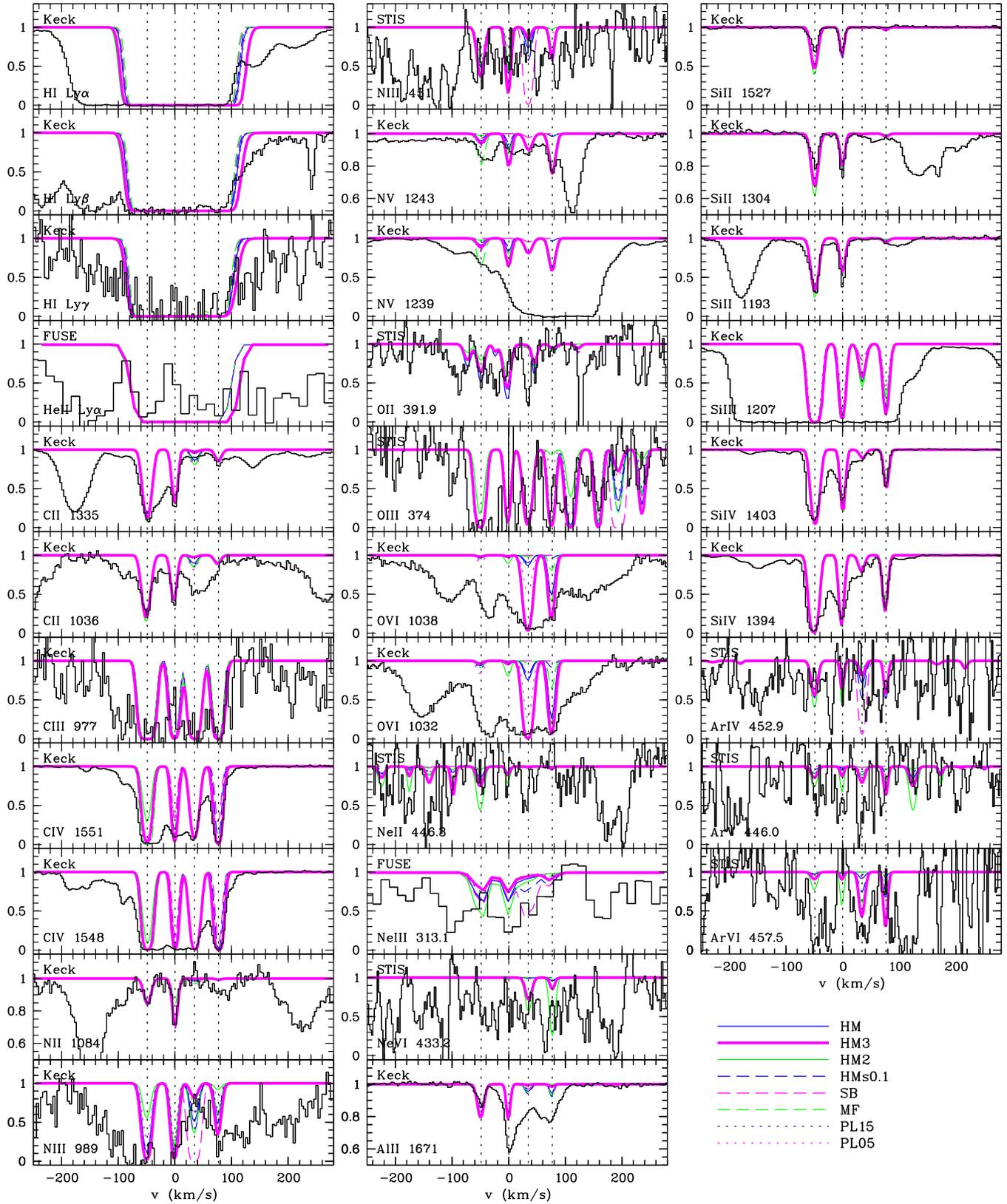}}
  \caption{Observed and modelled absorption lines of the system at $z = 2.3155$. The histogram-like lines give the observed flux at the expected position of the indicated transitions. The name of the observing spectrograph is indicated in the upper left corner of each panel. The plotted STIS data is unsmoothed, i.e., no Savitzky-Golay filter is applied. The preferred model HM3 is presented as the thick line. All other models are shown as thinner lines for a rough comparison. 
}
  \label{z2.3155fig}
\end{figure*}

The Lyman limit system at $z = 2.3155$ gives rise to a very complex structure of metal line absorption. 
We find 6 components spread over $\sim 130\,\mathrm{km\,s}^{-1}$, with 4 main components at $-49\,\mathrm{km\,s}^{-1}$, $0\,\mathrm{km\,s}^{-1}$, $35.1\,\mathrm{km\,s}^{-1}$, and $77\,\mathrm{km\,s}^{-1}$ (Fig. \ref{z2.3155fig}), which we consider individually in the following investigation.
The two blue components are visible in almost all detectable ions, while the red components are less strong.
Unfortunately, some important transitions are blended with Ly$\alpha$ forest lines, and an estimate of column densities is impossible for \ion{C}{iii} and \ion{Si}{iii}.
Also the \ion{O}{vi} profiles suffer from blending with forest lines and the derived values are biased with large systematic uncertainties.
\ion{Al}{ii} is blended with the $\lambda 1548.2$ component of \ion{C}{iv} arising from the system at $z = 2.5785$.
This makes the fitting procedure complicated, but since the doublet component $\lambda 1550.8$ of the $z=2.5785$ system is unblended, the derived values are supposed to be reliable.
\ion{Al}{iii} is located outside the observed spectral range, thus we adopt the value measured by \citet{trippetal1997}.
Nearly all \ion{C}{iv} absorption is saturated in both the doublet components, as well as the strongest subcomponent in \ion{Si}{iv}.
Possibly, we also detect \ion{N}{v}, which usually is not detected in intervening systems.

Analyzing this system, it is evident that it is a multi-phase absorber.
None of the models reaches a satisfactory description of both the low and highly ionized species (see Fig. \ref{z2.3155fig}). 
This is illustrated particularly by \ion{O}{vi}. 
Only weak features for the blue components are predicted, while a stronger line consistent with the observation is produced for the red component.
The reason is the way we fix the ionization parameter.
For both the blue components we use the ratio of \ion{Si}{ii}/\ion{Si}{iv}, which is more reliable than \ion{C}{ii}/\ion{C}{iv} since the column densities suffer less from saturation effects.
Whereas, for the red components, no \ion{Si}{ii} feature is observed and the ratio of the ions of carbon has to serve to fix the ionization parameter.
If the bulk of the \ion{C}{iv} absorption originates in a high ionization phase, but most of the \ion{Si}{iv} arises from low ionized gas, we probe two different gas phases whether the model is based on silicon or carbon.
Using two successive ionization stages should avoid this problem.
However, since both \ion{C}{iii} and \ion{Si}{iii} are blended and unresolved in the present data, there is no possibility to do so.
\citet{trippetal1997} made some efforts to model this system as well (with their data, they resolve 3 components, which correspond to 3 of the main components considered in this analysis).
They failed to derive a single phase model and suggest a multi-phase gas, too.

Comparing the models with the observational data (Fig. \ref{z2.3155fig}), the red wing of the \ion{H}{i} features (Ly$\alpha$ and Ly$\beta$) is poorly described by all models. The best are MF and SB since they produce the lowest temperature ($T \sim 10^{4.0}\,\mathrm{K}$) of the red component.
Concerning \ion{H}{i}, we distributed the measured total column density to the subcomponents according to carbon.
Since we only account for the 4 main components, the amount of \ion{H}{i} may be overestimated. 
But the error is supposed to be small since the contribution of the weaker components to the total carbon column density is negligible.

A reasonable $\chi^2$-value is difficult to define since only parts of the absorption complexes are modelled.
Therefore, we define velocity intervals for each subcomponent using lines we expect to have no systematic errors.
The $\chi^2$-value of the blue component considers \ion{C}{ii} $\lambda 1036$, \ion{N}{ii} $\lambda 1084$, \ion{Al}{ii}, \ion{Si}{ii} $\lambda\lambda\lambda 1527,1304,1193$, and \ion{Si}{iv} $\lambda 1403$.
\ion{Si}{iv} dominates the value but is similar for all models.
Best fitting models are both the power laws, HM, HMs0.1, and SB.

Considering the central component the situation gets even worse since \ion{Al}{ii} is blended and thus cannot be included in the $\chi^2$-estimate.
\ion{Si}{iv} is unusable as well since no reasonable velocity interval can be defined.
With these exceptions we use the same transitions as before. 
Then $\chi^2$ is dominated by the \ion{Si}{ii} lines, which differ only marginal in the appearance of the profiles.
The MF model underestimates \ion{N}{ii}, but no further results can be conclude from the $\chi^2$-values of the single lines since they strongly resemble for each model.
\ion{N}{v}, which is not considered in the $\chi^2$, is slightly overproduced by the HM3 model.

Regarding the component at $35.1\,\mathrm{km\,s}^{-1}$, only very few lines are appropriate to estimate a $\chi^2$-value.
We chose \ion{N}{ii} $\lambda 1084$, \ion{N}{v} $\lambda 1243$, and \ion{Si}{ii} $\lambda 1527$ and find that all HM-like models are preferred.
Additionally, both the power laws and the MF model overproduce \ion{C}{ii}.
SB severely overestimates \ion{N}{iii}, which is also overpredicted by MF, HMs0.1, and HM.
Thus, HM3 and HM2 are the best fitting models with respect to this component.

For the red component, the lines for the $\chi^2$-estimate have to be chosen considering the high ionization gas phase.
We use appropriate velocity intervals of \ion{C}{ii} $\lambda 1335$, \ion{C}{iv} $\lambda 1551$, \ion{N}{v} $\lambda 1243$, \ion{Si}{ii} $\lambda 1527$, and \ion{Si}{iv} $\lambda 1403$.
The $\chi^2$-value is dominated by \ion{C}{iv}.
The strongest deviations from the observed features are found for SB and HMs0.1, which also produce no \ion{N}{v}.

In order to decide which ionizing radiation field is optimal for reproducing the whole system, we determine a total $\chi^2$-value considering all the lines and velocity intervals contributing to the $\chi^2$-values of the subcomponents. 
Following the resulting numbers, HM3 produces the best fitting model (52.1).
Regarding ionization parameter, density, and temperature, the different gas phases can be recovered.
The red subcomponents are modelled to be the hottest ($10^{4.45}\,\mathrm{K}$) absorbers with the highest ionization parameter ($\log U \approx -1.02$ and $-1.38$), while both the blue absorbers are somewhat colder ($\sim 10^{4.23}\,\mathrm{K}$) revealing a lower ionization parameter ($\log U \sim -2.2$ for both components). 
The metallicity is set to $\mathrm{[M/H]} \sim -0.55$ corresponding to the silicon abundance of the bluest system.
We derive $\mathrm{[Si/C]} \sim 0.25$ in case of the blue components and $\sim 0.78$ for the red, highly ionized components.

The LLS at $z =2.3155$ is a very complex system.
The models, we have derived, are certainly oversimplifications and do not serve as a realistical description of the physical conditions in the absorbing gas.
According to the modelling strategy and the results reported above, it is evident that we model the low ionization phase of both the blue component and an intermediate ionization phase for the red components.
Improving the physical model by introducing additional ionization phases would probably modify the metallicity of the system.
Furthermore, the smooth observed line profiles of \ion{C}{iv} and \ion{O}{vi} indicate the presence of a third phase of a hot, highly ionized, diffuse gas component.
But since our main concern in the presented work is the derivation of the metal lines expected in the far-UV and the investigation of the ionizing UV background rather than exploring the physical state of Lyman limit systems, we quit the modelling of the system at this point and address a deeper analysis to future work.

Due to the multi-phase nature of this system, we likely underpredict the absorption of high or low ionization stages for the blue or red subcomponents, respectively, in the FUSE spectral range.
Except the \ion{C}{iv} $\lambda 312.4, 312.5$ doublet, only lines from low ionization stages are expected (\ion{C}{iii}, \ion{N}{iii}, \ion{O}{iii}, \ion{Ne}{iii}).
According to Fig. \ref{z2.3155fig} \ion{C}{iii}, \ion{N}{iii}, and \ion{O}{iii} are modelled confidently, while \ion{Ne}{iii} may by underestimated.
In case of a high $\eta$ value at the corresponding wavelength ($1037.9\,\mathrm{\AA}$), additional absorption due to \ion{Ne}{iii} should be considered.
The UV doublet of \ion{C}{iv} is certainly underpredicted since the doublet observed in the optical is saturated and underestimated by the model at least for the blue components.
However, the features are expected to arise close to the interstellar \ion{C}{ii} $\lambda 1036$ line.
Since saturated interstellar absorption of \ion{C}{ii} $\lambda 1334$ is observed in the STIS data, the $\lambda 1036$ component will be of comparable strength and therefore dominate the blend with \ion{C}{iv}.

\subsection{System $z = 2.3799$}

The system at $z = 2.3799$ shows absorption features of \ion{C}{iii}, \ion{C}{iv}, \ion{Si}{iii}, \ion{Si}{iv}, and \ion{O}{vi} in the optical data.
\ion{C}{ii} seems to be present, too, but is surely blended.
Thus, the derived column density suffers from systematical uncertainties and may serve as an upper limit.
Clear non-detections are \ion{Si}{ii} and \ion{N}{v}.
Furthermore, we could derive limits for \ion{N}{ii}, \ion{Al}{ii}, and \ion{Fe}{ii}. 
In the UV, we may detect several ionic transitions of argon.
The triplet of \ion{Ar}{iv} ($\lambda\lambda\lambda 451.2, 451.9, 452.9$) is obviously present, but appears to be shifted ($\sim +10\,\mathrm{km\,s}^{-1}$) with respect to the optical transitions.
\ion{Ar}{v} and \ion{Ar}{vi} appear to be present, too, but might be blended with sulphur and carbon features arising from the $z = 1.1573$ and $z = 1.8450$ LLS, respectively.
The \ion{H}{i} Lyman series is detected down to Ly$\delta$, which is strongly affected by noise, but Ly$\alpha$, Ly$\beta$, and Ly$\gamma$ are used to estimate the column density.

The observed features show no substructure. 
There may be a small shift of $\sim 5\,\mathrm{km\,s}^{-1}$ between the centroids of \ion{O}{vi} and those of carbon and silicon.
Blending with Ly$\alpha$ forest lines, however, might mimic a shifted position.
Furthermore, the \ion{O}{vi} features are unusually broad ($b = (18.6 \pm 1.2)\,\mathrm{km\,s}^{-1}$).
This might indicate a multi-phase nature of the absorber.
Given the apparent absence of features from low ionized species, a phase of intermediate ionization is modelled.
But an additional hot, high ionization phase might be present as well.

Fortunately, all other optical metal lines are apparently unblended, so that we can compute a $\chi^2$ with several ions. 
We use appropriate velocity intervals of \ion{H}{i} Ly$\beta$, \ion{C}{iv} (both components), \ion{N}{ii} $\lambda 1084$, \ion{N}{v} $\lambda 1243$, both components of \ion{O}{vi}, \ion{Al}{ii}, \ion{Si}{ii} $\lambda 1527$, \ion{Si}{iii}, and both components of \ion{Si}{iv}.
\ion{C}{iii} $\lambda 977$ is rejected for the $\chi^2$-estimation since the comparison with the $\lambda 386$ component in the STIS data suggests that it might be polluted by additional absorption. 
Regarding the \ion{H}{i} feature, we concentrate on the red wing of the profile, which is mismatched by the models based on the hard ionizing spectra (both the power laws, MF, but also HMs0.1), while the softer spectra fit well.
The dominating contribution comes from \ion{O}{vi}.
Here, the best fitting models are produced by HM3 and HM2.
\ion{Si}{iii} is slightly underestimated by HM and PL05.
Considering the overall fit and the actual numbers, HM3 is the preferred model ($\chi^2 = 2.14$,  $\log U = -1.04$), while HM2 fits only slightly worse ($\chi^2 = 2.28$, $\log U = -1.27$).
This suggests that a further optimized ionizing continuum might have a break slightly below 3\,Ryd.

\subsection{Systems $z = 2.4321$ and $z = 2.4331$}

\begin{figure}
  \centering
  \resizebox{\hsize}{!}{\includegraphics[bb=50 270 345 760,clip=]{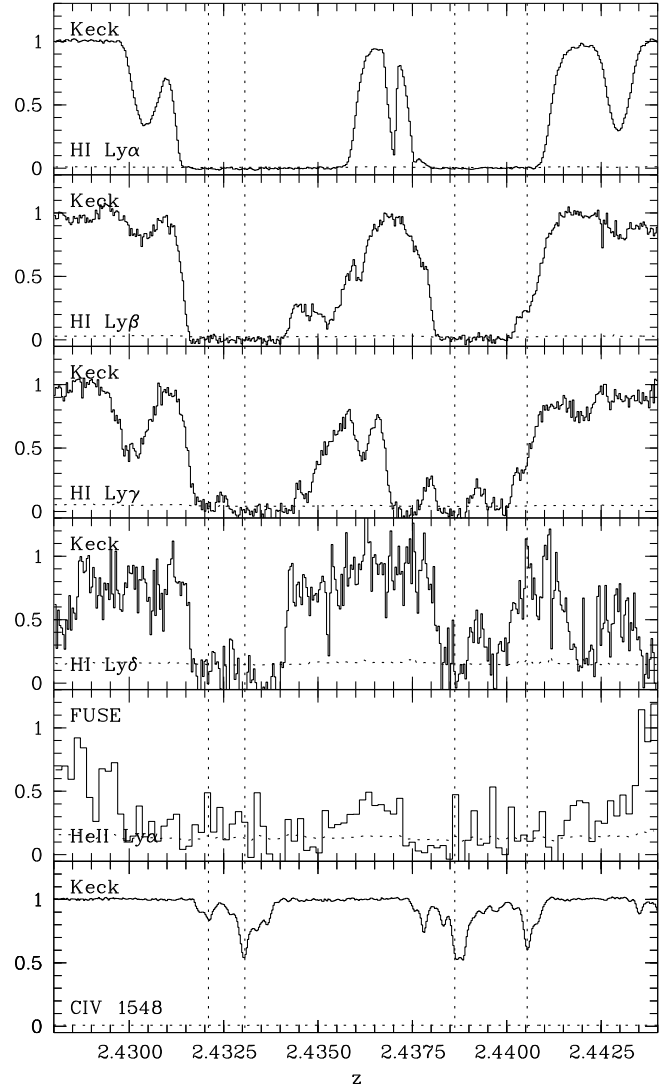}}
  \caption{Observed \ion{H}{i} absorption complex at $z \sim 2.435$. Presented are the features of the Lyman series down to Ly$\delta$ and \ion{He}{ii} Ly$\alpha$ \citep[the corresponding FUSE data is presented in][]{reimersetal_fuse}. The lower panel shows the corresponding \ion{C}{iv} $\lambda 1548$ absorption complexes. The vertical dotted lines indicate the position of the main component of the systems at $z = 2.4321, 2.4331, 2.4386, 2.4405$.
}
  \label{hi2.433}
\end{figure}

There is a strong absorption complex at $z \approx 2.433$, which we assume as two systems at $z = 2.4321$ and $z = 2.4331$, respectively.
The latter appears to cause the observed Lyman limit.
Considering the Lyman series up to Ly$\delta$, we estimate the column densities $\log N(\ion{H}{i}) = 15.53 \pm 0.03$ and $\log N(\ion{H}{i}) = 16.84 \pm 0.35$ for the $z = 2.4321$ and $z = 2.4331$ system, consistent with the value measured from the Lyman limit by \citet{vogelreimers1995}.
Additionally, the red wing of the Ly$\alpha$ absorption trough contains a significant amount of hydrogen without any metal line absorption.
The column density is hard to quantify because of blending with Ly$\alpha$ forest lines in Ly$\beta$ and $\gamma$ and poor data quality for Ly$\delta$.
The \ion{H}{i} absorption complex is presented in Fig. \ref{hi2.433}.
Besides the systems discussed in this Section it shows as well the features of the systems at $z = 2.4386$ and $z = 2.4405$, presented in the next Section. The narrow feature in between the two saturated absorption troughs visible in the Ly$\alpha$ panel is \ion{Si}{iii} of the $z = 2.3155$ LLS.
\ion{He}{ii} Ly$\alpha$ is shown as well. 
For a better orientation, one doublet component of \ion{C}{iv} is also presented since it illustrates the velocity structure visible in the associated metal lines.

Considering the $z = 2.4321$ system, the model parameters are constrained by the carbon ions, although \ion{C}{iv} $\lambda 1551$ is blended with the first \ion{C}{iv} doublet component of the $z = 2.4386$ system.
For the resulting models, we compute $\chi^2$-values based on the features of \ion{H}{i} Ly$\beta$ and Ly$\gamma$, \ion{C}{ii} $\lambda 1335$, \ion{C}{iii} $\lambda 977$, \ion{C}{iv} $\lambda 1548$, \ion{N}{v} $\lambda 1243$, \ion{Al}{ii}, and \ion{Si}{iv} $\lambda 1394$.
None of the lines dominates the $\chi^2$-value, even though the greatest differences are found for the \ion{H}{i} features.
While the HM3 and HM2 models fit well the blue wing of the features, both the power laws as well as MF produce lines too narrow to match the observations.
Regarding the transitions not included into the $\chi^2$-estimation, we have several models producing \ion{Si}{iii} and \ion{O}{vi} absorption consistent with the data.
Models predicting \ion{Si}{iii} are HM, SB, PL15, PL05, and MF, where the latter slightly overproduced the strength of the absorption features.
Whereas only HM3 and HM2 predict perceivable absorption of \ion{O}{vi}, which is still consistent with the data.
According to the $\chi^2$-estimation, the best fitting model is yielded with the HM3 background (1.13).
The derived ionization parameters $\log U \sim -1.0$, $-1.2$, $-1.4$ confirm an intermediate ionization phase.  

In case of the $z = 2.4331$ system, we decide on a line sample to compute $\chi^2$-values which consists of appropriate velocity intervals of \ion{H}{i} Ly$\delta$, \ion{C}{iv} $\lambda 1548$, \ion{N}{v} $\lambda 1239$, \ion{Al}{ii}, \ion{Si}{ii} $\lambda 1193$, \ion{Si}{iii}, \ion{Si}{iv} $\lambda 1394$, and \ion{S}{vi} $\lambda 945$.
PL05 provides a poor description of the observations.
\ion{C}{iv} is significantly underestimated, while \ion{Si}{ii} is overproduced. Furthermore, the SB model predicts too much \ion{N}{iii} absorption and no \ion{S}{vi} features at all.
Differences in the predicted strength of \ion{O}{vi} cannot serve as indicator for the quality of the models since we can only place an upper limit on its column density due to the observational data.
We detect a shift between the position of silicon and \ion{C}{iv} of $(2.66 \pm 0.28)\,\mathrm{km\,s}^{-1}$ in the blue component.
This might be interpreted as an indication of a multi-phase absorber with a more complicated structure than assumed here.
As our parameter constraints are based on the $\ion{Si}{iii}/\ion{Si}{iv}$ we presumable model the low to intermediate ionization phase.
The presence of an additional phase cannot be ruled out.
Therefore the resulting metallicities for the system may be biased.

While the estimated $\chi^2$-values are dominated by \ion{Si}{iii} as well as  \ion{C}{iv}, there are only little differences in the actual numbers for all models except PL05.
The \ion{C}{iv} is best fitted by the HM2 model since it produces a slightly broader profile than the other models, while \ion{Si}{iii} is matched slightly better by the HM3 model.
But at least, these are only marginal differences.
Considering the total $\chi^2$, HM3 ($20.04$) and HM2 ($20.12$) are the best fitting models.
HM3 produces too much \ion{Al}{ii} absorption in the middle component, while HM2 underestimates \ion{Si}{ii} worse than HM3.
The situation seems comparable to the $z = 2.3799$ absorber, where the optimal position of the break is apparently shifted to energies slightly below 3\,Ryd.
We prefer the HM3 model,
which yields the ionization parameters $\log U \sim -1.2$, $-1.5$, $-1.4$.

\subsection{Systems $z = 2.4386$ and $z = 2.4405$}

The absorption complex of the systems at $z = 2.4386$ and $z = 2.4405$ is shifted roughly $500\,\mathrm{km\,s}^{-1}$ from the LLS discussed in the previous Section.
Ly$\alpha$ shows a broad absorption trough, but considering higher Lyman series, distinct features are visible and the column densities can be measured (Fig. \ref{hi2.433}).
We estimate $\log N(\ion{H}{i}) = 15.783 \pm 0.066$ at $z = 2.4386$ and $\log N(\ion{H}{i}) = 14.45 \pm 0.24$ at $z = 2.4405$. 
The large error bar for the latter value arises since it is estimated using only Ly$\beta$.
In between at $z = 2.4397$, another strong \ion{H}{i} system ($\log N(\ion{H}{i}) = 15.490 \pm 0.037$) is detected.
Maybe, the metal line absorption belongs to the stronger \ion{H}{i} but is shifted by $\sim 75\,\mathrm{km\,s}^{-1}$. 
However, weak absorption features of carbon (\ion{C}{iii}, \ion{C}{iv}) and \ion{O}{vi} are apparently associated to this strong \ion{H}{i} absorber.
According to \ion{C}{iv}, 4 subcomponents can be identified for the $z=2.4386$ system, clearly present in \ion{Si}{iii} and \ion{Si}{iv} as well.
The system at $z=2.4405$ spread up in 3 subcomponents considering \ion{C}{iv}.
The corresponding silicon features are very weak and only measurable for the strongest component, which we concentrate on in course of the modelling procedure.

Considering the system at $z=2.4386$, the amount of neutral hydrogen distributed to the bluest subcomponent is overestimated.
Thus, the derived metallicities for this component are definitely too low, but the main conclusions remain unaffected.
We define appropriate velocity intervals to compute $\chi^2$-values based on \ion{H}{i} Ly$\delta$, \ion{C}{ii} $\lambda 1335$, \ion{C}{iii} $\lambda 977$, \ion{C}{iv} $\lambda 1551$, \ion{N}{v} $\lambda 1243$, \ion{Al}{ii}, \ion{Si}{ii} $\lambda 1527$, \ion{Si}{iii}, and both the doublet components of \ion{Si}{iv}.
The interval including \ion{H}{i} is concentrated on the red wing of the feature, and only the two blue components are considered regarding \ion{C}{iii}.
\ion{O}{vi} and \ion{S}{vi} are not included since we can only place upper limits on the presence of these transitions.
The $\chi^2$-values are dominated by \ion{Si}{iii} since it is modelled to be narrower than observed in all models, but MF, HM2, PL05, and HMs0.1 match the observational profile best.
The model PL05 fails because it predicts a significant \ion{Si}{ii} feature, which is not observed, and clearly underestimates the \ion{C}{iv} absorption of the reddest component.  

According to the derived numbers, the MF model fits the observations best (4.46).
Both the red components in \ion{C}{iii} are rather weak, but since the $\lambda 977$ feature is definitely affected by blending with a Ly$\alpha$ forest line, even weak absorption is consistent with the data. 
A further constraint on the \ion{C}{iii} column density of these two components could be provided by the $\lambda 386$ line, but at the corresponding wavelength, the STIS data suffer from insufficient $S/N$.
However, the ionization parameters of the components are based on the column density ratio of silicon.
With $\log U = -2.46$, $-2.56$, $-1.71$, $-2.00$ the intermediate ionization phase is modelled consistently with the absence of strong lines from species in low or high ionization stages.
The estimated overall metallicity is $\mathrm{[M/H]} = -1.29$. 
Relative abundances are $\mathrm{[Si/C]} = -0.77$, $-0.16$, $0.86$, $0.69$.

Even though the system at $z = 2.4405$ shows 3 subsystems with respect to \ion{C}{iv}, we concentrate the analysis on the main component.
This means again that we probably overestimate the amount of neutral hydrogen associated with the absorber.
For this reason, we neglect the \ion{H}{i} profile in the computation of $\chi^2$.
The models HM and PL05 result in low temperatures, e.g. narrow \ion{H}{i} profiles, which are too deep in the line core regarding Ly$\beta$.
But following the argumentation above, we cannot conclude on a shortcoming of these models.

We compute $\chi^2$-values based on appropriate velocity intervals of \ion{C}{ii} $\lambda 1335$, both doublet components of \ion{C}{iv}, \ion{N}{iii} $\lambda 989$, \ion{N}{v} $\lambda 1243$, \ion{Si}{ii} $\lambda 1527$, \ion{Si}{iii}, and both the components of \ion{Si}{iv}.
The latter ions illustrate the goodness of the models.
While the models HM3 and both the power laws overestimate the amount of silicon absorption, HM and HM2 fit well.
\ion{Si}{iii} is well described by SB, but \ion{Si}{iv} is overproduced.
Considering also the contributions of the other lines to $\chi^2$, we find the best fitting model to be HM (1.64).
The SB model produces a good description of the observations as well (1.83).
We find rather high ionization parameters $\log U_{\mathrm{HM}} = -1.24$ and $\log U_{\mathrm{SB}} = -0.54$.
Since only very weak features of silicon are detected but pronounced absorption arising from \ion{C}{iv} and \ion{O}{vi}, we probably observe a high ionization phase.
If this is true, the SB model may be inappropriate since it leads to unrealistic results for ions with high ionization potentials ($\gtrsim$ 4\,Ryd).
Thus, we reject the SB model and, therefore, the unmodified HM background provides the best model.

The system at $z = 2.4397$ (or at $-71.1\,\mathrm{km\,s}^{-1}$ with respect to $z = 2.4405$) is also modelled best using the HM ionizing radiation.
The derived metallicity is $\mathrm{[C/H]} = -2.86$,
which is more than 1\,dex lower than in case of the $z = 2.4405$ central component, indicating that the enrichment history of the IGM is highly inhomogeneous.

\subsection{System $z = 2.4965$}

The system at $z = 2.4965$ shows only weak \ion{C}{iv} and probably \ion{O}{vi}.
If \ion{C}{iii} is present as well, its detection is very uncertain because of blending.
For the few different ions observed the models are constrained using the \ion{C}{iv}/\ion{O}{vi} column density ratio, i.e.\ a high ionization phase is modelled which seems to be the predominating phase for this system.
But since we cannot be sure, whether \ion{O}{vi} is really detected, the model interpretation has to be considered carefully.
 
The main discrepancies of the models are found in the \ion{H}{i} profile.
The best fitting model with respect to neutral hydrogen is then HM3.
This is confirmed by the computed $\chi^2$-value based on the lines \ion{H}{i} Ly$\alpha$ and Ly$\beta$, \ion{C}{ii} $\lambda 1335$, both components of \ion{C}{iv}, \ion{N}{iii} $\lambda 989$, \ion{N}{v} $\lambda 1243$, \ion{O}{vi} $\lambda 1038$, \ion{Si}{ii} $\lambda 1527$, and \ion{Si}{iv} $\lambda 1394$.
Of course, the dominating profiles are those of \ion{H}{i}.
The numbers concerning the continuum windows are comparable, and there are only little differences for \ion{C}{iv}. 
Considering \ion{C}{iv}, the HM3 model slightly underestimated the observed column density.
Whereas the feature, we identified as \ion{O}{vi}, is modelled best by HM3 and HM2.
Both models also predict weak \ion{Ne}{v} and \ion{Ne}{vi} absorption consistent with the data.
Therefore, HM3 is the preferred model ($\chi^2 = 3.86$) leading to an ionization parameter of $\log U = -0.33$.

\subsection{System $z = 2.5683$}

Rather weak metal line features arise from the system at $z = 2.5683$.
\ion{C}{iv} shows 2 subcomponents close to another ($\Delta v = 17\,\mathrm{km\,s}^{-1}$), also present in \ion{C}{iii}.
The observed, weak \ion{O}{vi} features cannot be resolved into two subcomponents.
Thus, the measured values can be considered as an upper limit for both components.
For silicon the weaker, red component is below the detection limit.
The higher order Lyman series lines of hydrogen are blended with forest lines.
Therefore, the most confident column density results from the Ly$\alpha$ fit, which we performed simultaneously with Ly$\beta$.

Comparing the photoionization models with the data, the blue wing of Ly$\beta$ and the red wing of Ly$\gamma$ are supposed to be appropriate, even though all models overestimate \ion{H}{i}.
This may be due to a wrong distribution of the total amount of neutral hydrogen to the subcomponents, or only part of the measured \ion{H}{i} column density is associated to the metal line absorption.
Another possibility is that all models underestimate the temperature of the absorber since the comparison of the artificial profile to the observed data reveals that the observed profiles have much smoother and broader wings indicating a higher temperature.
Also a multi-phase approach might lead to better models since the smooth profiles of \ion{O}{vi} indicate an additional high ionization phase.
However, within our simple approach the broadest and thus best fitting profiles are produced by the models HM2 and HM3.

We compute the $\chi^2$-values adopting appropriate velocity intervals of \ion{H}{i} Ly$\beta$, \ion{C}{ii} $\lambda 1135$, \ion{C}{iii} $\lambda 977$, both components of \ion{C}{iv}, \ion{N}{v} $\lambda 1239$, \ion{Si}{iii}, and both doublet components of \ion{Si}{iv}.
As discussed above, the $\chi^2$-values are dominated by \ion{H}{i}, but also \ion{Si}{iii} contributes above average.
Regarding \ion{Si}{iii}, the best fits are yielded using the PL15, MF, or HMs0.1 radiation backgrounds, which are the best fitting models if \ion{H}{i} is ignored, while HM overestimated the observed column density significantly.
The decision for the best fitting model depends on whether the \ion{H}{i} profile is considered or not.
If we include the profile of Ly$\beta$, HM2 and HM3 are the best fitting models, whereas PL15 fits best if \ion{H}{i} is neglected.
However, only HM3 and HM2 provide enough \ion{O}{vi} absorption to fit the observed features ($\mathrm{[O/C]} \approx 0.5$).
But one should keep in mind that a multi-phase approach where the oxygen abundance possibly can be left unchanged, might result in an improvement of the models.
We claim HM2 to be the best fitting model (4.19) yielding the ionization parameters $\log U = -1.43$ and $-1.18$.
Because the quality of the HM3 fit is only slightly inferior, we may conclude that the energy distribution of ionizing radiation could be optimized positioning a break at a low  energy slightly above 2\,Ryd.

\subsection{System $z = 2.5785$}

The broad Ly$\alpha$ absorption trough of the system at $z = 2.5785$ separates into two distinct features at $-50\,\mathrm{km\,s}^{-1}$ and $25\,\mathrm{km\,s}^{-1}$ visible in Ly$\gamma$ and higher order Lyman series.
Each of the subsystems can be separated into 2 components, with respect to \ion{C}{iv}.
The complex of the \ion{C}{iv} $\lambda 1548$ component is blended with \ion{Al}{ii} of the system at $z =2.3155$.
The center of the blue \ion{H}{i} features mismatches the center of the blue subcomponents observed in the metal line transitions. 
They appear to be shifted by roughly $20\,\mathrm{km\,s}^{-1}$.
This leads to an overproduction of neutral hydrogen in the blue components for all models.
Thus, the derived metallicities are supposed to be slightly underestimated.

We compute $\chi^2$ considering \ion{H}{i} Ly$\epsilon$, \ion{C}{ii} $\lambda 1335$, \ion{C}{iv} $\lambda 1551$, \ion{Si}{ii} $\lambda 1527$, both doublet components of \ion{Si}{iv}, and \ion{S}{vi} $\lambda 945$.
The dominating contributions are given by \ion{C}{iv} and \ion{Si}{iii}, which are observed to be broader than predicted by the models.
Best fitting models with regard to \ion{C}{iv} are HM2 and HM3, and HM and MF with respect to \ion{Si}{iii}.
The feature of \ion{C}{ii} is overestimated by the models PL05 and HM2.
PL05 also predicts too much absorption in \ion{Si}{iv}.
Considering the red component of neutral hydrogen in Ly$\epsilon$, the modified HM models are slightly too broad, while MF, PL15 and SB fit well the observed profile.

Regarding the transitions not included into the $\chi^2$-computation, HM3 severely overestimates \ion{Ne}{v}.
The reddest feature of \ion{C}{iii} might be underpredicted by all models except PL05 and SB. 
But possibly it is blended with the low redshift Ly$\alpha$ forest. 
Thus, it is not clear, if the total feature is due to \ion{C}{iii}.
Unfortunately, the $\lambda 386$ transition is blended as well and therefore cannot answer the question.
A significant amount of \ion{O}{vi} is produced by the models HM, HM3, HM2, and PL15.
But HM and PL15 are then severely overpredicting the blue features of \ion{O}{iii}.

Considering all arguments given above, the best fitting model is HM3, even though it underestimates the \ion{Si}{iii} component at $-18.5\,\mathrm{km\,s}^{-1}$.
According to the HM3 model, silicon is overabundant by more than 1\,dex ($\mathrm{[Si/C]} = 2.98, 1.93, 0.63, 1.76$) in nearly all subcomponents.
Test calculation showed, slight changes in the position of the break of the ionizing radiation would cause significant variations for [Si/C].
A further refinement of the UV background might possibly decrease the silicon overabundance.
Considering HM3, the ionization parameter is rather high for the blue components ($\log U = -0.78$ and $-0.99$) and slightly lower for the red components ($-1.26$ and $-1.17$).
This indicates that we model a high ionization phase, which is also consistent with the presence of \ion{O}{vi}.
However, introducing an additional phase of highly ionized, diffuse gas might improve the models.

\section{Discussion}\label{discussion}

\begin{figure}
  \centering
  \resizebox{\hsize}{!}{\includegraphics[bb=40 515 360 755,clip=]{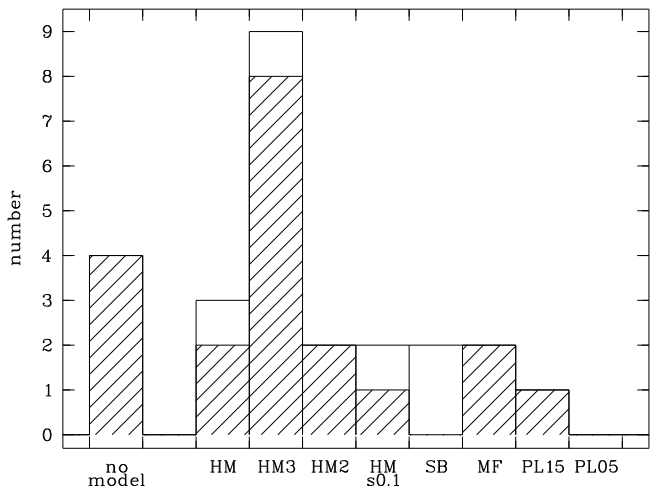}}
  \caption{Numbers of the best fitting models for all observed metal line systems except the associated systems. ``no model'' summerizes the systems for which no model can be derived because of the lack of observed ions. The hatched area contains systems with unique solutions, while the blank area includes ambiguous models as well (see Table \ref{summary}). The total number of systems considered is 22.
  }
  \label{numbers}
\end{figure}

From the 25 observed metal line systems we were able to derive models for 18 systems with the purpose to predict the metal line spectrum in the far-UV.
Simple photoionization models were evaluated with the aim to determine the ionizing UV background which leads to the best description of the observed lines and of the predicted far-UV spectrum.
The investigation is based on the comparison of 8 different background shapes.
A summary of the results is given in Table \ref{summary}, which is presented graphically in Fig. \ref{numbers}, where we exclude the systems associated to the QSO. 
Thus, the total number of systems considered here is 22.
Note that systems at $z \leq 1.1573$ (except the $z=0.7222$ system) show only very few lines in the optical.
Realistic error estimates are therefore not possible and the models of the low redshift systems have been evaluated using only qualitative arguments.
In case of the high redshift systems a $\chi^2$-estimate supplements the visual inspection of the models.

Evaluating the sample reveals HM3 to be the model leading largely to the best description in the framework of our simple models.
HM3 denotes a UV background based on the \citet{haardtmadau2001} energy distribution, where the break at 4\,Ryd is shifted to 3\,Ryd.
7 of 11 systems with $z > 2$ find HM3 to be the favoured radiation background, and one leads to HM2.
As discussed in the previous Section there are hints that for several systems the break of a more optimized ionizing continuum would be somewhere in between $2-3\,\mathrm{Ryd}$.
This is in line with recent results from \citet{agafonovaetal2005}, who find a significant intensity decrease between 3 and 4\,Ryd compare to the UV background of \citet{haardtmadau1996} at $z \sim 3$ investigating 4 metal line systems with a more sophisticated method.

The classical HM background was only found in 3 cases ($z = 0.7772,\, 0.8643,\, 2.4405$), where the $z=0.8643$ system can be equivalently modelled with the SB ionizing radiation.
The spectral energy distribution of a starburst galaxy, adopted from \cite{bruzualcharlot1993}, is found to be 2 times among the preferred models as well as the hard radiation of the quasar-like SED as derived by \cite{mathewsferland1987}.
The HMs0.1 model, a HM background with reduced intensity for energies $< 1.0\,\mathrm{Ryd}$, is preferred twice for the systems at $z \sim 1$.
Note that only the system at $z = 1.4941$ leads to the pure power law model with $\nu^{-1.5}$ (PL15).
As discussed above, the model is extremely uncertain, due to few observed transitions.

\begin{table*}
  \caption[]{Summary of the radiation backgrounds that best model the metal line systems. The more objective procedure of computing $\chi^2$-values to estimate the goodness of a fit could be applied to the systems at redshifts $z \geq 1.4941$ and $z=0.7222$.
Also given are the derived mean metallicity [M/H] and the abundance of [Si/C] for the modelled ionization phases (last column) as well as the ionization parameter $\log U$.
  }
  \label{summary}
  $$ 
  \begin{array}{c l r r l l}
    \hline
    \hline
    \noalign{\smallskip}
    z & \mathrm{model} & \mathrm{[M/H]} & \mathrm{[Si/C]} & \ \ \log U & \mbox{modelled phase} \\
    \noalign{\smallskip}
    \hline
    \noalign{\smallskip}
    0.2140 & \mathrm{HM2}    &  0.14 & 0.17 & -2.97 & \mathrm{low} \\
    0.7222 & \mathrm{HM}     &  0.13 & 0.0^{\mathrm{a}} & -3.70,-3.70,-3.73 & \mathrm{low} \\
    0.8643 & \mathrm{HM}     & -0.35 & 0.0^{\mathrm{a}} & -2.11,-1.77,-1.74,-2.73,-1.55,-2.40 & \mathrm{intermediate} \\
           & \mathrm{HMs0.1} & -0.68 & 0.0^{\mathrm{a}} & -1.83,-1.42,-1.36,-2.45,-1.15,-2.09 & \mathrm{intermediate} \\
           & \mathrm{SB}     & -0.79 & 0.0^{\mathrm{a}} & -1.96,-1.63,-1.60,-2.55,-1.44,-2.24 & \mathrm{intermediate} \\
    1.1573 & \mathrm{HMs0.1} & -1.09 & 0.0^{\mathrm{b}} & -3.55 & \mathrm{low} \\ 
    1.4941 & \mathrm{PL15}   & -0.97 & 0.0^{\mathrm{b}} & -2.25,-2.06,-2.47 & \mathrm{intermediate} \\ 
    1.7241 & \mathrm{HM3}    & -2.82 & -0.09 & -0.52 & \mathrm{high} \\
    1.8450 & \mathrm{MF}     & -0.48 & -0.68 & -2.12,-2.47,-2.65 & \mathrm{intermediate}^{\mathrm{e}} \\
    2.0211 & \dots           & \dots & \dots & \dots & \dots \\
    2.1287 & \dots           & \dots & \dots & \dots & \dots \\
    2.1680 & \mathrm{HM3}    & -3.23 &  0.90 & -0.98,-0.70 & \mathrm{intermediate} \\
           & \mathrm{SB}     & -2.80 & -0.43 & -1.22,-1.25 & \mathrm{intermediate} \\
    2.1989 & \dots           & \dots & \dots & \dots & \dots \\
    2.2896 & \mathrm{HM3}    & -3.31 &  0.89 & -1.46 & \mathrm{intermediate} \\
    2.3079 & \dots           & \dots & \dots & \dots & \dots \\
    2.3155 & \mathrm{HM3}    & -0.55 &  0.25, 0.78^{\mathrm{c}} & -2.23,-2.14,-1.02,-1.38 & \mathrm{low}, \mathrm{intermediate}^{\mathrm{e}} \\
    2.3799 & \mathrm{HM3}    & -1.40 &  1.07 & -1.04 & \mathrm{intermediate} \\
    2.4321 & \mathrm{HM3}    & -2.33 &  0.0^{\mathrm{d}} & -0.97,-1.16,-1.36 & \mathrm{intermediate} \\
    2.4331 & \mathrm{HM3}    & -1.73 &  1.61 & -1.17,-1.46,-1.39 & \mathrm{intermediate} \\
    2.4386 & \mathrm{MF}     & -1.29 & -0.16, 0.69^{\mathrm{c}} & -2.46,-2.59,-1.71,-2.00 & \mathrm{intermediate} \\
    2.4405 & \mathrm{HM}     & -0.37 &  1.01 & -1.24 & \mathrm{high} \\
    2.4965 & \mathrm{HM3}    & -2.15 &  0.0^{\mathrm{d}} & -0.33 & \mathrm{high} \\
    2.5683 & \mathrm{HM2}    & -1.71 &  1.49 & -1.43,-1.18 & \mathrm{high} \\
    2.5785 & \mathrm{HM3}    & -2.78 &  2.45, 1.20^{\mathrm{c}} & -0.78,-0.99,-1.26,-1.17 & \mathrm{high} \\
    \noalign{\smallskip}
    \hline
  \end{array}
  $$ 
\begin{list}{}{}
  \item[$^{\mathrm{a}}$] silicon outside observed spectral range
  \item[$^{\mathrm{b}}$] uncertain detection of \ion{Si}{ii} or \ion{Si}{iv}, respectively
  \item[$^{\mathrm{c}}$] several subcomponents (see text)
  \item[$^{\mathrm{d}}$] silicon not detected
  \item[$^{\mathrm{e}}$] multi-phase system
\end{list}
\end{table*}

One of the basic assumptions of our modelling strategy is that each absorber can be considered as single phase medium.
This is certainly an oversimplification.
More realistic models for most of the absorption systems towards this QSO are provided by \citet{simcoeetal2005}.
The authors take into account the multi-phase nature of the absorbers but consider only one type of ionizing radiation.
However, with our crude models it is possible to get information about one single ionization phase of the probably multi-phase medium.
This is illustrated in particular by the system at $z=2.3155$ discussed in detail in the previous Section (Fig.\ \ref{z2.3155fig}). 
For two of the four main components we model the low ionization phase where all lines arising from ions with low ionization potential are well modelled.
But it is clear from Fig.\ \ref{z2.3155fig} that there is an additional high ionization phase providing \ion{O}{vi} absorption which is not included in our model.
The red components of this system show only weak features of low ionization stages and the model describes a higher ionization phase.
Thus, our models are supposed to be appropriate for one single ionization phase even if an improved model of the system would require multiple ionization phases.

The modelled ionization phase is given in the last column of Table \ref{summary}.
Phases with features of \ion{C}{iv} and \ion{Si}{iv} as dominating lines are referred to as ``intermediate'' ionization phases.
For these systems it is possible that part of the absorption possibly arise from a low ionization phase while another part may arise from a high ionization phase.
Therefore, the results for systems in an intermediate ionization phase may be biased and have to be interpreted carefully.
As discussed in the previous Section, the absence of features from particular low or high ionization stages seems to justify the assumption of an intermediate ionization phase for most of the systems. 
However, unusual values given in Table \ref{summary} may result from this problem.
The low [Si/C] for the system at $z = 1.8450$ is certainly due to multiphase absorption.
Since the \ion{C}{iv} features will arise partly from the intermediate but also partly from a high ionization phase, our model (based on the intermediate ionization phase) overestimates the carbon abundance.
Thus, we recommend to consider the results given in Table \ref{summary} very carfully.
Our method is certainly unappropriate to derive unbiased elemental abundances.
However, investigating the metallicity of the intervening systems is not the main purpose of this paper but predicting the metal lines in the FUSE spectral range.

A further bias of the results listed in Table \ref{summary} may arise from unresolved components.
An example is the system at $z = 1.1573$ where asymmetries in the \ion{Mg}{ii} profiles indicate two unresolved components.
In order to perform the analysis we assume one single component even though this is probably incorrect.
If the blended components had different ionization parameters, the resulting models would be biased.
Similar assumptions are made for the systems at $z = 0.7222$, $2.1680$, and $2.2896$.
For other absorption systems ($z = 0.8643$, $1.7241$, $2.3155$, and $2.4404$) the metallicity might be underestimated since we neglect weak components that show an insufficient number of different ions to constrain a reasonable model.

Taking into account all the sources of errors the numbers presented in Table \ref{summary} should be considered as preliminary results.
However, general tendencies should be correct as also indicated by the comparison with the results from \citet{simcoeetal2005}.
Introducing the shape of the ionizing radiation as an additional free parameter may lead to further ambiguities.
Therefore, our results present only a first attempt to constrain the detailed shape of the UV background systematically \citep[for another approach see e.g.][]{agafonovaetal2005}.

The results suggest that a background similar to HM3 dominates at $z \gtrsim 2$, while at $z \lesssim 2$ no predominant shape of the ionizing radiation is found (see Table \ref{summary}).
This might indicate a major contribution of local sources at lower redshift because of less filtering of their radiation by the IGM due to advanced structure formation.
Unfortunately, the systems at $z \lesssim 2$ are mainly observed in the UV spectral range where the data quality is insufficient to draw solid conclusions.

The usually adopted unmodified \citet{haardtmadau2001} UV background is found only 3 times among the favoured models ($z = 0.7222$, $0.8643$, and $2.4405$).
It would be interesting to study systematic differences between the systems that prefer the HM background and those who do not.
But no particular characteristics of the HM-favouring systems are found.
Apparently, no ionization phase is preferred since the three systems are modelled in low ($z = 0.7222$), high ($z = 2.4405$), and intermediate ($z = 0.8643$) ionization state.
The systems are complex ($0.7222$ and $0.8643$ which are Lyman limit systems) or simple ($2.4405$), and neither low nor high redshift is preferred.
However, with three systems favouring the HM model the statistics is rather limited.
A detailed investigation of the properties would require the investigation of additional metal line systems. 

A possible modification of the ionizing background would have implications for other quantities related to the shape of the background radiation like the measurement of the metallicity of the IGM.
Exact estimates of the metallicity from observations are important e.g. to constrain the history of enrichment of the IGM \citep[e.g.][]{qianwasserburg2005}.  
A rough comparison of the metallicity as traced by carbon derived for the preferred models of each system to the values found using the \citet{haardtmadau2001} radiation field,
reveals the trend that the models based on the softer radiation fields (HM3, HM2, HMs0.1, and SB) lead to increased ionization parameters and lower metallicity by roughly 0.5\,dex in comparison to HM, while the harder MF radiation produces slightly lower ionization parameters and enhanced metallicity by about 0.4\,dex.
This means, a modification of the UV background would affect the determination of the metallicity of the IGM in the sense that abundances would decrease if the radiation softens in comparison to \citet{haardtmadau2001}, or increase with harder radiation. 
This is in line when comparing the results from our best fit models to the metallicities estimated by \citet{simcoeetal2005}.
They analyzed 9 of the systems selected by \ion{O}{vi} or \ion{N}{v} absorption (Actually, they report on 6 systems but consider our systems at $z = 2.4321$, $2.4331$, $2.4386$, and $2.4405$ as only one absorption system).
Using a composite spectrum of \citet{haardtmadau2001} and the energy distribution of a starburst, they find on average a mean metallicity roughly $0.4\,\mathrm{dex}$ higher than derived by our preferred HM3 models.

We find wide spread values for the silicon enhancement inferred from our models in the range $-0.7 \lesssim \mathrm{[Si/C]} \lesssim 1.6$ (summarized in Table \ref{summary}).
An unrealistically high value is found for the blue components of the system at $z = 2.5785$.
Comparing to the results of \citet{simcoeetal2005} we find at least the same tendencies, i.e. one silicon deficient system at $z=1.8450$ and silicon enhancement for all other absorbers.
This provides further arguments for HM3 being the preferred model in case of the $z=2.1680$ system since the SB model reveals Si deficiency. 
As discussed in Sect. \ref{limitations}, [Si/C] depends on the enrichment history, but also on the adopted UV background \citep{aguirreetal2004}.
Since \citet{simcoeetal2005} inferred a more smooth distribution of [Si/C], the differences are most likely caused by the adopted ionizing energy distribution.
If we only consider the HM3/HM2 models, the silicon enhancement is on average $\mathrm{[Si/C]} = 1.18 \pm 0.62$ (or $1.13\pm 0.32$, if both the extreme outliers, which are single subcomponents of observed systems, are neglected).
This is still about $1\,\mathrm{dex}$ higher than estimated by \citet{simcoeetal2005} and illustrates the importance of the assumed UV background for the determination of the metallicity of intergalactic absorbers \citep[see also][]{aguirreetal2004}.

\subsection{Implications on the \ion{He}{ii}/\ion{H}{i} ratio}

FUSE observations of the \ion{He}{ii} Ly$\alpha$ forest towards HE~2347-4342 \citep{krissetal2001, shulletal2004, zhengetal2004} and HS~1700+6416 \citep{reimersetal_fuse} found that the column density ratio $\eta = N(\ion{He}{ii})/N(\ion{H}{i})$ is apparently fluctuating over roughly 2\,dex on small scales ($\sim$ 1\,Mpc).
These results indicate the importance of local sources on the photoionization of the absorbing material.

The column density ratio $\eta$ depends on the radiation field and therefore can be used to diagnose its fluctuations.
In photoionization equilibrium the density of hydrogen and helium is
\begin{equation}
  n_{\ion{H}{i}} = \frac{n_{\mathrm{e}} n_{\ion{H}{ii}} \alpha_{\ion{H}{i}}^{(A)}}{\Gamma_{\ion{H}{i}}} \qquad\mbox{and}\qquad n_{\ion{He}{ii}} = \frac{n_{\mathrm{e}} n_{\ion{He}{iii}} \alpha_{\ion{He}{ii}}^{(A)}}{\Gamma_{\ion{He}{ii}}} \mbox{,}
\end{equation}
respectively,
with the case A recombination rate coefficients 
$\alpha_{\ion{H}{i}}^{(A)} = 2.51\cdot 10^{-13}\,T_{4.3}^{-0.755}\,\mathrm{cm}^{3}\,\mathrm{s}^{-1}$ and 
$\alpha_{\ion{He}{ii}}^{(A)} = 1.36\cdot 10^{-12}\,T_{4.3}^{-0.70}\,\mathrm{cm}^{3}\,\mathrm{s}^{-1}$.
The photoionization rates $\Gamma$ are 
\begin{equation}\label{gamma}
  \Gamma_{\ion{H}{i}} = \int_{\nu_{\mathrm{LL},\,\ion{H}{i}}}^{\infty}4\,\pi \frac{J_{\nu}}{h\nu}\,\sigma_{\ion{H}{i}}(\nu)\,d\nu\,\mbox{,} 
\end{equation}
in case of hydrogen and equivalent in case of \ion{He}{ii}.
$\nu_{\mathrm{LL}}$ denotes the ionizing threshold for \ion{H}{i} (1\,Ryd) or \ion{He}{ii} (4\,Ryd), respectively.
The cross section for photoionization decreases roughly with $\nu^{-3}$ above the ionizing threshold, thus it is
$\sigma(\nu) = \sigma^{0}\,(\nu/\nu_{\mathrm{LL}})^{-3}$,
with $\sigma^{0}_{\ion{H}{i}} = 6.30\cdot 10^{-18}\,\mathrm{cm}^{2}$ and 
$\sigma^{0}_{\ion{He}{ii}} = 1.58\cdot 10^{-18}\,\mathrm{cm}^{2}$.
Adopting primordial helium abundance \citep[$Y=0.244\pm 0.002$,][]{burlesetal2001} and assuming that the gas is almost fully ionized ($n_{\ion{H}{ii}} \simeq n_{\mathrm{H}}$ and $n_{\ion{He}{iii}} \simeq n_{\mathrm{He}}$), the ratio $\eta$ can be written as
\begin{equation}\label{eta_approx}
  \eta = \frac{n_{\ion{He}{ii}}}{n_{\ion{H}{i}}} \approx 0.437\cdot T_{4.3}^{0.055}\cdot\frac{\Gamma_{\ion{H}{i}}}{\Gamma_{\ion{He}{ii}}}\,\mbox{.}
\end{equation}
The integrals given in Eq. \ref{gamma}, which are still contained in Eq. \ref{eta_approx}, can be further approximated since the photoionization rates depend mainly on the specific intensity $J_{\nu}$ near the ionizing threshold $\nu_{\mathrm{LL}}$ due to the $\nu^{-3}$-dependence of the photoionization cross section.
The result of this approximation can be found e.g. in \citet{fardaletal1998}.

\begin{figure}
  \centering
  \resizebox{\hsize}{!}{\includegraphics[bb=35 540 290 710,clip=]{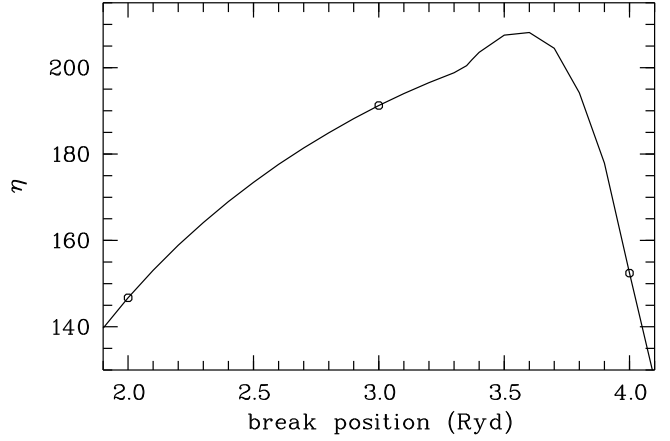}}
  \caption{Value of the column density ration $\eta$ computed according to Eq. \ref{eta_approx} versus the position of the break, which occurs in case of the classical \citet{haardtmadau2001} background at 4\,Ryd. The values corresponding to the models HM, HM3, and HM2 are marked with circles.
  }
  \label{eta_break}
\end{figure}

However, we perform a numerical integration to estimate $\eta$ according to Eq. \ref{eta_approx} for each of the ionizing spectra used in the analysis.
Energy distributions depending on redshift (all HM-like spectra) are considered at $z\sim 2$. 
The HM background leads to $\eta \approx 150$, which is higher than the value usually referred to \citep[$\sim 45$,][]{haardtmadau1996}.
The reason is the inclusion of radiation of galaxies in the new version of the UV background radiation, which increases $\eta$ \citep[illustrated in Fig. 8 of][]{haardtmadau2001}.
In case of the HM spectrum used here, the escape fraction of Lyman limit photons from galaxies is 10\,\%.
The spectrum with reduced flux at $E < 1\,\mathrm{Ryd}$ produces the same $\eta$ since the energy range included by the integrals of the photoionization rates remain unaffected by the modification. 

The dependence of $\eta$ on the position of the break, which we shifted from its actual position at 4\,Ryd to 3 (HM3) and 2\,Ryd (HM2), respectively, is illustrated in Fig. \ref{eta_break}.
The shape of the curve is determined by the integrals of the photoionization rates $\Gamma_{\ion{H}{i}}$ and $\Gamma_{\ion{He}{ii}}$ (Eq. \ref{gamma}).
For break positions above 3.4\,Ryd the ratio is dominated by $\Gamma_{\ion{He}{ii}}$, which gets smaller with decreasing break position, i.e. $\eta$ decreases with the maximum at $\sim 3.6\,\mathrm{Ryd}$.
If the position of the break is at energies $<3.4\,\mathrm{Ryd}$, the spectral energy distribution is flat at the \ion{He}{ii} ionization edge and the integral  $\Gamma_{\ion{He}{ii}}$ stays roughly constant.
Whereas the photoionization rate of \ion{H}{i}, and therefore $\eta$, decreases with decreasing position of the break.

The different ionizing spectra we used, lead to very different $\eta$-values, as expected.
The adopted energy distribution of a starburst galaxy \citep[][SB]{bruzualcharlot1993} produces a high value $\eta \approx 700$ due to the lack of helium ionizing photons.
The AGN-like spectrum of \citet[][MF]{mathewsferland1987} as well as the power laws with $\alpha = -1.5$ (PL15) and $-0.5$ (PL05), respectively, are very hard, and thus lead to low $\eta$-values.
While PL15 reveals $\eta\approx 15$, MF ($\eta\approx 10$) and PL05 ($\eta\approx 5$) produce even lower values.
Within our sample of metal line systems, the power laws are practically of no importance.
But the MF and SB spectra lead each in 11\,\% of the investigated systems to a preferred model, even though there are equivalent descriptions based on a different energy distribution in case of SB.
Thus, some of the extremely low and high $\eta$-values found analyzing the \ion{He}{ii} observations towards HE~2347-4342 \citep{krissetal2001, shulletal2004, zhengetal2004} and HS~1700+6416 \citep{reimersetal_fuse}, might be induced by ionizing radiation similar to the MF and SB energy distributions used in this analysis.

\subsection{Prediction of metal lines in the FUSE spectral range}

\begin{figure*}
  \centering
  \resizebox{\hsize}{!}{\includegraphics[bb=60 50 570 795,clip=,angle=-90]{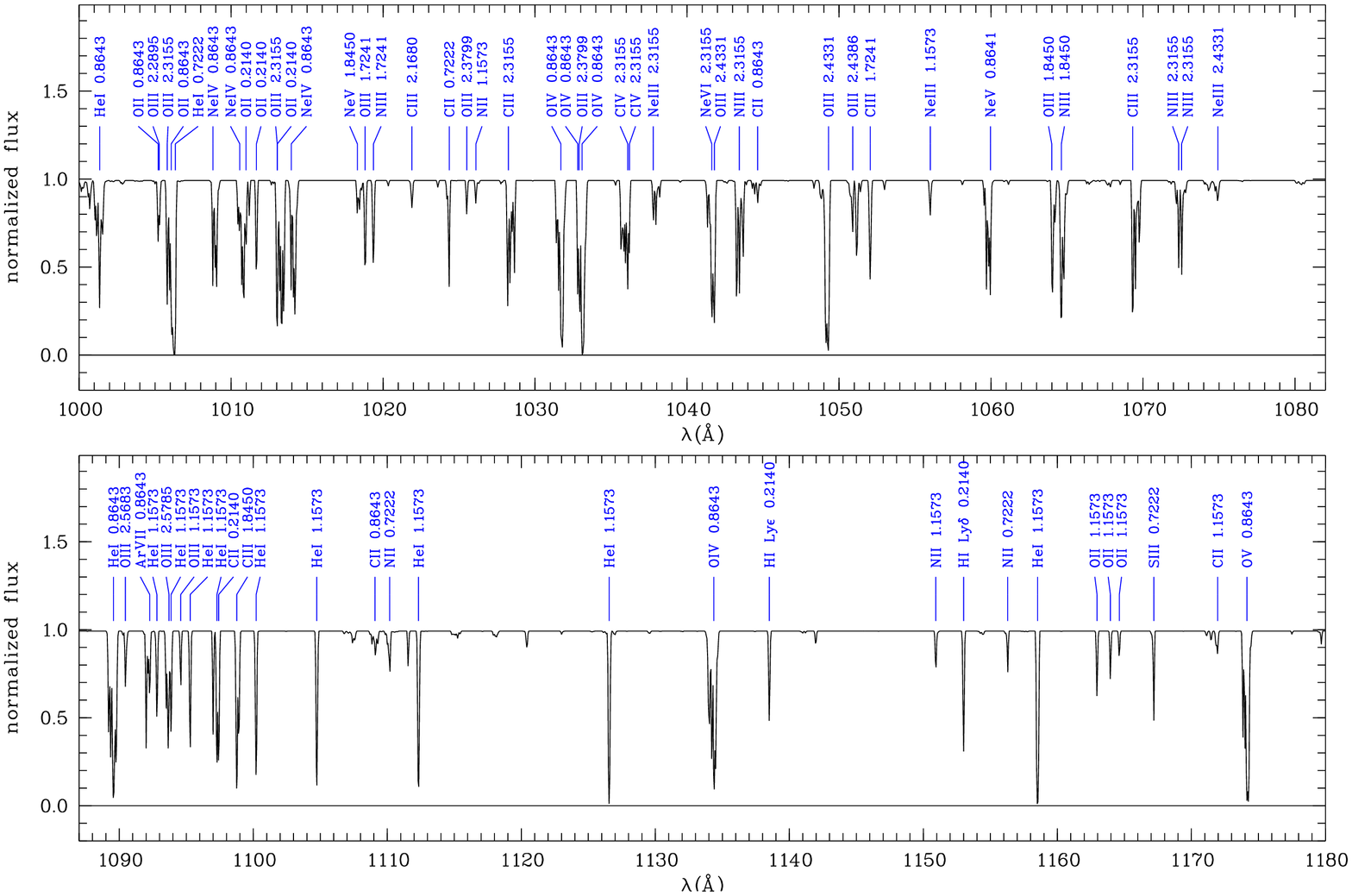}}
  \caption{Prediction of the metal line content in the FUSE spectral range based on the preferred ionizing energy distributions estimated in Section \ref{systems}. In case of equivalent models, the result based on the HM model is adopted. Prominent features are labeled with ion and system redshift.
  }
  \label{fuse_prognose}
\end{figure*}

\begin{figure}
  \centering
  \resizebox{\hsize}{!}{\includegraphics[bb=40 505 350 770,clip=]{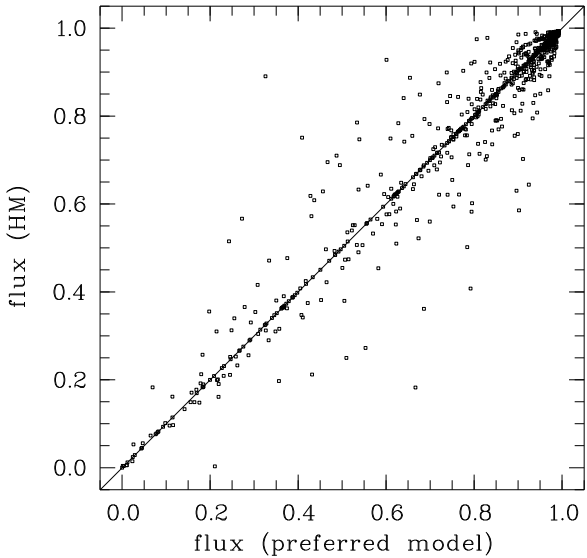}}
  \caption{Comparison of the fluxes predicted by the preferred models with those predicted by the HM models.
The flux level differs in 32\,\% of the pixels. The root mean square deviation of these pixel is 0.056.
  }
  \label{compare_prognose}
\end{figure}

The primary aim of the presented investigation was to derive a prediction of the metal lines arising in the FUSE spectral range.
The prediction of the metal line spectrum is important for the analysis of the \ion{He}{ii} features since pollution by unrecognized metal lines may lead to biased results concerning the \ion{He}{ii}/\ion{H}{i} ratio $\eta$ \citep{reimersetal_fuse, fechnerreimers_fuse}.
Thus, we summarize all line parameters produced by the preferred models discussed in Section \ref{systems} to compute a prediction for the appearance of the metal line spectrum in the FUSE spectral range ($1000 - 1180\,\mathrm{\AA}$).
For the 2 systems where we found several equivalent solutions, we adopt the HM or HM3 model, respectively, which lead to a preferred description of the respective systems.
The resulting expectation is presented in Fig. \ref{fuse_prognose}. 
Besides metal lines, the prediction also includes higher order Lyman series of \ion{H}{i} ($z=0.2140$) as well as \ion{He}{i} features of the systems $z=0.8643$ and $1.1573$.
A more conservative approach would be a prediction derived only using the \citet{haardtmadau2001} UV background.
The result would be qualitatively the same as shown in Fig. \ref{fuse_prognose} since the prominent features are similar.
A comparison between both models is shown in Fig. \ref{compare_prognose}.
In the FUSE spectral range 32\,\% of the pixels have different flux values if the preferred model or the HM model is considered.
The average root mean square deviation is 0.056 considering only those pixels with different fluxes.

For the systems only seen in \ion{C}{iv}, the $\lambda 312$ features can be computed directly from the fitted line parameters.
Only the system at $z = 2.3079$ contributes to the absorption in the FUSE spectral range and the expected features are weak ($\tau_0 \lesssim 0.2$).
Furthermore, they are located at $\sim 1033.5\,\mathrm{\AA}$ blended with the much stronger \ion{O}{iv} $\lambda 554$ feature ($\tau_0 \approx 151$) of the system at $z = 0.8643$.

Considering the multi-phase nature of several systems discussed in Section \ref{systems}, we certainly have preferred low ionized species on average (outstanding examples are the systems at $z=1.8450$ and $2.3155$).
The predicted line spectrum contains more features arising from low ionization stages (predominately \ion{O}{ii}, \ion{O}{iii}, \ion{C}{ii}, \ion{C}{iii}, \ion{N}{ii}, \ion{N}{iii}) than higher ionized ions like \ion{Ne}{iv}\,--\,\ion{Ne}{vi}, \ion{O}{v} and \ion{Ar}{vii}.
Of course, this is also due to the restwavelengths of the considered ions in combination with the redshifts of the absorbing systems, e.g. transitions with restwavelengths $< 400\,\mathrm{\AA}$ enter the FUSE spectral range only for $z \gtrsim 1.5$.
However, since the derived models often reveal problems to produce significant features of high ionization stages in the optical (like \ion{O}{vi}), the far-UV prediction probably misses highly ionized lines rather than transitions of low ionized elements. 
As discussed in the corresponding Sections, there are only two cases, where possible underestimated absorption features may bias the \ion{He}{ii} column density.
These are \ion{Ne}{vii} $\lambda 465$ of the $z=1.1573$ system and \ion{Ne}{iii} of the system at $z=2.3155$.
The absorption lines are expected to arise at $1003.6\,\mathrm{\AA}$ and $1037.9\,\mathrm{\AA}$, respectively.
Thus, these wavelengths should be kept in mind to be possible biased when analyzing the \ion{He}{ii} Ly$\alpha$ forest.

A comparison to the observed FUSE data and the inclusion of the prediction into the analysis of the \ion{He}{ii} Ly$\alpha$ forest can be found in \citet{reimersetal_fuse} and will be addressed more extensively in a future paper.

\section{Conclusions}\label{conclusions}

In the spectrum of the QSO HS~1700+6416, we detect 25 metal line systems. 
18 systems show absorption features in the present observational data taken with Keck/HIRES and HST/STIS, providing a sufficient number of ions in order to derive photoionization models.
The adopted systems (7 of them are optical thin Lyman limit systems) are located at redshifts $0.2 < z < 2.6 $.
For each system we derive 8 simple photoionization models based on different radiation backgrounds.
As shape of the ionizing radiation we adopt the \citet{haardtmadau2001} intergalactic background radiation and 3 different modifications of it, as well as the energy distribution of a starburst galaxy \citep{bruzualcharlot1993}, an AGN \citep{mathewsferland1987}, and two power laws with $\alpha = -1.5$ and $-0.5$.
Among these models, the best one is chosen by applying a $\chi^2$-test to supplement the visual inspection if possible.

For 9 systems (50\,\%) the preferred energy distribution is a modification of the UV background of \citet{haardtmadau2001}, where the break, usually located at 4\,Ryd, is shifted to 3\,Ryd.
Additionally, the HM-like spectrum with the break shifted to 2\,Ryd is found twice.
Thus, more than 50\,\% of the systems can be described with modified \citet{haardtmadau2001} spectra, where the helium break is shifted to energies $< 4\,\mathrm{Ryd}$.
For 3 absorbers the unchanged \citet{haardtmadau2001} radiation is preferred, where one can be equivalently modelled on the basis of other ionizing backgrounds.
Furthermore, the energy distribution of starburst galaxies \citep{bruzualcharlot1993} or quasars \citep{mathewsferland1987} lead to an appropriate description of the absorbers in 2 cases each.
All systems modeled with the ionizing radiation of starburst galaxies can be described alternatively with a different radiation background as well.
There is a clear tendency for modified HM shapes prevailing at higher redshift ($z > 2$), while at lower redshifts individual sources like AGN and starburst galaxies appear to become more important.

These results imply that the assumption of a uniform intergalactic ionization background with a shape according to \citet{haardtmadau2001} to investigate the metallicity of the IGM is at least problematic.
A comparison between the derived carbon abundance using the preferred radiation field to that derived from the standard HM energy distribution reveals, that assuming the HM background, the metallicity would be overestimated (underestimated), if the real ionizing radiation was softer (harder).
However, the individual models may improve by introducing multi-phase models.
In this case the choice of a different ionizing spectrum might be compensated by a more complicated model with several ionization phases.
Thus, our results are only preliminary and a future study adopting more sophisticated models is required.

Different shapes of the ionizing radiation for different absorbers imply that the \ion{He}{ii}/\ion{H}{i} ratio should vary when observed at distinct locations. 
Assuming a spectrum like that of \citet{haardtmadau2001} and varying the position of the break, which is actually located at 4\,Ryd, leads to values of \ion{He}{ii}/\ion{H}{i} in the range of $140 - 210$. 
The maximum is reached when the position of the break is $\sim 3.6\,\mathrm{Ryd}$.
The apparent correlation between the strength of the \ion{H}{i} absorption and the column density ratio $\eta$ in the sense that $\eta$ is higher in \ion{H}{i} voids \citep{shulletal2004, reimersetal_fuse} cannot be addressed here since we predominately probe dense gas in the vicinity of galaxies.

The presented analysis illustrates the potential of using metal line absorption systems to trace the shape of the intergalactic UV background using column density ratios.
A more sophisticated method would require that the ionizing background can be derived self-consistently from the observed absorption features of a metal line system also taking into account the multi-phase nature of most of the complex systems.
A possible approach introduced by \citet{agafonovaetal2005} and applied by \citet[][appendix A]{reimersetal2005}, which parameterizes an initial shape of the background, e.g. that of \citet{haardtmadau2001}, and fits the parameters in the course of the analysis using a response function.

Regarding the multitude of ions of elements like sulphur, oxygen, or neon with observable transitions in the spectral range covered by the STIS data, we emphasize the potential of high quality UV spectra, which would make it possible to obtain further constraints on the models. 
High resolution, high signal-to-noise UV data of quasars with numerous metal line systems in the optical would probably reveal many intrinsic EUV transitions as well, as the example of HS~1700+6416 shows.
Ratios of different ions from one element, for example $\ion{S}{ii} - \ion{S}{vi}$, would make the investigation more independent of the metallicity and possible deviations from the solar abundance pattern, which would help to estimate the shape of the ionizing background at the location of the absorber.

The derived photoionization models are used to compute the expected appearance of the metal line spectrum (as well as contribution of low redshift \ion{H}{i} absorption of higher order Lyman series and \ion{He}{i}) in the spectral range covered by FUSE ($1000 - 1180\,\mathrm{\AA}$).
In a future study, the predicted lines will be included in the analysis of the \ion{He}{ii} Ly$\alpha$ forest to avoid pollution of the \ion{He}{ii}/\ion{H}{i} ratio by unrecognized metal line absorption.

\begin{acknowledgements}
This work has been supported by the Verbundforschung (DLR) of the BMBF under Grant No. 50 OR 0203 and by the DFG under RE 353/49-1.
\end{acknowledgements}

\bibliographystyle{aa}
\bibliography{/data/hspc29/st1b305/text/bibtex/papers}

\begin{table*}
  \caption[]{Measured column densities for the metal absorption systems at $z < 2.0$.
Value marked with a colon (:) are derived from the STIS data and suffer from large uncertainties.
If no value is given, the corresponding feature is not present or severely blended, and no upper limit can be estimated.
  }
  \label{coldens}
  $$ 
  \begin{array}{l c c c c c c c c c }
    \hline
    \hline
    \noalign{\smallskip}
    \log N_{\mathrm{ion}} & 0.7222^{\mathrm{a}} & 0.7222 & 0.7222 & 0.8643^{\mathrm{b}} & 0.8643 & 0.8643 & 0.8643 & 0.8643 & 0.8643 \\
      & -46.5\,\mathrm{km\,s}^{-1} & -20.2\,\mathrm{km\,s}^{-1} & 0.0\,\mathrm{km\,s}^{-1} & -39.7\,\mathrm{km\,s}^{-1} & 0.0\,\mathrm{km\,s}^{-1} & 40.3\,\mathrm{km\,s}^{-1} & 57.2\,\mathrm{km\,s}^{-1} & 75.2\,\mathrm{km\,s}^{-1} & 111.3\,\mathrm{km\,s}^{-1} \\
    \noalign{\smallskip}
    \hline
    \noalign{\smallskip}
\ion{H}{i}   & 14.92  & 15.24  & 16.12  & 15.19  & 15.82  & 15.51  & 15.61  & 15.71  & 15.26  \\
\ion{N}{iii} & 13.91: & 15.16: & 14.49: & 14.13: & \dots  & 14.11: & 13.49: & 13.93: & \dots  \\
\ion{N}{iv}  & 13.80: & 13.97: & 12.72: & 13.59: & 13.20: & 13.31: & 13.47: & 12.85: & 13.17: \\
\ion{O}{i}   & 13.37: & 13.51: & 14.11: & \dots  & \dots  & \dots  & \dots  & \dots  & \dots  \\
\ion{O}{ii}  & 13.70: & 13.60: & 13.90: & \dots  & \dots  & \dots  & \dots  & \dots  & \dots  \\
\ion{O}{iii} & 14.70: & 15.77: & 14.39: & 14.50: & 14.57: & 14.95: & \dots  & 15.69: & 14.43: \\
\ion{O}{iv}  & 13.68: & 16.41: & 14.83: & 15.40: & 15.01: & 15.52: & \dots  & 15.12: & \dots   \\
\ion{Mg}{i}  & <10.40 & <10.34 & <10.32 & <10.18 & <10.18 & <10.17 & <10.17 & <10.17 & <10.17 \\
\ion{Mg}{ii} & 11.70\pm0.01 & 12.03\pm0.01 & 12.91\pm0.01 & 11.19\pm0.04 & 11.82\pm0.01 & 11.51\pm0.03 & 11.61\pm0.02 & 11.71\pm0.02 & 11.26\pm0.06 \\
\ion{Al}{iii}& \dots  & \dots  & \dots  & <11.76 & <11.76 & <11.84 & <11.76 & <11.76 & <11.76 \\
\ion{S}{ii}  & 12.87: & 12.81: & 12.60: & \dots  & \dots  & \dots  & \dots  & \dots  & \dots  \\
\ion{S}{iii} & 12.90: & 13.22: & 13.16: & 12.89: & 13.30: & 13.09: & 13.99: & 12.92: & 13.02: \\
\ion{S}{iv}  & 13.43: & 13.29: & 13.75: & 12.95: & 13.69: & 13.56: & 13.20: & 13.52: & 12.75:  \\
\ion{S}{v}   & 12.63: & 12.61: & 12.65: & 12.19: & 12.71: & 13.75: & 12.53: & 14.73: & 12.68:  \\
\ion{Fe}{ii} & <12.00 & <12.00 & 12.60\pm0.02 & <11.22 & <11.22 & <11.22 & <11.23 & <11.23 & <11.30 \\
\ion{Fe}{iii}& 13.29: & 14.29: & 14.21: & \dots  & \dots  & \dots  & \dots  & \dots  & \dots  \\
    \noalign{\smallskip}
    \hline
    \hline
    \noalign{\smallskip}
    \log N_{\mathrm{ion}} & 0.2140 & 1.1573^{\mathrm{c}} & 1.4941^{\mathrm{d}} & 1.4941 & 1.4941 & 1.7241^{\mathrm{e}} & 1.8450^{\mathrm{f}} & 1.8450 & 1.8450 \\
     & 0.0\,\mathrm{km\,s}^{-1} & 0.0\,\mathrm{km\,s}^{-1} & -22.9\,\mathrm{km\,s}^{-1} & 0.0\,\mathrm{km\,s}^{-1} & 20.5\,\mathrm{km\,s}^{-1} & 0.0\,\mathrm{km\,s}^{-1} & -20.8\,\mathrm{km\,s}^{-1} & 0.0\,\mathrm{km\,s}^{-1} & 42.7\,\mathrm{km\,s}^{-1} \\
    \noalign{\smallskip}
    \hline
    \noalign{\smallskip}
\ion{H}{i}   & 15.05: & 16.85  & 15.22  & 15.34  & 14.14  & 17.05  & 15.71  & 15.89  & 15.51  \\
\ion{He}{i}  & \dots  & 15.03: & \dots  & 14.22: & \dots  & 15.61: & 14.94: & 15.12: & 14.73: \\
\ion{C}{ii}  & 13.91: & \dots  & <12.48 & <12.45 & <12.50 & <13.00 & <12.03 & 13.864\pm0.014 & 13.400\pm0.020 \\
\ion{C}{iv}  & \dots  & \dots  & 13.42\pm0.02 & 13.54\pm0.01 & 12.34\pm0.11 & 13.23\pm0.01 & 13.45: & 14.446\pm0.003 & 13.769\pm0.003 \\
\ion{N}{ii}  & \dots  & 13.13: & \dots  & \dots  & \dots  & \dots  & \dots  & \dots  & \dots  \\
\ion{N}{iii} & \dots  & 13.77: & \dots  & \dots  & \dots  & \dots  & \dots  & \dots  & \dots \\
\ion{N}{v}   & \dots  & \dots  & \dots  & \dots  & \dots  &  \dots  & <12.47 & 13.20\pm0.05 & <12.70 \\
\ion{O}{iii} & \dots  & 14.63: & \dots  & \dots  & \dots  & 14.52: & 14.20: & 16.36: & 15.95: \\
\ion{O}{iv}  & \dots  & \dots  & 14.38: & 15.06: & 13.95: & 14.04: & \dots  & \dots  & \dots \\
\ion{O}{v}   & \dots  & 14.44: & \dots  & \dots  & \dots  & \dots  & \dots  & \dots  & \dots \\
\ion{Ne}{iii}& \dots  & \dots  & \dots  & \dots  & \dots  & 13.70: & \dots  & \dots  & \dots  \\
\ion{Ne}{iv} & \dots  & \dots  & \dots  & \dots  & \dots  & \dots  & 14.12: & 14.53: & 14.80: \\
\ion{Ne}{v}  & \dots  & \dots  & \dots  & \dots  & \dots  & \dots  & \dots  & \dots  & 13.47: \\
\ion{Ne}{vi} & \dots  & \dots  & \dots  & \dots  & \dots  & \dots  & 14.89: & \dots  & 14.99: \\
\ion{Ne}{vii}& \dots  & \dots  & \dots  & \dots  & \dots  & \dots  & 13.57: & \dots  & 13.86: \\
\ion{Mg}{i}  & 11.32\pm0.05 & \dots  & \dots  & \dots  & \dots & \dots & \dots  & \dots & \dots \\
\ion{Mg}{ii} & 12.76\pm0.02 & 12.680\pm0.004 & \dots  & \dots  & \dots & \dots & \dots & 12.65\pm0.10 & \dots  \\
\ion{Al}{ii} & \dots  & 12.11\pm0.03 & <10.50 & <10.50 & <10.50 & \dots  & <10.40 & 11.02\pm0.04 & <10.39 \\
\ion{Al}{iii}& \dots  & \dots & 10.70 & <10.69 & < 10.69 & <11.02 & <10.98 & 11.65\pm0.03 & <11.01 \\
\ion{Si}{ii} & 12.89: & \dots  & \dots  & \dots  & \dots & <11.45 & <11.39 & 12.293\pm0.021 & <11.39 \\
\ion{Si}{iii}& 12.98: & \dots  & \dots  & \dots  & \dots & \dots  & 12.57: &12.983\pm0.019 & 12.841\pm0.018 \\
\ion{Si}{iv} & \dots  & \dots  & 12.00\pm0.11 & 11.79\pm0.15 & 11.37: & 12.77\pm0.05 &  12.50: & 13.169\pm0.006 & 12.600\pm0.012\\
\ion{S}{iii} & \dots  & 13.96: & \dots  & \dots  & \dots & \dots  & \dots  & \dots  & \dots \\
\ion{S}{iv}  & \dots  & 12.96: & \dots  & \dots  & \dots & \dots  & \dots  & \dots  & \dots \\
\ion{Fe}{ii} & \dots  & 11.88\pm0.04 & <12.30 & <12.25 & <12.30 & \dots  & <11.67 & <11.67 & <11.68 \\
    \noalign{\smallskip}
    \hline
  \end{array}
  $$ 
\begin{list}{}{}
  \item[$^{\mathrm{a}}$] The total \ion{H}{i} measured from the Lyman break by \citet{vogelreimers1995} is $\log N_{\mathrm{LLS}} = 16.20$. The given distribution follows magnesium (\ion{Mg}{i} and \ion{Mg}{ii}).
  \item[$^{\mathrm{b}}$] The total \ion{H}{i} measured from the Lyman break by \citet{vogelreimers1995} is $\log N_{\mathrm{LLS}} = 16.35$. The given distribution follows \ion{Mg}{ii}.
  \item[$^{\mathrm{c}}$] The \ion{H}{i} feature is located outside the considered spectral portion. The column density is therefore adopted from \citet{vogelreimers1995}.
  \item[$^{\mathrm{d}}$] The \ion{H}{i} feature is located outside the used spectral portion. Therefore, the column density is adopted from \citet{vogelreimers1995}, who find $\log N(\ion{H}{i}) = 15.60$. It is distributed to the subsystems according to \ion{C}{iv}.
  \item[$^{\mathrm{e}}$] The \ion{H}{i} column density is adopted from \citet{vogelreimers1995}.
  \item[$^{\mathrm{f}}$] The total \ion{H}{i} fitted to the Lyman series is $\log N_{\mathrm{LLS}} = 16.21 \pm 0.11$. The given distributions follows silicon (\ion{Si}{ii}, \ion{Si}{iii}, and \ion{Si}{iv}). The column density of \ion{Mg}{ii} is adopted from \citet{trippetal1997}.
\end{list}
\end{table*}

\begin{table*}
  \caption[]{The same as Table \ref{coldens} but for the metal absorption systems at $2.0 < z < 2.4$.
  }
  \label{coldens2}
  $$ 
  \begin{array}{l c c c c c c c c }
    \hline
    \hline
    \noalign{\smallskip}
    \log N_{\mathrm{ion}} & 2.1680^{\mathrm{a}} & 2.1680 & 2.3155^{\mathrm{b}} & 2.3155 & 2.3155 & 2.3155 & 2.3155 & 2.3155 \\
       & -37.9\,\mathrm{km\,s}^{-1} & 0.0\,\mathrm{km\,s}^{-1} & -48.8\,\mathrm{km\,s}^{-1} & -33.3\,\mathrm{km\,s}^{-1} & 0.0\,\mathrm{km\,s}^{-1} & 35.1\,\mathrm{km\,s}^{-1} & 50.6\,\mathrm{km\,s}^{-1} & 76.9\,\mathrm{km\,s}^{-1} \\
    \noalign{\smallskip}
    \hline
    \noalign{\smallskip}
\ion{H}{i}   & 15.58  & 16.83  & 16.38  & \dots  & 15.91  & 15.53  & \dots  & 15.61   \\
\ion{C}{ii}  & 12.94\pm0.43 & 12.83\pm0.48 & 13.77\pm0.01 & \dots  & 13.56\pm0.01 & \dots  & \dots  & 12.69\pm0.10  \\
\ion{C}{iii} & \dots  & \dots  & 14.56: & 14.28: & 14.49: & 14.14: & 14.01: & 13.63\pm0.12 \\
\ion{C}{iv}  & 12.766\pm0.014 & 14.007\pm0.003 & 15.07\pm0.06 & \dots & 14.28\pm0.05 & 14.04\pm0.05 & 13.65\pm0.12 & 14.35\pm0.03 \\
\ion{N}{ii}  & <12.55 & <12.54 & 12.93\pm0.06 & <12.42 & 13.14\pm0.04 & <12.41 & <12.41 & <11.49  \\
\ion{N}{iii} & \dots  & \dots  & 14.76: & \dots  & 13.77: & \dots  & \dots  & \dots  \\
\ion{N}{v}   & <11.84 & <11.8  & 13.05\pm0.07 & 12.69\pm0.09 & 13.09\pm0.06 & 12.82\pm0.14 & 12.57\pm0.20 & 13.17\pm0.03 \\
\ion{O}{vi}  & \dots  & \dots  & 14.35\pm0.03 & \dots  & 14.45\pm0.18 & \dots  & \dots  & 13.84\pm0.08 \\
\ion{Ne}{iii}& 14.10: & 14.45: & \dots  & \dots  & \dots  & \dots  & \dots  & \dots \\
\ion{Ne}{v}  & 14.80: & 14.53: & \dots  & \dots  & \dots  & \dots  & \dots  & \dots \\
\ion{Ne}{vi} & \dots  & \dots  & 14.75: & \dots  & 14.48: & \dots  & \dots  & \dots \\
\ion{Ne}{vii}& \dots  & \dots  & 13.87: & \dots  & 13.45: & \dots  & \dots  & \dots \\
\ion{Al}{ii} & <10.49 & 10.90\pm0.09 & 11.43\pm0.02 & <10.51 & 11.44\pm0.04 & \dots  & \dots  & \dots  \\
\ion{Al}{iii}& <10.89 & <10.93 & 12.20\pm0.18 & \dots  & 11.95\pm0.18 & \dots  & \dots  & \dots \\
\ion{Si}{ii} & <11.02 & 11.73\pm0.07 & 13.00\pm0.01 & <11.51 & 12.73\pm0.03 & <11.51 & <11.49 & <11.48 \\
\ion{Si}{iii}& <10.93 & 13.056\pm0.046 & \dots  & \dots  & \dots & \dots  & \dots  & \dots \\
\ion{Si}{iv} & <11.02 & 13.241\pm0.002 & 13.748\pm0.004 & 12.965\pm0.09 & 13.435\pm0.013 & 12.251\pm0.019 & 11.918\pm0.018 & 12.930\pm0.005 \\
\ion{Fe}{ii} & <11.90 & <11.90 & \dots  & \dots  & \dots  & \dots  & \dots  & \dots  \\
    \noalign{\smallskip}
    \hline
    \hline
    \noalign{\smallskip}
    \log N_{\mathrm{ion}} & 2.0211^{\mathrm{c}} & 2.0211 & 2.0211 & 2.0211 & 2.1278^{\mathrm{c}} & 2.1989^{\mathrm{c}} & 2.1989 & 2.1989 \\
       & -63.5\,\mathrm{km\,s}^{-1} & -35.3\,\mathrm{km\,s}^{-1} & 0.0\,\mathrm{km\,s}^{-1} & 17.2\,\mathrm{km\,s}^{-1} & 0.0\,\mathrm{km\,s}^{-1} & -73.5\,\mathrm{km\,s}^{-1} & -41.6\,\mathrm{km\,s}^{-1} & 0.0\,\mathrm{km\,s}^{-1} \\
    \noalign{\smallskip}
    \hline
    \noalign{\smallskip}
\ion{H}{i}   & \dots  & \dots  & 15.29\pm0.04 & \dots  & 14.073\pm0.004 & \dots  & \dots  & 15.437\pm0.024 \\
\ion{C}{iv}  & 12.17\pm0.03 & 11.45\pm0.19 & 12.71\pm0.02 & 12.05\pm0.08 & 12.692\pm0.012 & 11.72\pm0.11 & 12.05\pm0.05 & 12.87\pm0.01 \\
\ion{Al}{ii} & <10.48 & <10.48 & <10.48 & <10.57 & \dots & <11.45 & <11.44 & <11.44 \\
\ion{Si}{ii} & <11.23 & <11.23 & <11.23 & <11.23 & <11.36 & <11.37 & <11.37 & <11.37 \\
\ion{Si}{iii}& <11.05 & <11.26 & <11.05 & <11.08 & <10.88 & \dots  & \dots  & \dots  \\
\ion{Si}{iv} & <11.14 & <11.13 & <11.13 & <11.13 & <11.05 & <11.28 & <11.30 & <11.27 \\
\ion{Fe}{ii} & <12.02 & <12.02 & <12.02 & <12.04 &  <11.92 & <11.95 & <11.95 & <11.94 \\
    \noalign{\smallskip}
    \hline
    \hline
    \noalign{\smallskip}
    \log N_{\mathrm{ion}} & 2.2895 & 2.2895 & 2.2895 & 2.3079^{\mathrm{e}} & 2.3079 & 2.3079 & 2.3799 \\
       & -22.4\,\mathrm{km\,s}^{-1} & 0.0\,\mathrm{km\,s}^{-1} & 4.3\,\mathrm{km\,s}^{-1} & -38.0\,\mathrm{km\,s}^{-1} & 0.0\,\mathrm{km\,s}^{-1} & 15.9\,\mathrm{km\,s}^{-1} & 0.0\,\mathrm{km\,s}^{-1} \\
    \noalign{\smallskip}
    \hline
    \noalign{\smallskip}
\ion{H}{i}   & \dots  & 16.00\pm0.07 & \dots  & \dots  & 16.43\pm0.02 & \dots  & 15.41\pm0.17 \\
\ion{C}{ii}  & <12.58 & <12.45 & <12.48 & <11.64 & <11.63 & <11.64 & 13.36\pm0.65 \\
\ion{C}{iii} & \dots & 13.19\pm 0.09 & \dots & \dots  & \dots  & \dots  & 13.50\pm0.19 \\
\ion{C}{iv}  & 12.19\pm0.83 & 12.51\pm0.34 & 12.37\pm0.15 & 12.54\pm0.05 & 12.73\pm0.08 & 13.23\pm0.03 & 13.059\pm0.005 \\
\ion{N}{ii}  & <12.40 & <12.40 & <12.40 & <12.40 & <12.38 & <12.40 & <12.38 \\ 
\ion{N}{v}   & \dots  & \dots  & \dots  & <12.06 & <12.07 & <12.07 & <11.71 \\
\ion{O}{vi}  & \dots  & \dots  & \dots  & \dots  & \dots  & \dots  & 14.497\pm0.025  \\
\ion{Al}{ii} & <10.40 & <10.40 & <10.40 & <10.60 & <10.59 & <10.60 & <10.38 \\
\ion{Si}{ii} & <11.36 & <11.36 & <11.36 & <11.49 & <11.49 & <11.49 & \dots  \\
\ion{Si}{iii}& \dots  & \dots  & \dots  & \dots  & \dots  & \dots  & 11.689\pm0.035 \\
\ion{Si}{iv} & 11.13\pm0.16 & 11.30\pm0.09 & \dots  & <11.17 & <11.17 & <11.17 & 11.689\pm0.037 \\
\ion{Fe}{ii} & <12.01 & <12.01 & <12.01 & <11.89 & <11.89 & <11.89 & <12.01 \\
    \noalign{\smallskip}
    \hline
  \end{array}
  $$ 
\begin{list}{}{}
  \item[$^{\mathrm{a}}$] The total column density of hydrogen $\log N_{\mathrm{LLS}} = 16.85$ is adopted from \citet{vogelreimers1995} and distributed to the subsystems according to \ion{C}{iv}.
  \item[$^{\mathrm{b}}$] The total \ion{H}{i} fitted to the Lyman series is $\log N_{\mathrm{LLS}} = 16.56 \pm 1.06$. The given distribution follows the portions of carbon (\ion{C}{ii}, \ion{C}{iii}, and \ion{C}{iv}). \ion{Al}{iii} arises outside the observed spectral portion, the corresponding values are taken from \citet{trippetal1997}.
  \item[$^{\mathrm{c}}$] For these systems no models are derived due to the lack of detected transitions.
\end{list}
\end{table*}

\begin{table*}
  \caption[]{The same as Table \ref{coldens} but for the metal absorption systems at $2.4 < z < 2.5$.
  }
  \label{coldens3}
  $$ 
  \begin{array}{l c c c c c c c }
    \hline
    \hline    
    \noalign{\smallskip}
    \log N_{\mathrm{ion}} & 2.4321^{\mathrm{a}} & 2.4321 & 2.4321 & 2.4331^{\mathrm{b}} & 2.4331 & 2.4331 & \\
      & -23.2\,\mathrm{km\,s}^{-1} & 0.0\,\mathrm{km\,s}^{-1} & 52.8\,\mathrm{km\,s}^{-1} & 0.0\,\mathrm{km\,s}^{-1} & 32.4\,\mathrm{km\,s}^{-1} & 50.6\,\mathrm{km\,s}^{-1} & \\
    \noalign{\smallskip}
    \hline
    \noalign{\smallskip}
\ion{H}{i}   & 15.16  & 14.96  & 15.02  & 16.41  & 16.31  & 16.37  & \\
\ion{C}{ii}  & <11.45 & <11.45 & <11.45 & 12.808\pm0.011 & 12.489\pm0.019 & 11.71: & \\
\ion{C}{iii} & 12.98\pm0.07 & 12.01\pm0.20 & \dots & 13.780\pm0.098 & \dots  & 14.242\pm0.165 & \\
\ion{C}{iv}  & 12.25\pm0.05 & 12.78\pm0.02 & 12.70\pm0.04 & 13.192\pm0.024 & 13.069\pm0.039 & 12.667\pm0.045 & \\
\ion{N}{ii}  & <12.23 & <12.20 & <12.23 & <12.25 & <12.29 & <12.21 & \\
\ion{N}{v}   & <11.98 & <11.98 & <12.00 & <11.68 & <11.68 & <11.68 & \\
\ion{O}{iii} & \dots  & \dots  & \dots  & 15.61: & \dots  & 14.83: & \\
\ion{O}{vi}  & <12.38 & <12.30 & <12.38 & <13.10 & <13.10 & <13.00 & \\
\ion{Ne}{vi} & 14.09: & 13.94: & 14.61: & 14.91: & \dots  & 14.19: & \\
\ion{Al}{ii} & <10.33 & <11.39 & <11.29 & 10.977\pm0.037 & <11.50 & 11.003\pm0.047 & \\
\ion{Si}{ii} & <11.33 & <11.39 & <11.39 & 12.357\pm0.039 & \dots  & 12.138\pm0.076 & \\
\ion{Si}{iii}& \dots  & 12.69\pm0.26 & 12.30\pm1.11 & 12.824\pm0.027 & 12.820\pm0.088 & 12.799\pm0.047 & \\
\ion{Si}{iv} & <11.14 & <11.10 & <11.15 & 12.609\pm0.008 & 12.429\pm0.011 & 12.476\pm0.012 & \\
\ion{S}{vi}  & \dots  & \dots  & \dots  & 13.85\pm1.34 & <12.00 & 13.21\pm0.21 & \\
\ion{Ar}{iv} & 13.00: & \dots  & 12.33: & 12.93: & 12.82: & \dots  & \\
\ion{Ar}{v}  & \dots  & \dots  & 12.44: & 12.14: & 12.64: & \dots  & \\
\ion{Fe}{ii} & <12.13 & <12.13 & <12.13 & <12.04 & <12.11 & <12.08 & \\
    \noalign{\smallskip}
    \hline
    \hline
    \noalign{\smallskip}
    \log N_{\mathrm{ion}} & 2.4386^{\mathrm{c}} & 2.4386 & 2.4386 & 2.4386 & 2.4405^{\mathrm{d}} & 2.4405 & 2.4965 \\
        & -72.3\,\mathrm{km\,s}^{-1} & -28.6\,\mathrm{km\,s}^{-1} & 0.0\,\mathrm{km\,s}^{-1} & 19.1\,\mathrm{km\,s}^{-1} & -71.1\,\mathrm{km\,s}^{-1} & 0.0\,\mathrm{km\,s}^{-1} & 0.0\,\mathrm{km\,s}^{-1}\\
    \noalign{\smallskip}
    \hline
    \noalign{\smallskip}
\ion{H}{i}   & 14.32  & 15.13  & 15.30  & 15.40  & 15.490 \pm 0.037 & 14.45\pm0.24 & 14.536\pm0.011 \\
\ion{C}{ii}  & <11.47 & <11.48 & 11.89\pm0.13 & 11.98\pm0.10 & <11.42 & <11.48 & <11.63 \\
\ion{C}{iii} & 12.84\pm0.03 & 13.06\pm0.04 & 14.50\pm0.37 & \dots & 12.45\pm0.06 & 13.64: & 13.27: \\
\ion{C}{iv}  & 12.67\pm0.01 & 12.71\pm0.02 & 12.96\pm0.18 & 12.85\pm0.10 & 12.589\pm0.018 & 13.065\pm0.017 & 12.177\pm0.028 \\
\ion{N}{iii} & <12.56 & <12.57 & <12.36 & <12.56 & <12.55 & <12.55 & <12.51 \\
\ion{N}{v}   & <11.69 & <11.69 & <11.69 & <11.69 & <12.07 & <12.07 & <11.94 \\
\ion{O}{iii} & \dots  & 14.36: & 14.29: & \dots  & 13.90: & \dots  & 13.27: \\
\ion{O}{vi}  & \dots  & \dots  & \dots  & \dots  & 13.484\pm0.065 & 13.20\pm0.11 & 13.739\pm0.007 \\
\ion{Al}{ii} & <10.31 & <10.30 & <10.30 & <10.30 & <10.32 & <10.31 & <10.34 \\
\ion{Si}{ii} & <11.29 & <11.29 & <11.29 & <11.29 & <11.30 & <11.30 & <11.41 \\
\ion{Si}{iii}& \dots  &  12.69\pm0.26 & 12.30\pm1.11 & \dots & <10.68 & <10.70 & \dots\\
\ion{Si}{iv} & \dots  & \dots  & \dots  & \dots  & <11.12 & 11.59\pm0.09 & <11.10 \\
\ion{S}{vi}  & \dots  & \dots  & \dots  & \dots  & \dots  & \dots  & <12.60 \\
\ion{Fe}{ii} & <12.10 & <12.11 & <12.11 & <12.03 & <12.04 & <12.05 & <12.01 \\
    \noalign{\smallskip}
    \hline
  \end{array}
  $$ 
\begin{list}{}{}
  \item[$^{\mathrm{a}}$] The total hydrogen column density measured for the Lyman series is $\log N (\ion{H}{i}) = 15.530 \pm 0.029$. The values for the subsystems are distributed according to carbon (\ion{C}{iii} and \ion{C}{iv}).
  \item[$^{\mathrm{b}}$] The total hydrogen column density measured for the Lyman series is $\log N (\ion{H}{i}) = 16.84 \pm 0.35$. The values for the subsystems are distributed according to silicon (\ion{Si}{ii}, \ion{Si}{iii}, and \ion{Si}{iv}).
  \item[$^{\mathrm{c}}$] The total hydrogen column density measured for the Lyman series is $\log N (\ion{H}{i}) = 15.783 \pm 0.066$. The values for the subsystems are distributed according to silicon (\ion{Si}{iii} and \ion{Si}{iv}).
  \item[$^{\mathrm{d}}$] The component at $-71.1\,\mathrm{km\,s}^{-1}$ may also be considered as a distinct system at $z = 2.4397$.
\end{list}
\end{table*}

\begin{table*}
  \caption[]{The same as Table \ref{coldens} but for metal absorption systems at $z > 2.5$.
  }
  \label{coldens4}
  $$ 
  \begin{array}{l c c c c c c c }
    \hline
    \hline
    \noalign{\smallskip}
    \log N_{\mathrm{ion}} & 2.5683^{\mathrm{a}} & 2.5683 & 2.5785^{\mathrm{b}} & 2.5785 & 2.5785 & 2.5785 \\
        & 0.0\,\mathrm{km\,s}^{-1} & 16.9\,\mathrm{km\,s}^{-1} & -30.7\,\mathrm{km\,s}^{-1} & -18.5\,\mathrm{km\,s}^{-1} & 18.5\,\mathrm{km\,s}^{-1} & 40.2\,\mathrm{km\,s}^{-1} \\
    \noalign{\smallskip}
    \hline
    \noalign{\smallskip}
\ion{H}{i}   & 14.28  & 14.09  & 15.04  & 15.39  & 14.93  & 15.58  \\
\ion{C}{ii}  & <11.63 & <11.63 & <11.68 & <11.68 & <11.67 & <11.68 \\
\ion{C}{iii} & 12.82\pm0.11 & 12.45\pm0.26 & 12.54\pm0.10 & 12.18\pm0.17 & 12.77\pm9.59 & 13.30\pm0.02 \\
\ion{C}{iv}  & 12.54\pm0.13 & 12.53\pm0.15 & 12.68\pm0.06 & 13.23\pm0.01 & 12.06\pm0.16 & 13.05\pm0.02 \\
\ion{N}{ii}  & <12.10 & <12.08 & \dots  & \dots  & \dots  & \dots  \\
\ion{N}{iii} & \dots  & \dots  & 13.35: & \dots  & 12.97: & \dots  \\
\ion{N}{v}   & <11.59 & <11.59 & <11.86 & <11.86 & <11.86 & <11.86 \\
\ion{O}{iii} & \dots  & \dots  & 13.76: & 14.43: & \dots  & 13.94: \\
\ion{O}{vi}  & 13.27\pm0.14 & \dots  & \dots  & 14.18\pm0.01 & \dots  & 12.88\pm0.08  \\
\ion{Al}{ii} & <10.67 & <10.67 & <10.96 & <10.96 & <10.94 & <10.94 \\
\ion{Si}{ii} & <11.51 & <11.51 & <11.57 & <11.57 & <11.57 & <11.57 \\
\ion{Si}{iii}& 11.654\pm0.033 & \dots  & 11.63\pm0.05 & 11.90\pm0.02 & <10.65 & 11.76\pm0.02 \\ 
\ion{Si}{iv} & 11.28\pm0.11 & <11.11 & 11.63\pm0.13 & <11.20 & <11.20 & 11.69\pm0.05 \\
\ion{Fe}{ii} & <11.80 & <11.80 & <11.78 & <11.78 & <11.78 & \dots  \\
    \noalign{\smallskip}
    \hline
    \hline
    \noalign{\smallskip}
    \log N_{\mathrm{ion}} & 2.7124^{\mathrm{c}} & 2.7124 & 2.7124 & 2.7164^{\mathrm{c}} & 2.7164 & 2.7164 \\
        & -36.5\,\mathrm{km\,s}^{-1} & 0.0\,\mathrm{km\,s}^{-1} & 77.5\,\mathrm{km\,s}^{-1} & -87.0\,\mathrm{km\,s}^{-1} & 0.0\,\mathrm{km\,s}^{-1} & 90.8\,\mathrm{km\,s}^{-1} \\
    \noalign{\smallskip}
    \hline
    \noalign{\smallskip}
\ion{H}{i}   & 11.915\pm0.034 & 13.385\pm0.003 & 13.231\pm0.004 & 12.597\pm0.016 & 12.627\pm0.022 & 13.002\pm0.045 \\
\ion{C}{ii}  & <11.99 & <11.99 & <11.99 & <11.73 & <11.75 & <11.75 \\
\ion{C}{iii} & <11.57 & \dots  & \dots  & \dots  & <11.60 & \dots  \\
\ion{C}{iv}  & 12.513\pm0.100 & 13.256\pm0.017 & 12.414\pm0.082 & <11.32 & 12.069\pm0.148 & <11.31 \\
\ion{N}{ii}  & <12.00 & <12.00 & <12.00 & \dots  & <12.50 & <12.48 \\
\ion{N}{iii} & <12.30 & <12.28 & <12.26 & \dots  & <12.27 & <12.27 \\
\ion{N}{v}   & 12.927\pm0.218 & 13.584\pm0.053 & 12.379\pm0.044 & 12.289\pm0.035 & 13.178\pm0.007 & <11.49 \\
\ion{O}{vi}  & \dots  & \dots  & \dots  & 13.659\pm0.009 & 14.227\pm0.007 & 13.517\pm0.013 \\
\ion{Ne}{vi} & 13.95: & 14.35: & 14.51: & \dots  & \dots  & \dots  \\
\ion{Si}{ii} & <10.74 & <10.73 & <10.72 & <11.70 & <11.72 & <11.78 \\
\ion{Si}{iii}& <10.50 & <10.46 & <10.47 & \dots  & \dots  & \dots  \\
\ion{Si}{iv} & <11.00 & <11.00 & <11.00 & <11.00 & <11.01 & <11.01 \\
\ion{S}{vi}  & <12.14 & 12.211\pm0.088 & <12.18 & <12.10 & <12.41 & <12.43 \\
\ion{Fe}{ii} & <12.19 & <12.18 & <12.20 & <12.41 & <12.41 & <12.43 \\
    \noalign{\smallskip}
    \hline
  \end{array}
  $$ 
\begin{list}{}{}
  \item[$^{\mathrm{a}}$] The total hydrogen column density estimated by fitting simultaneously Ly$\alpha$ and Ly$\beta$ is $\log N (\ion{H}{i}) = 14.50 \pm 0.01$. The values for the subcomponents are distributed according to carbon (\ion{C}{iii} and  \ion{C}{iv}).
  \item[$^{\mathrm{b}}$] The total hydrogen column density estimated by fitting the Lyman series is $\log N (\ion{H}{i}) = 15.549 \pm 0.022$ at $v \approx -50\,\mathrm{km\,s}^{-1}$ and $\log N (\ion{H}{i}) = 15.663 \pm 0.037$ at $v \approx 25\,\mathrm{km\,s}^{-1}$. The values for the subcomponents are distributed according to carbon (\ion{C}{iii} and  \ion{C}{iv}).
  \item[$^{\mathrm{c}}$] System appears to be associated to the QSO and is not modelled. 
\end{list}
\end{table*}

\clearpage
\appendix
\section{Figures}

\begin{figure*}
  \centering
  \resizebox{\hsize}{!}{\includegraphics[bb=35 380 545 777,clip=]{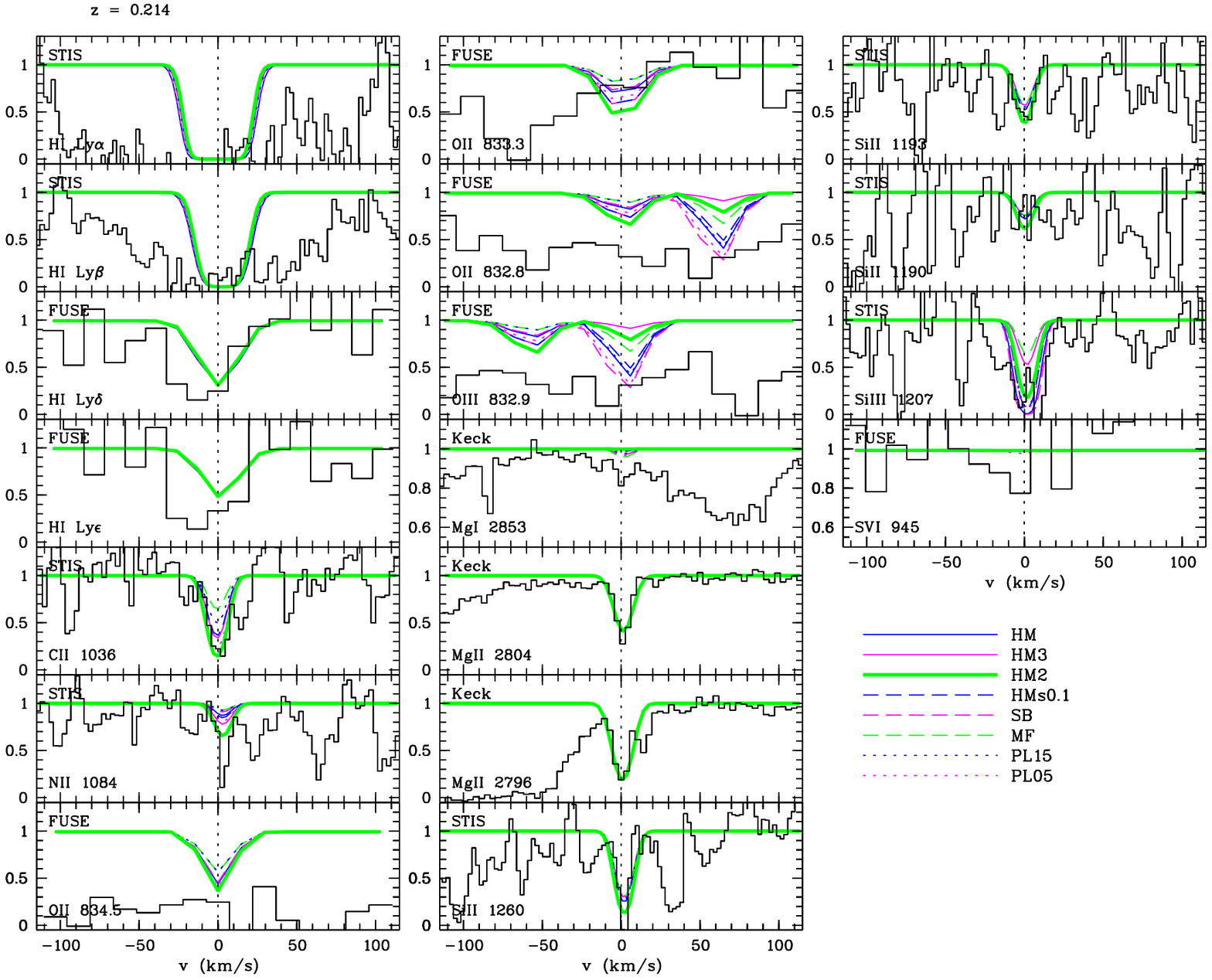}}
  \caption{Observed and modelled absorption lines of the system at $z = 0.2140$. The preferred model is HM2.
}
  \label{z0.2140fig}
\end{figure*}

\begin{figure*}
  \centering
  \resizebox{\hsize}{!}{\includegraphics[bb=35 275 545 780,clip=]{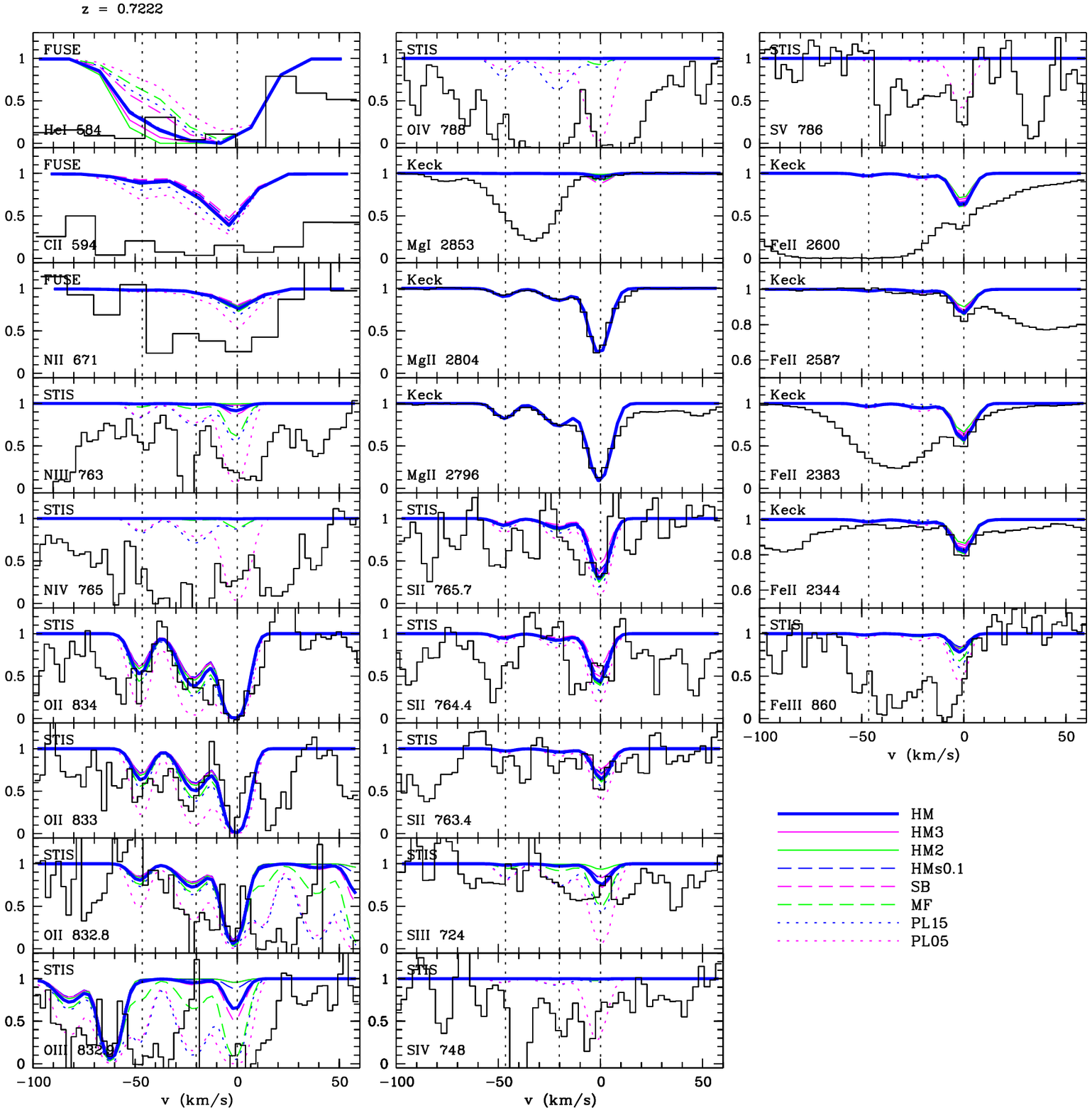}}
  \caption{Observed and modelled absorption lines of the system at $z = 0.7222$. The preferred model is HM.
}
  \label{z0.7222fig}
\end{figure*}

\begin{figure*}
  \centering
  \resizebox{\hsize}{!}{\includegraphics[bb=35 275 545 780,clip=]{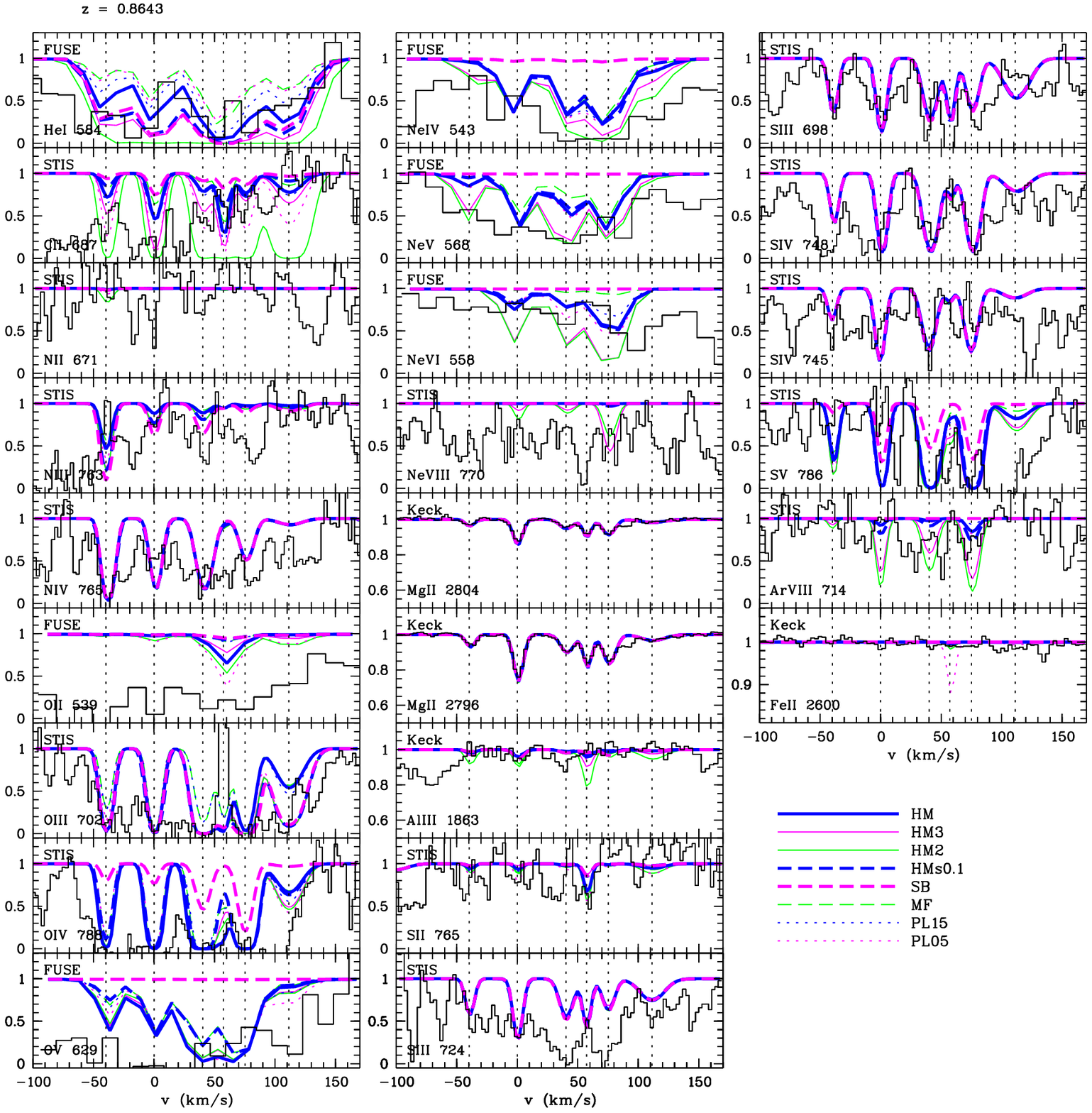}}
  \caption{Observed and modelled absorption lines of the system at $z = 0.8643$. The preferred models are HM, HMs0.1, and SB.
}
  \label{z0.8643fig}
\end{figure*}

\begin{figure*}
  \centering
  \resizebox{\hsize}{!}{\includegraphics[bb=35 275 545 780,clip=]{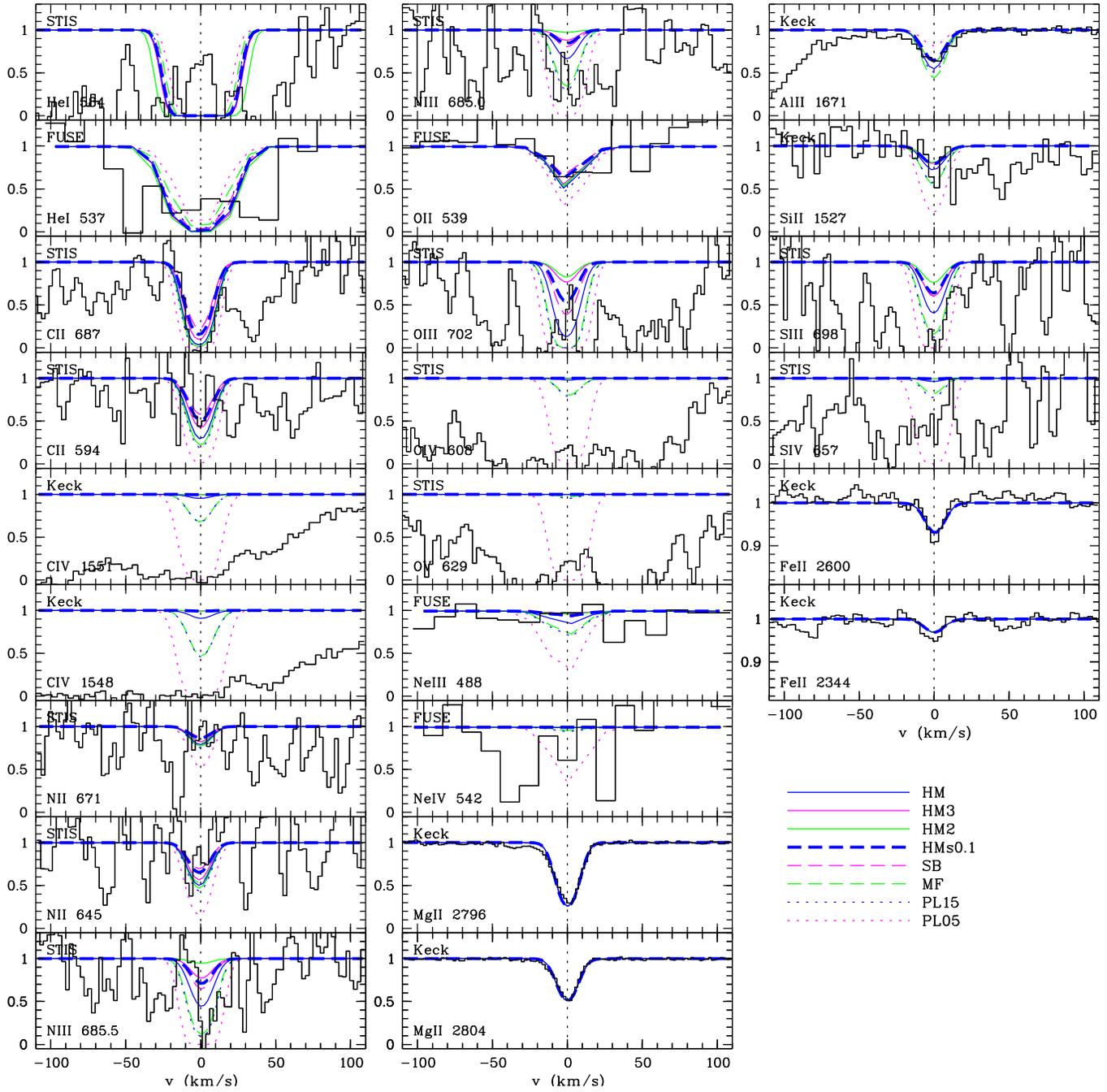}}
  \caption{Observed and modelled absorption lines of the system at $z = 1.1573$. The preferred model is HMs0.1.
}
  \label{z1.1573fig}
\end{figure*}

\begin{figure*}
  \centering
  \resizebox{\hsize}{!}{\includegraphics[bb=35 435 545 780,clip=]{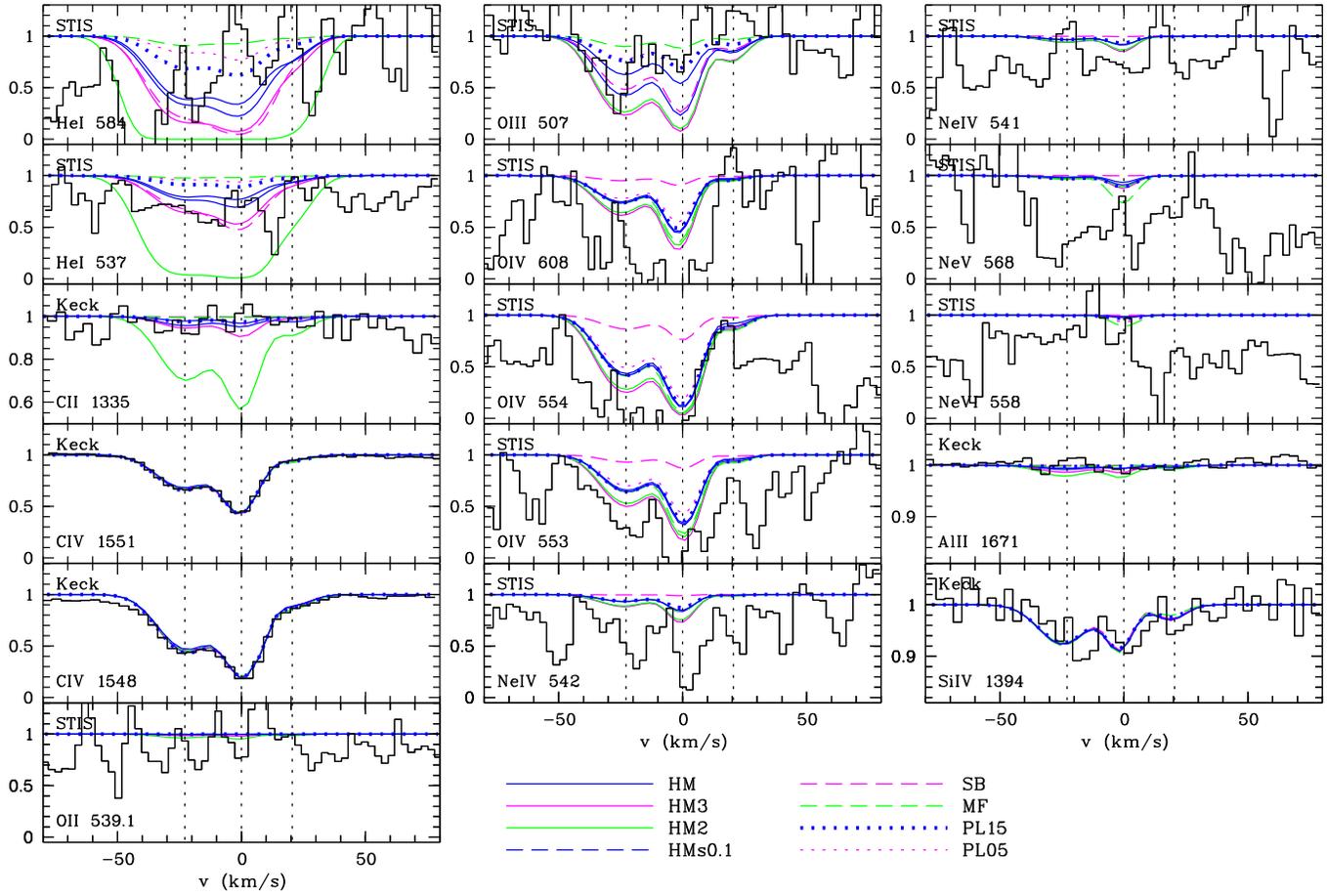}}
  \caption{Observed and modelled absorption lines of the system at $z = 1.4941$. The preferred model is PL15.
}
  \label{z1.4941fig}
\end{figure*}

\begin{figure*}
  \centering
  \resizebox{\hsize}{!}{\includegraphics[bb=35 275 545 780,clip=]{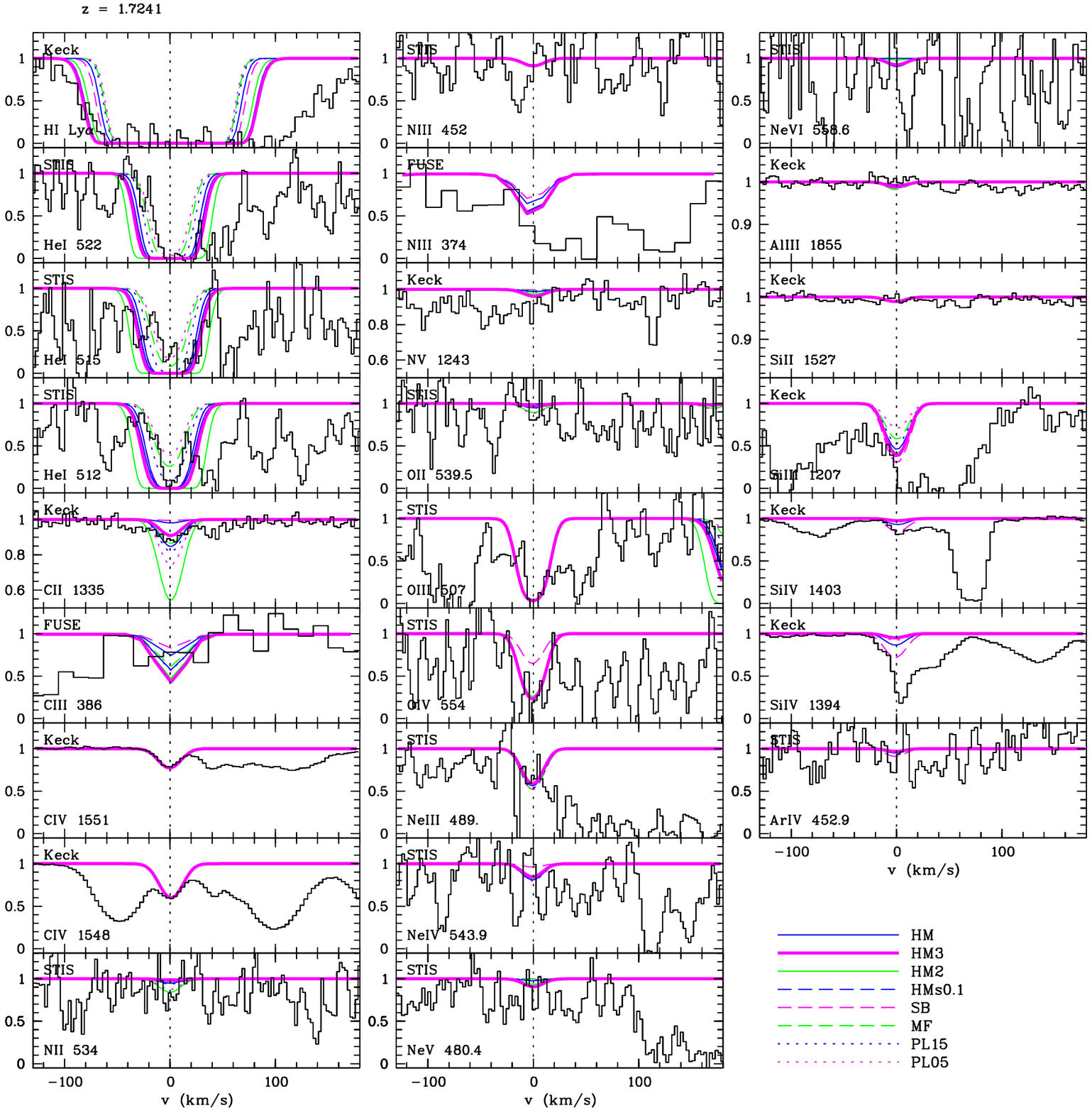}}
  \caption{Observed and modelled absorption lines of the system at $z = 1.7241$. The preferred model is HM3.
}
  \label{z1.7241fig}
\end{figure*}

\begin{figure*}
  \centering
  \resizebox{\hsize}{!}{\includegraphics[bb=35 110 545 780,clip=]{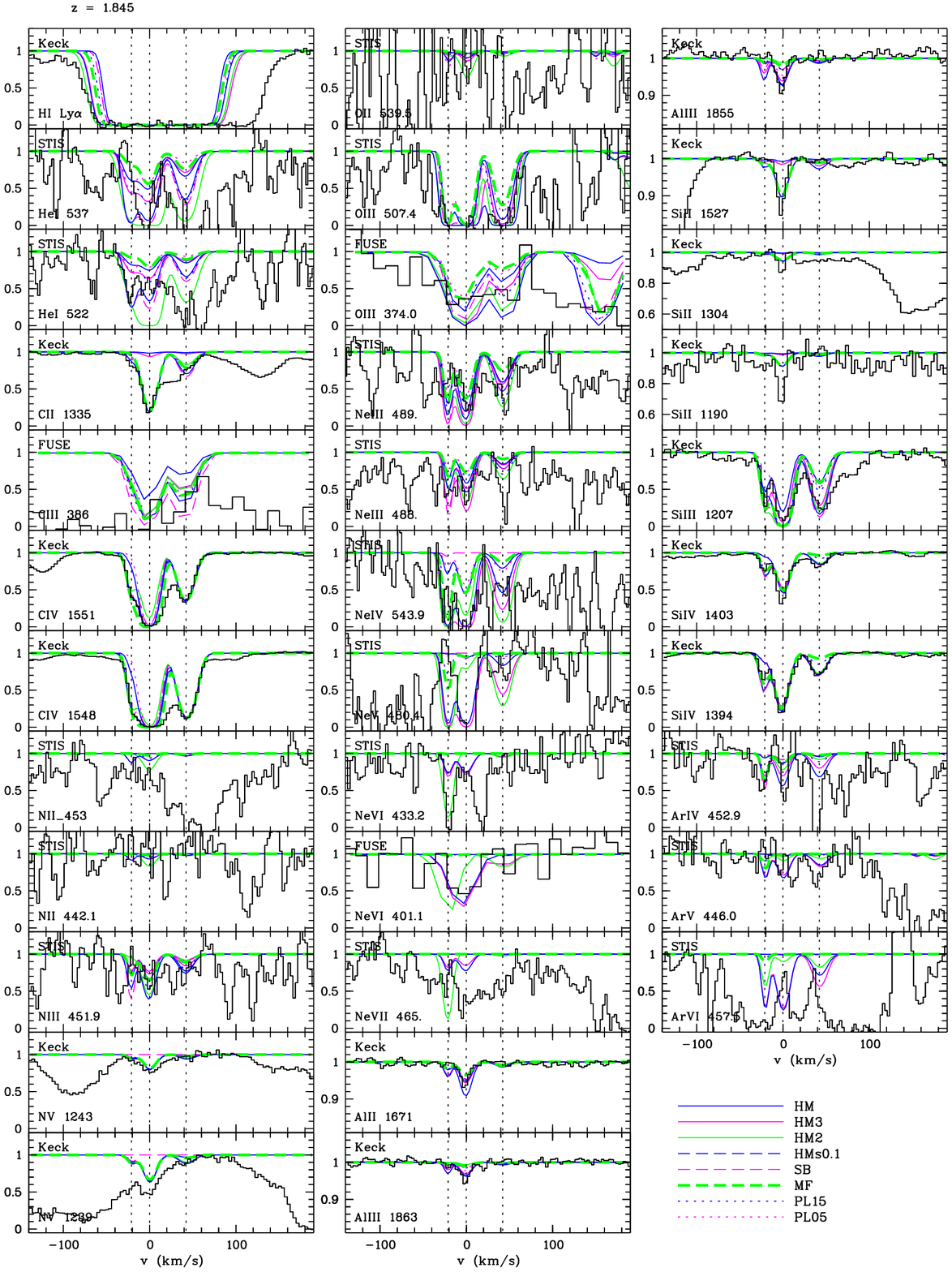}}
  \caption{Observed and modelled absorption lines of the system at $z = 1.8450$. The preferred model is MF.
}
  \label{z1.8450fig}
\end{figure*}

\begin{figure*}
  \centering
  \resizebox{\hsize}{!}{\includegraphics[bb=35 270 545 780,clip=]{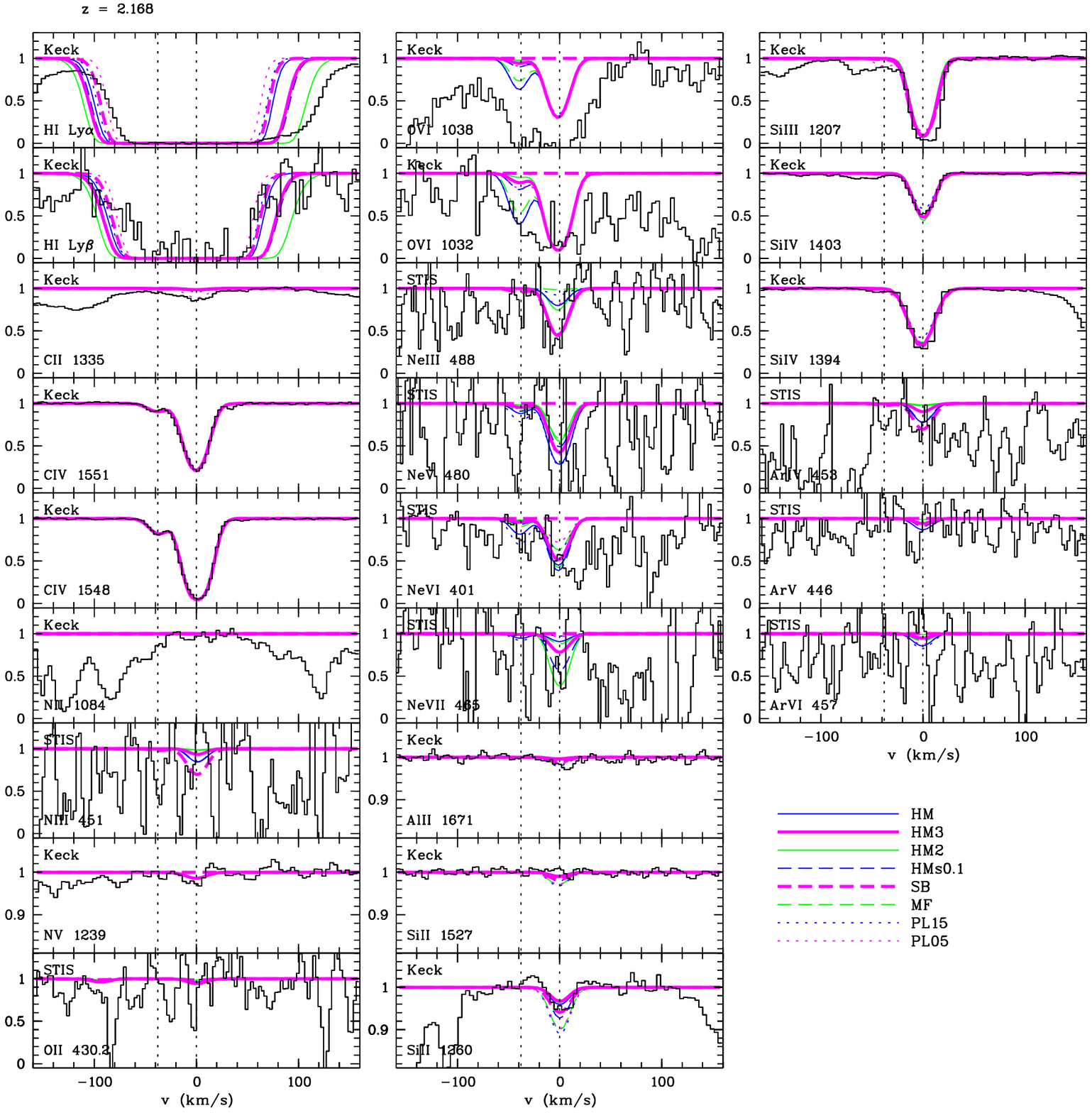}}
  \caption{Observed and modelled absorption lines of the system at $z = 2.1680$. The preferred models are HM3 and SB.
}
  \label{z2.1680fig}
\end{figure*}

\begin{figure*}
  \centering
  \resizebox{\hsize}{!}{\includegraphics[bb=35 380 545 780,clip=]{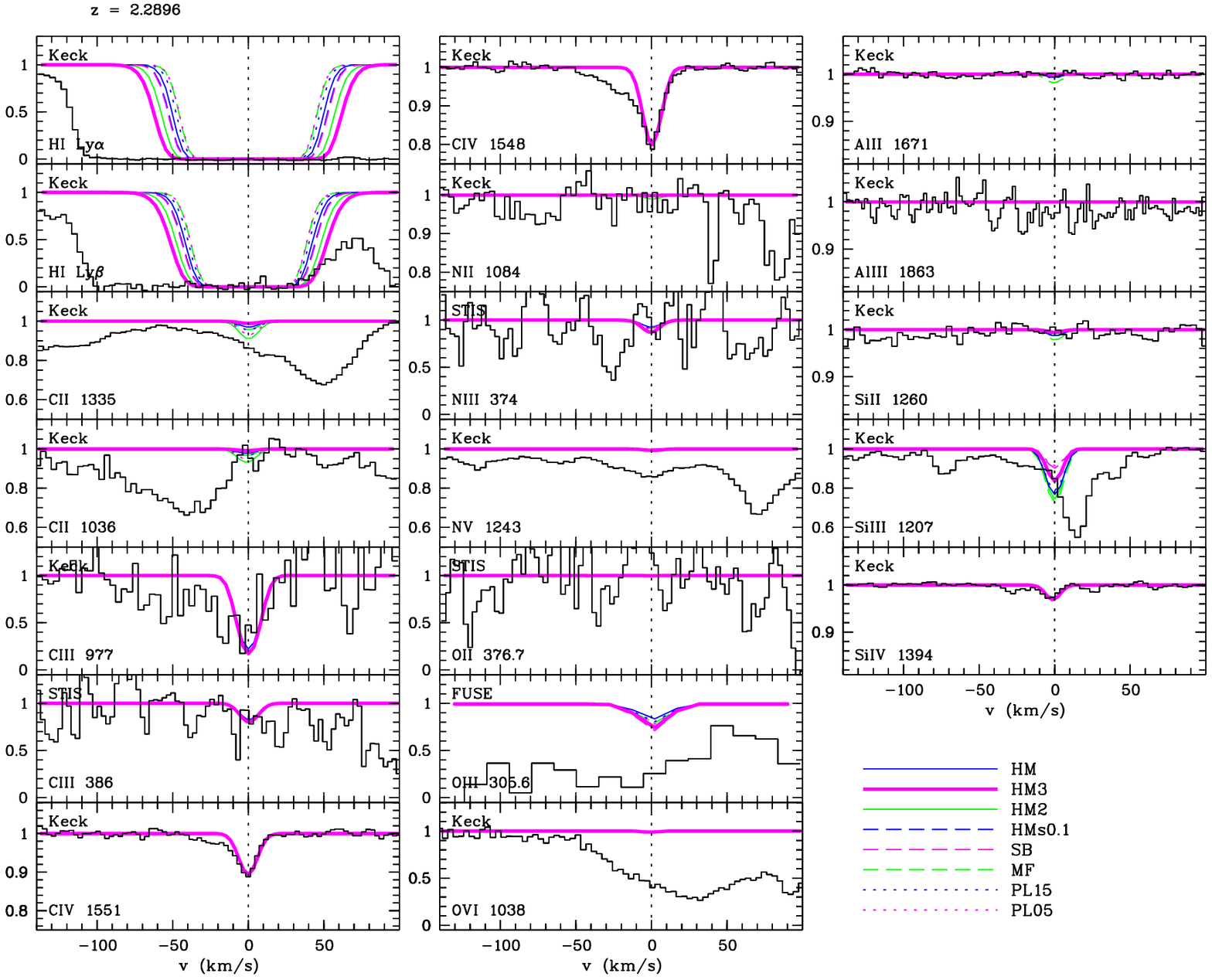}}
  \caption{Observed and modelled absorption lines of the system at $z = 2.2895$. The preferred model is HM3.
}
  \label{z2.2896fig}
\end{figure*}

\begin{figure*}
  \centering
  \resizebox{\hsize}{!}{\includegraphics[bb=35 325 545 780,clip=]{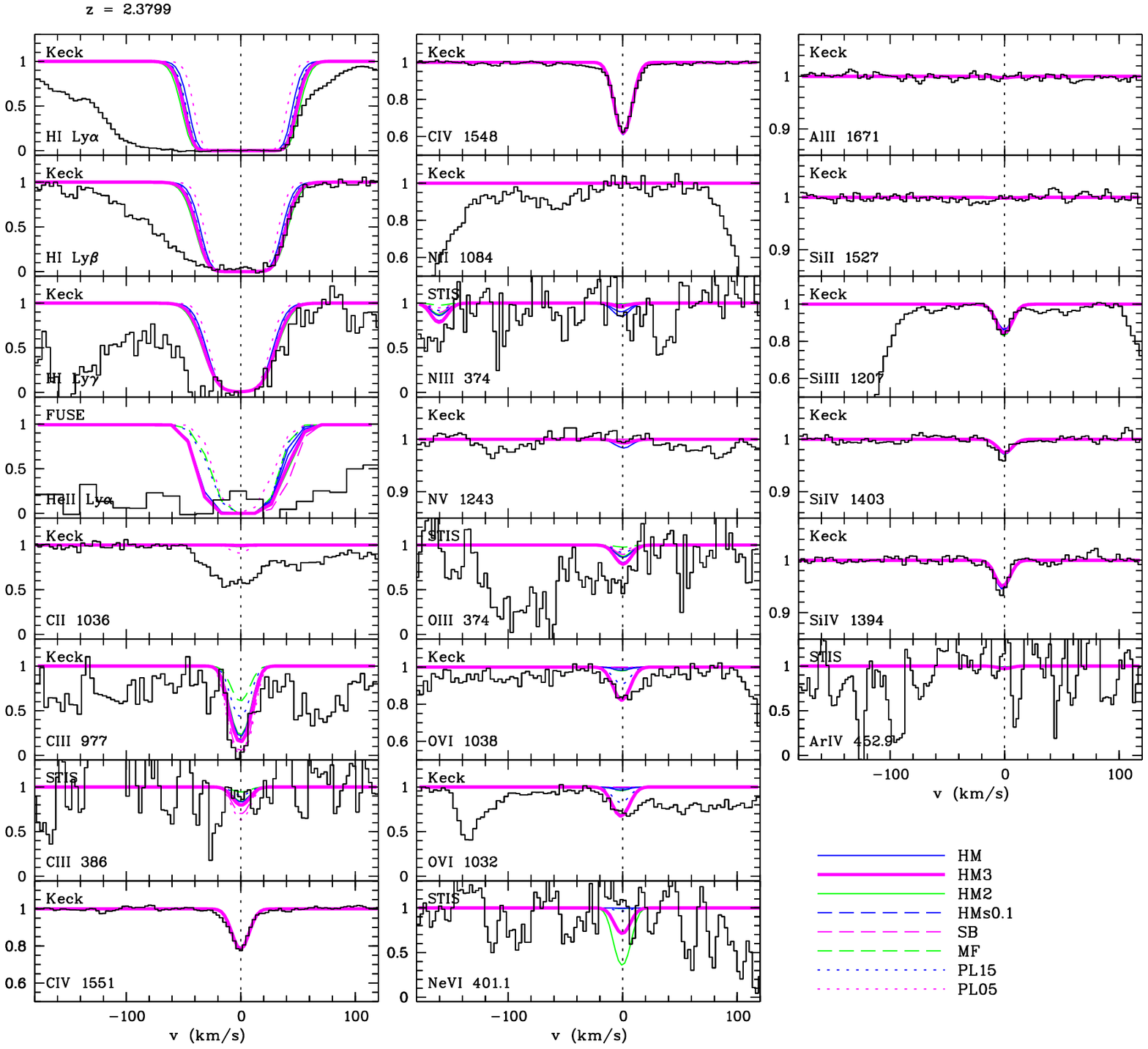}}
  \caption{Observed and modelled absorption lines of the system at $z = 2.3799$. The preferred model is HM3.
}
  \label{z2.3800fig}
\end{figure*}

\begin{figure*}
  \centering
  \resizebox{\hsize}{!}{\includegraphics[bb=35 435 545 780,clip=]{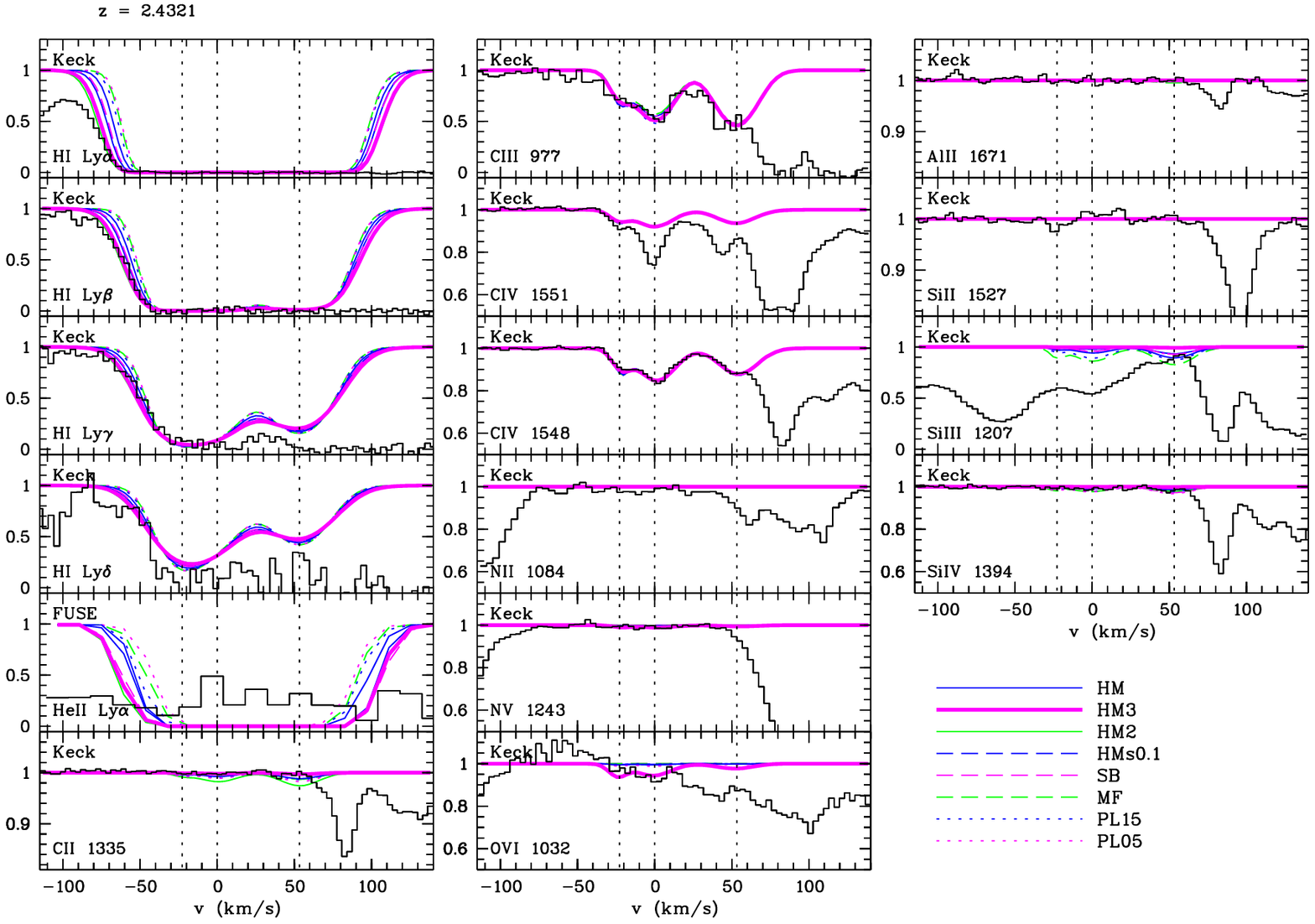}}
  \caption{Observed and modelled absorption lines of the system at $z = 2.4321$. The preferred model is HM3.
}
  \label{z2.4321fig}
\end{figure*}

\begin{figure*}
  \centering
  \resizebox{\hsize}{!}{\includegraphics[bb=35 325 545 780,clip=]{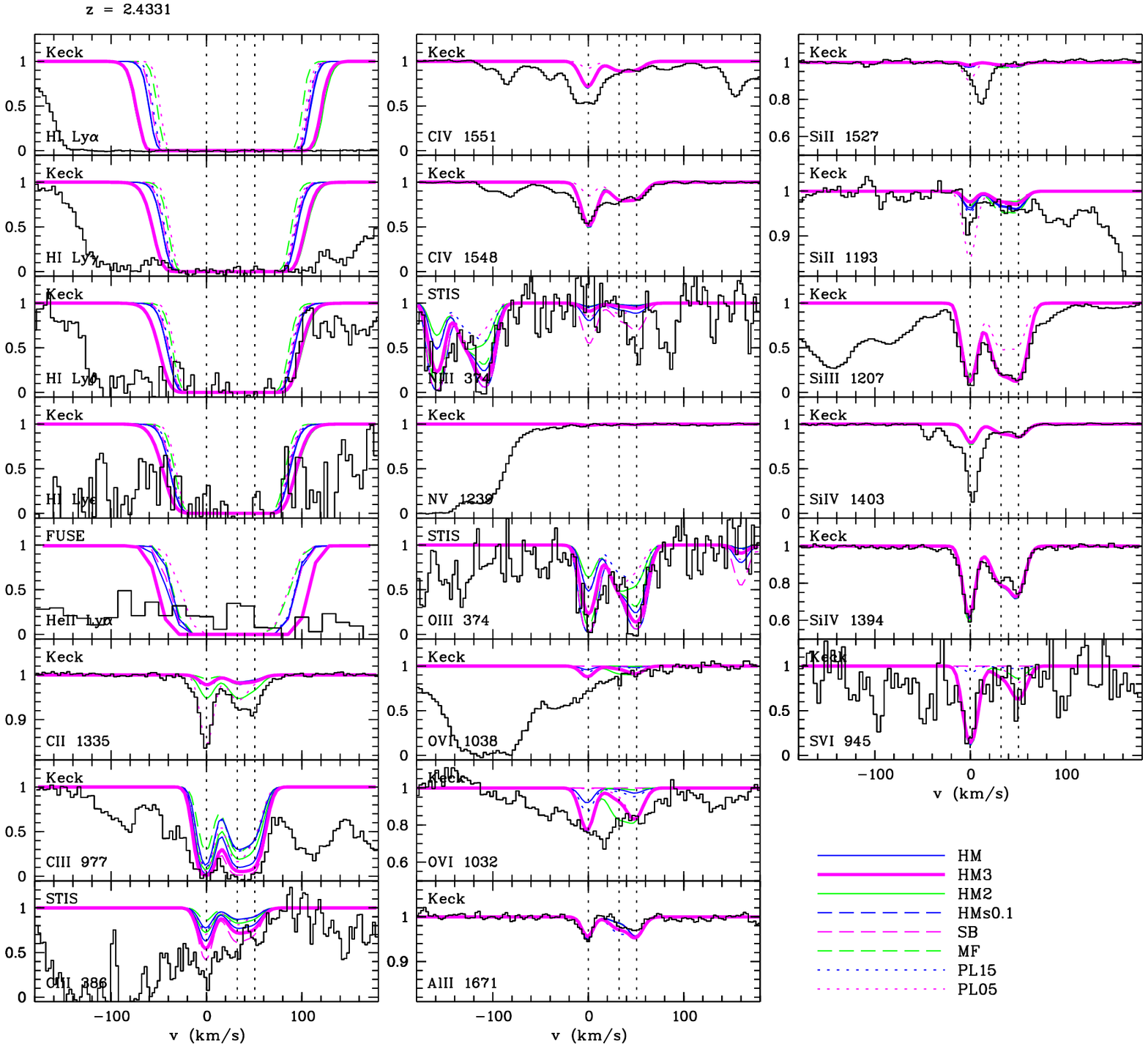}}
  \caption{Observed and modelled absorption lines of the system at $z = 2.4331$. The preferred model is HM3.
}
  \label{z2.4331fig}
\end{figure*}

\begin{figure*}
  \centering
  \resizebox{\hsize}{!}{\includegraphics[bb=35 305 545 780,clip=]{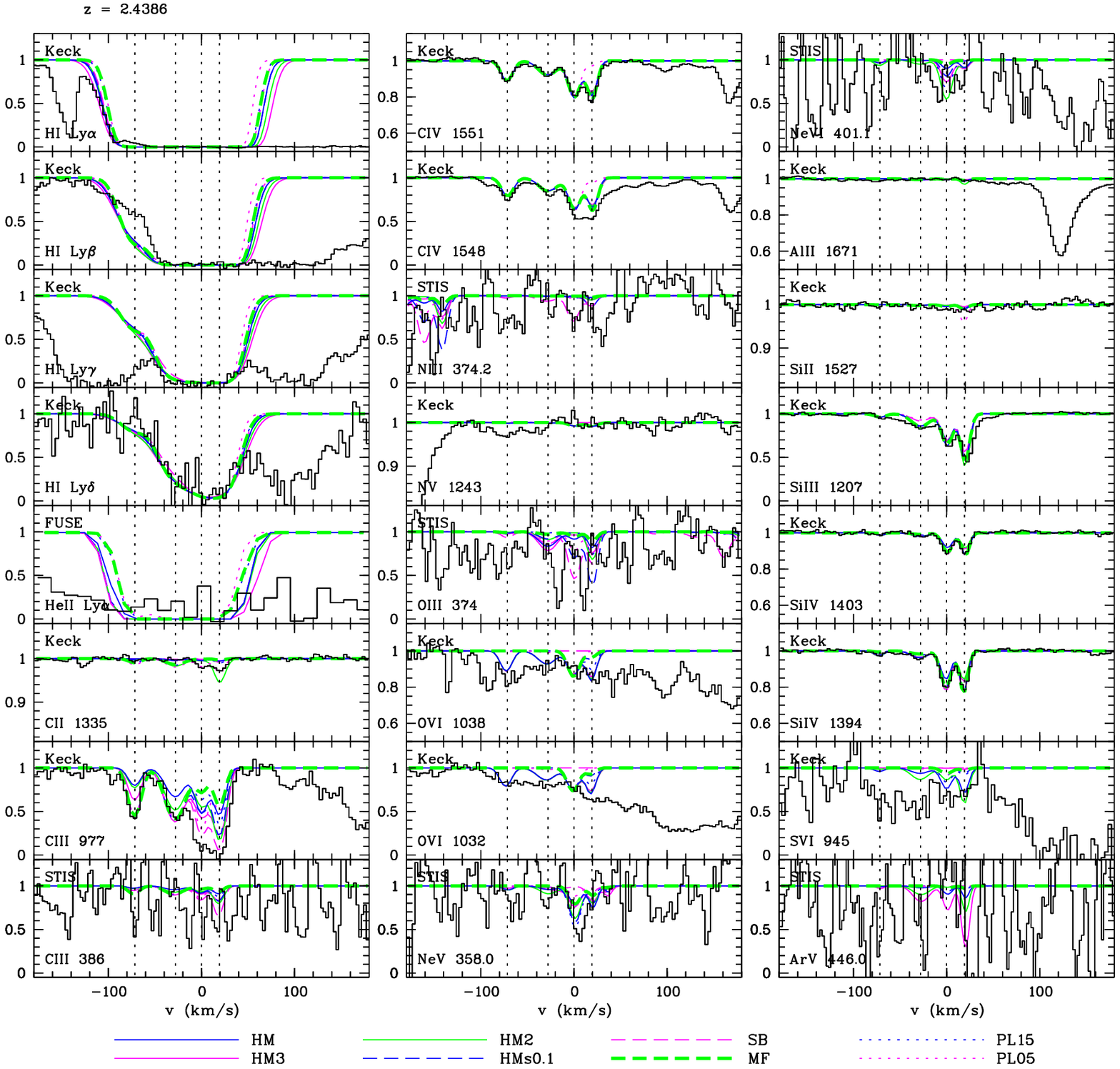}}
  \caption{Observed and modelled absorption lines of the system at $z = 2.4386$. The preferred model is MF.
}
  \label{z2.4386fig}
\end{figure*}

\clearpage

\begin{figure*}
  \centering
  \resizebox{\hsize}{!}{\includegraphics[bb=35 380 545 780,clip=]{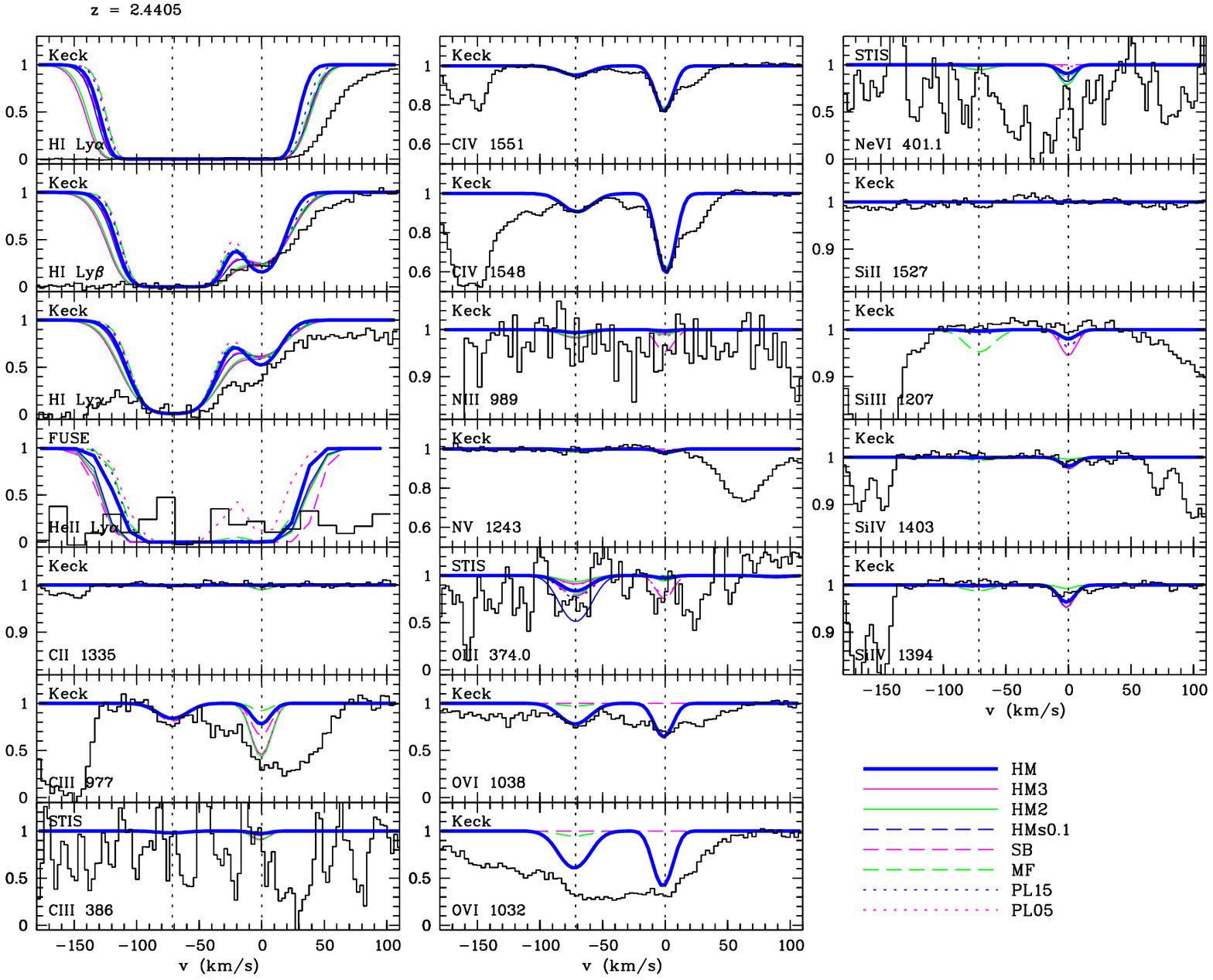}}
  \caption{Observed and modelled absorption lines of the system at $z = 2.4405$. The preferred models are HM and SB.
}
  \label{z2.4405fig}
\end{figure*}

\begin{figure*}
  \centering
  \resizebox{\hsize}{!}{\includegraphics[bb=35 410 545 780,clip=]{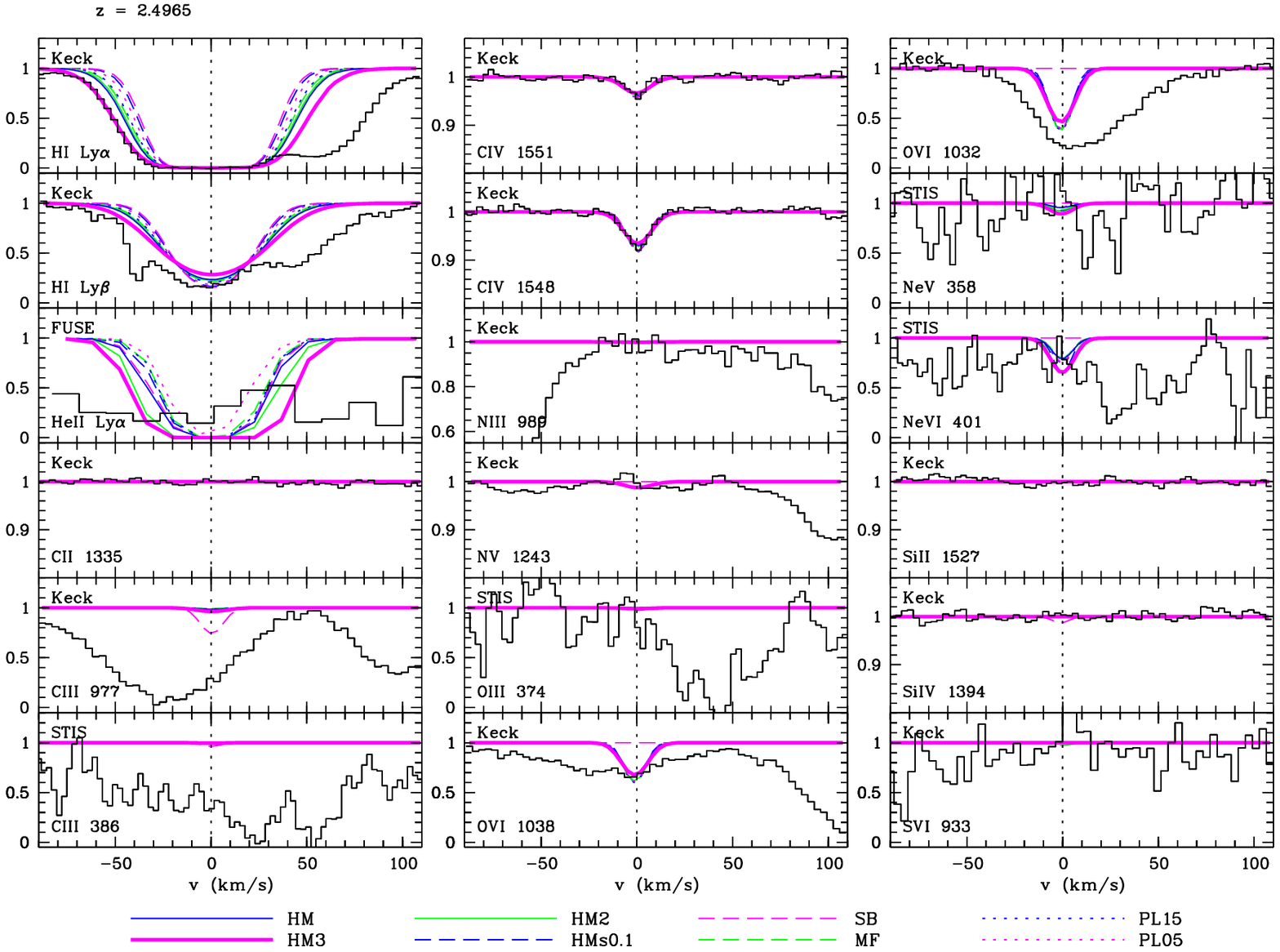}}
  \caption{Observed and modelled absorption lines of the system at $z = 2.4965$. The preferred model is HM3.
}
  \label{z2.4965fig}
\end{figure*}

\begin{figure*}
  \centering
  \resizebox{\hsize}{!}{\includegraphics[bb=35 470 545 780,clip=]{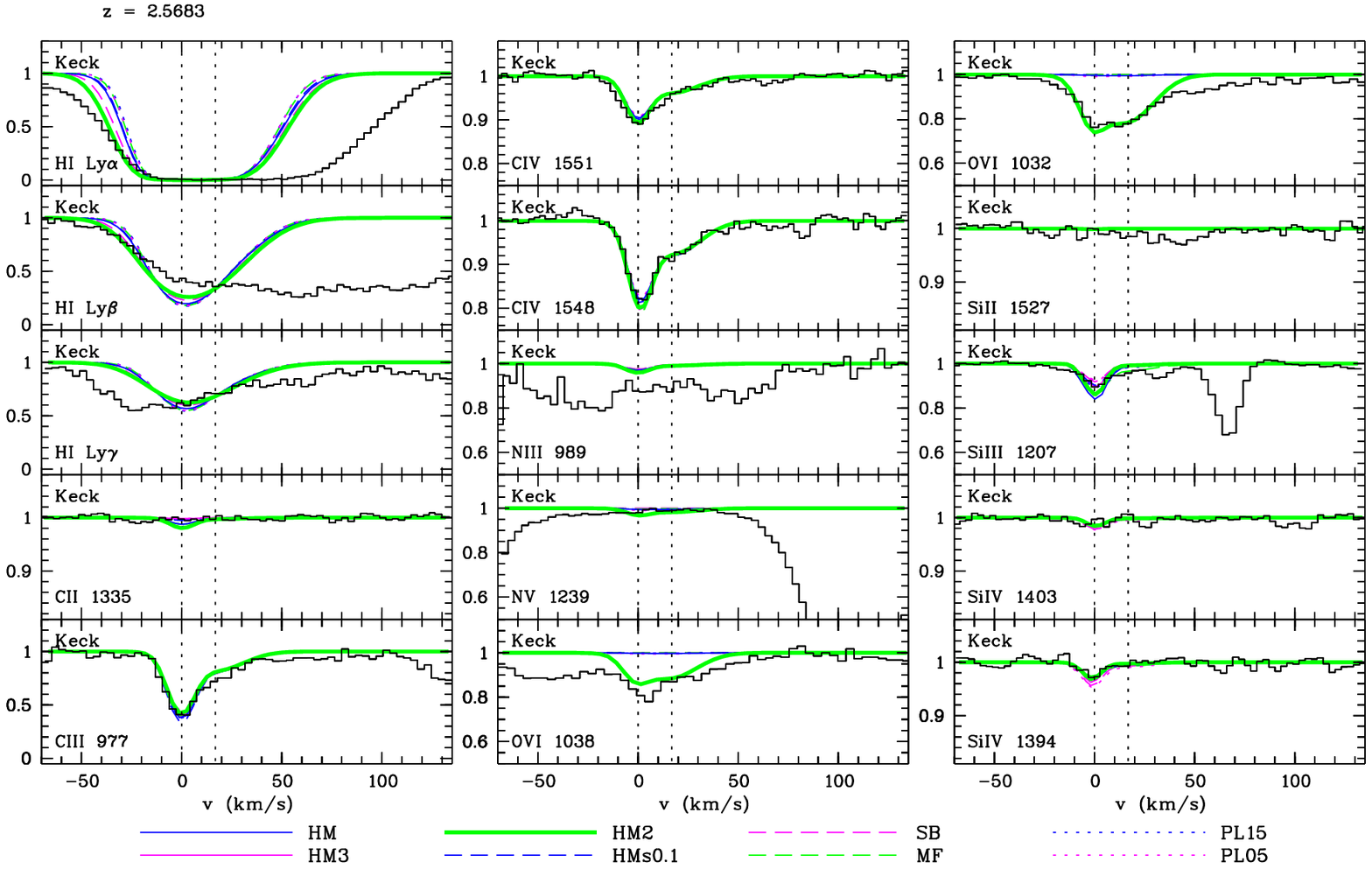}}
  \caption{Observed and modelled absorption lines of the system at $z = 2.5683$. The preferred model is HM2.
}
  \label{z2.5683fig}
\end{figure*}

\begin{figure*}
  \centering
  \resizebox{\hsize}{!}{\includegraphics[bb=35 305 545 780,clip=]{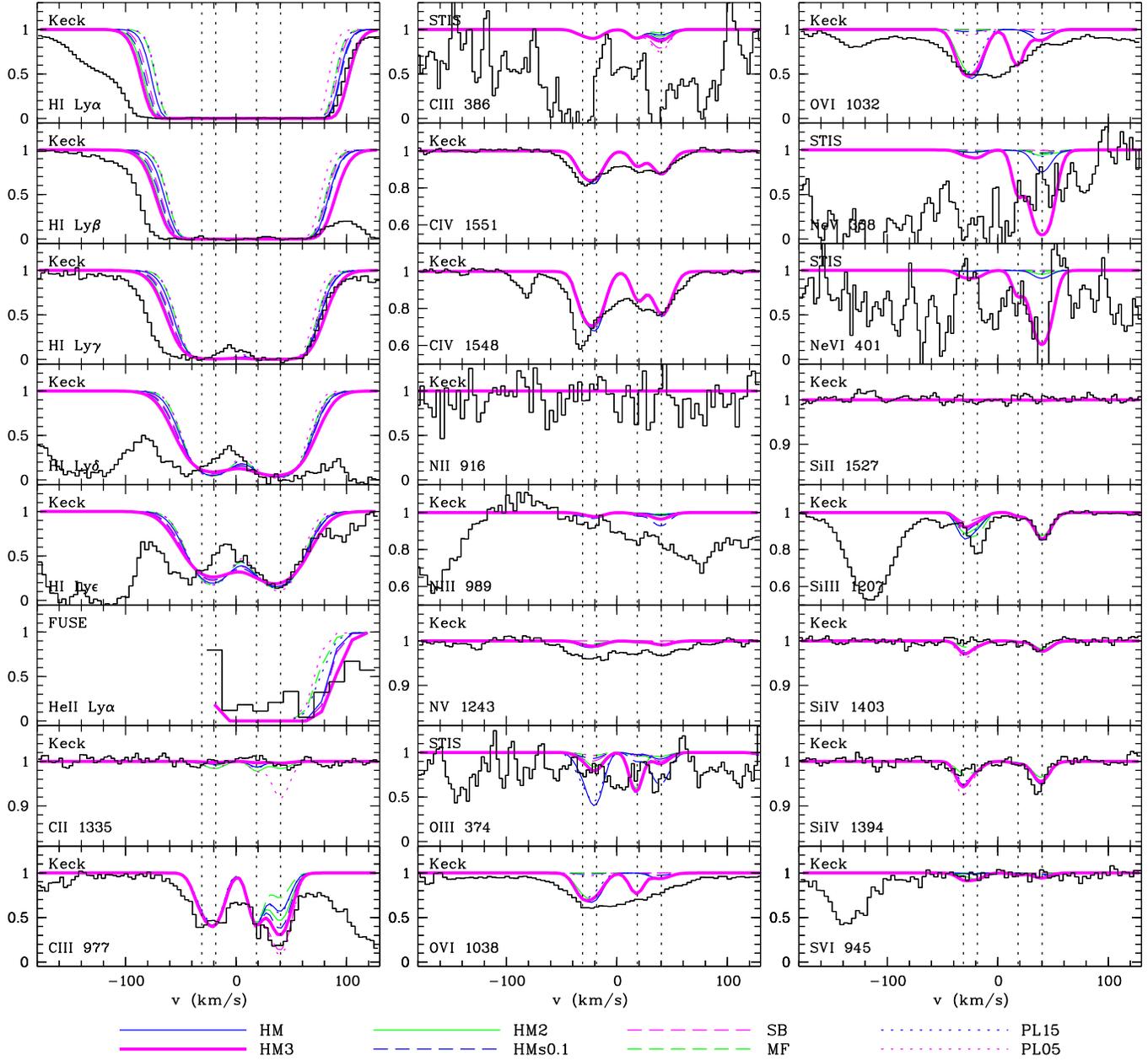}}
  \caption{Observed and modelled absorption lines of the system at $z = 2.5785$. The preferred model is HM3. The blue component of the \ion{He}{ii} profile is affected by the FUSE detector gap and therefore not observed.
}
  \label{z2.5786fig}
\end{figure*}

\end{document}